\documentclass[A4paper]{aa}  
\usepackage[english]{babel}  
\usepackage{graphicx}  
\usepackage[utf8]{inputenc}  
\usepackage{microtype}  
\usepackage{booktabs}  
\usepackage{rotating}  
\usepackage{lscape}
\usepackage{longtable}
\usepackage[squaren]{SIunits}
\usepackage{natbib}
\usepackage{amsmath}
\usepackage{amssymb}
\usepackage{pstricks}
\usepackage{txfonts}
\usepackage{tikz}
\usepackage{color}
\usepackage{hyperref}

\newcommand{\tex}{T_\mathrm{{ex}}}

\newcommand{\tbg}{T_\mathrm{{BG}}}

\newcommand{\tk}{T_\mathrm{{K}}}

\newcommand{\tb}{T_\mathrm{{B}}}

\newcommand{\pot}[1]{10^{#1}}
\newcommand{\expo}[1]{\mathrm{e}^{#1}}
\newcommand{\vlsr}{V_\mathrm{{LSR}}}

\newcommand{\cm}{\usk\centi \metre}
\newcommand{\cmtab}{\centi \metre}

\newcommand{\pc}{\usk\mathrm{pc}}

\newcommand{\kpc}{\usk\mathrm{kpc}}

\newcommand{\hii}{H\textsc{ii}}

\newcommand{\msun}{\usk\mathrm{M_\odot}} 
\newcommand{\msuntab}{\mathrm{M_\odot}} 
\newcommand{\jy}{\usk\mathrm{Jy}}

\newcommand{\lsun}{\usk\mathrm{L_\odot}}
\newcommand{\lsuntab}{\mathrm{L_\odot}}
\newcommand{\jansky}{\usk\mathrm{Jy}}
\newcommand{\janskytab}{\mathrm{Jy}}
\newcommand{\beam}{\usk\mathrm{beam}}

\newcommand{\rsun}{\usk\mathrm{R_\odot}}

\newcommand{\kms}{\usk\kilo\metre\usk\second^{-1}}
\newcommand{\kmstab}{\kilo\metre\usk\second^{-1}}
\newcommand{\kel}{\usk\kelvin}
\newcommand{\mum}{\usk\micro\metre}
\newcommand{\mm}{\usk\milli\metre}

\newcommand{\an}{CH$_{3}$CN}
\newcommand{\ma}{CH$_{3}$CCH}
\newcommand{\mt}{CH$_{3}$OH}

\newcommand{\koenigSed}{\citep{Koenig+17_aap599_139}}
\newcommand{\navareteMidJCO}{\citet{Navarete_llum_subm}}
\newcommand{\koenigSedt}{\citet{Koenig+17_aap599_139}}

\title{ATLASGAL-selected massive clumps in the inner Galaxy: \\ V. Temperature structure and evolution\thanks{Tables from \ref{tab:source_properties} to \ref{tab:fit_res_mt_hot} are only available in electronic format the CDS via anonymous ftp to \url{cdsarc.u-strasbg.fr} (130.79.128.5) or via \url{http://cdsweb.u-strasbg.fr/cgi-bin/qcat?J/A+A/}}}
\author{
A. Giannetti \inst{\ref{mpi},\ref{ira}}
\and S. Leurini \inst{\ref{mpi},\ref{oac}}
\and F. Wyrowski \inst{\ref{mpi}}
\and J. Urquhart \inst{\ref{mpi},\ref{ukent}}
\and T. Csengeri \inst{\ref{mpi}}
\and K.~M. Menten \inst{\ref{mpi}}
\and C. K\"onig \inst{\ref{mpi}}
\and R. G\"usten \inst{\ref{mpi}}
}
\institute{
Max-Planck-Institut f\"ur Radioastronomie, auf dem H\"ugel 69, D-53121, Bonn, Germany \label{mpi}
\and INAF - Istituto di Radioastronomia \& Italian ALMA Regional Centre, Via P. Gobetti 101, I-40129 Bologna, Italy\label{ira}
\and INAF-Osservatorio Astronomico di Cagliari, Via della Scienza 5, I-09047, Selargius (CA), Italy \label{oac}
\and School of Physical Sciences, University of Kent, Ingram Building, Canterbury, Kent CT2\,7NH,\,UK \label{ukent}
}

\abstract{
Observational identification of a solid evolutionary sequence for high-mass star-forming regions is still missing. Spectroscopic observations give the opportunity to test possible schemes and connect the phases identified to physical processes.
}
{
We aim to use the progressive heating of the gas caused by the feedback of high-mass young stellar objects to prove the statistical validity of the most common schemes used to observationally define an evolutionary sequence for high-mass clumps, and characterise the sensitivity of different tracers to this process.
}
{
From the spectroscopic follow-ups carried out towards submillimeter continuum (dust) emission-selected massive clumps (the ATLASGAL TOP100 sample) with the IRAM 30\,m, Mopra, and APEX telescopes between $84\usk\giga\hertz$ and $365\usk\giga\hertz$, we selected several multiplets of \an, \ma, and \mt\ emission lines to derive and compare the physical properties of the gas in the clumps along the evolutionary sequence, fitting simultaneously the large number of lines that these molecules have in the observed band.
Our findings are compared with results obtained from optically thin CO isotopologues, dust, and ammonia from previous studies on the same sample.
}
{
The chemical properties of each species have a major role on the measured physical properties. Low temperatures are traced by ammonia, methanol, and CO (in the early phases), the warm and dense envelope can be probed with \an, \ma, and, in evolved sources where CO is abundant in the gas phase, via its optically thin isotopologues. \mt\ and \an\ are also abundant in the hot cores, and we suggest that their high-excitation transitions are good tools to study the kinematics in the hot gas associated with the inner envelope surrounding the young stellar objects that these clumps are hosting.
All tracers show, to different degrees according to their properties, progressive warming with evolution. The relation between gas temperature and the luminosity-to-mass ($L/M$) ratio is reproduced by a simple toy model of a spherical, internally heated clump. 
}
{
The evolutionary sequence defined for the clumps is statistically valid and we could identify the physical processes dominating in different intervals of $L/M$. For $L/M\lesssim2\lsun\msun^{-1}$ a large quantity of the gas is still accumulated and compressed at the bottom of the potential well. Between $2\lsun\msun^{-1}\lesssim L/M \lesssim 40\lsun\msun^{-1}$ the young stellar objects gain mass and increase in luminosity; the first hot cores hosting intermediate- or high-mass ZAMS stars appear around $L/M\sim10\lsun\msun^{-1}$. Finally, for $L/M\gtrsim40\lsun\msun^{-1}$ \hii\ regions become common, showing that dissipation of the parental clump dominates.
}

\offprints{A. Giannetti (agiannetti@mpifr-bonn.mpg.de)}

\begin{document}  
    \maketitle

    \section{Introduction}\label{sec:intro}
        High-mass stars ($M\gtrsim10\msun$) are the main drivers of physics and chemistry of the interstellar medium. Through powerful molecular outflows, winds, strong UV radiation, as well as supernova explosions, they influence the environment, regulate star formation, and dominate the energy budget, from their immediate surroundings to entire galaxies.
        
        Despite the prominent role that massive stars have for the evolution of their host galaxies \citep{KennicuttEvans12_araa50_531}, the physical processes connected to their formation are still far from being clear. This limits the advancements in star formation and galaxy evolution research \citep[e.g.][]{Scannapieco+12_mnras423_1726}. 
        
        The observational characterisation of the initial conditions (temperature, density, magnetic field, degree of fragmentation, ionisation fraction, etc.) of the process and how they change with time due to the feedback of the newly-formed stars constitutes a fundamental step in this direction. Such a study requires the identification of an evolutionary sequence.
        The process of high-mass star formation can be divided into four broad phases, described in \citet{ZinneckerYorke07_ARAA45_481}:
        
        \begin{enumerate}
            \item Compression: cold, dense cores and filaments are formed through gravoturbulent fragmentation \citep{MacLowKlessen04_RvMP76_125}, marking the location where star-formation will take place.
            \item Collapse: as gravity takes over, these dense pockets of gas collapse, forming protostellar embryos that start warming up their local environment.
            \item Accretion: the protostars continue to gain mass. Numerical simulations show that, due to the high-accretion rates, these objects become very large \citep[$R\sim100\rsun$,][]{HosokawaOmukai09_apj691_823}, until they reach $M\sim 10\msun$. Then the protostar starts contracting, and begins burning hydrogen before reaching its final mass. 
            \item Disruption: The combined mechanical and radiative feedback from massive stars, during and at the end of their lives, progressively ionise, dissociate, and ultimately completely disperse the parent molecular cloud.
        \end{enumerate}
        
        From an observational perspective, two main schemes are used to separate evolutionary stages, based on the infrared- (IR) and radio-continuum properties, and on the luminosity-to-mass ratio, respectively. The former classification rests on the idea that the dust, whose emission is initially not visible at wavelengths $<100\mum$ due to the low temperatures expected in quiescent sources, is warmed-up by the feedback from young stellar objects (YSOs) and becomes detectable in more evolved stages. With time, the YSOs become more luminous and ionise more and more material, which is detected in radio-continuum emission as an \hii\ region. Alternatively, the $L/M$ ratio was successfully used in the low-mass regime to define an evolutionary sequence \citep{Saraceno+96_aap309_827}; \citet{Molinari+08_aap481_345} proposed to use the same indicator for high-mass sources. Two phases are identified: the accelerated accretion phase, in which the YSO accretes mass and increases in $L$, leaving the mass of the parent clump nearly unchanged, and the clump dispersal phase, where the YSO has reached its final mass and starts dissociating the nearby gas. Therefore $L/M$ should increase with time. The first method is purely phenomenological and can be severely influenced by extinction and different distances, whereas the second borrows an efficient distance-independent indicator of evolution from the low-mass regime and applies it to a completely different scale. The robustness of the evolutionary sequence defined must therefore be independently tested, and a link between the observational and theoretical phases is needed.
        
        One of the expected effects of the copious amount of energetic photons emitted by high-mass stars, in combination with their mechanical feedback from outflows and winds, is the progressive heating of the surrounding gas, both directly by the photons and through collisions with dust grains. Hot molecular cores (or hot cores) are the direct result of this process \citep{Cesaroni05_IAUS227_59}. The warm-up is not instantaneous, with important consequences for the molecular abundances \citep{Viti+04_mnras354_1141,Garrod+08_apj682_283,GarrodWeaver13_ChRv113_8939}. The progressive increase in temperature can be used to statistically validate the evolutionary schemes discussed in the previous paragraph.
        
        Previous work made use of ammonia and dust continuum emission to investigate variations in the physical conditions of the clumps as a function of evolution, finding evidence for only a marginal increase in the average temperature and central density \citep[e.g.][]{Wienen+12_aap544_146,Giannetti+13_aa556_16,Elia+13_apj772_45,Koenig+17_aap599_139}. 
        The most dramatic changes happen in the dense inner layers of the clumps, and molecules that selectively trace this gas are likely to be more sensitive to changes in the physical conditions, allowing a clear distinction between the phases.
        \citet{Molinari+16_apjl826_8} observed \ma\ $J=12\rightarrow11$ in a sample of 51 high-mass sources spanning $0.01\lsun\msun^{-1}\lesssim L/M \lesssim 200\lsun\msun^{-1}$. The authors show that the temperature increases by a factor $\sim2$ from $L/M\sim1\lsun\msun^{-1}$ to $L/M\sim200\lsun\msun^{-1}$, confirming that this molecule is sensitive to the warm-up. 
		However, of the 23 detected sources, the vast majority have temperatures in the range $30-45\kel$, and the statistics are poor for high $L/M$ ratios, where the change in temperature is evident. The evolutionary trend must therefore be confirmed with a larger sample and a more extensive set of molecules and transitions, probing a larger range in temperature.
		
        In this work we compare the physical properties derived with commonly-used temperature tracers for a statistically significant sample of high-mass star-forming regions to test the statistical validity of the evolutionary sequence defined by the IR and radio-continuum properties of the source and by $L/M$, identify the physical process (compression, collapse, accretion, disruption) dominating each phase, and determine the best tracers and transitions to follow the warm-up process, and characterise the emitting region of different species.

        \begin{table*}
        \caption[]{Summary of the observations.}\label{tab:table_obs}
            \begin{center}
                \begin{tabular}{lrccccc}
                    \hline
                    \hline
                    Line series         &Frequency                 &Beam FWHM                    &$\eta_{MB}$ & $\Delta V$    &r.m.s. noise\tablefootmark{a} & Telescope     \\
                                        &\multicolumn{1}{c}{[GHz]} &\multicolumn{1}{c}{[\arcsec]}&            &[$\kmstab$]    &[K]                           &               \\
                    \hline                                                                                                                                   
                    CH$_3$CN\,(5--4)    &92                        &26\farcs7, 35\arcsec         & 0.81,0.49  & 0.64,0.88     & 0.04,0.04                    &IRAM-30\,m, MOPRA\\
                    CH$_3$CN\,(6--5)    &110                       &22\farcs4                    & 0.81       & 0.53          & 0.06                         &IRAM-30\,m       \\
                    CH$_3$CN\,(19--18)  &349                       &17\farcs9                    & 0.73       & 1.00          & 0.03                         &APEX           \\
                    CH$_3$CCH\,(5--4)   &86                        &28\farcs6, 35\arcsec         & 0.81,0.49  & 0.69,0.95     & 0.04,0.05                    &IRAM-30\,m, MOPRA\\
                    CH$_3$CCH\,(6--5)   &103                       &23\farcs9                    & 0.81       & 0.57          & 0.05                         &IRAM-30\,m       \\
                    CH$_3$CCH\,(20--19) &342                       &18\farcs2                    & 0.73       & 0.53          & 0.07                         &APEX           \\
                    CH$_3$OH\,(7--6)    &338                       &18\farcs5                    & 0.73       & 1.00          & 0.05                         &APEX           \\
                    \hline
                \end{tabular}
            \end{center}
            \tablefoot{\tablefoottext{a}{The temperatures are reported on the main-beam temperature scale.}
            }
        \end{table*}

    \section{The sample}\label{sec:sample}
    
        The inner $\pm60^\circ$ of the Galactic plane hosts most of the massive stars and clusters in our Galaxy as well as most of its molecular gas \citep{Urquhart+14_mnras437_1791}. The APEX Telescope Large Area Survey of the Galaxy \citep[ATLASGAL; ][]{Schuller+09_aap504_415} is the first complete survey at $870\mum$ of the inner Galactic plane, covering $420$ square degrees in the region $-80\degree \leq l \leq -60\degree$, between $-2\degree \leq b \leq 1\degree$ and $|l| \leq 60\degree$ with $|b| \leq 1.5\degree$. 
        ATLASGAL delivered the most comprehensive inventory of objects at $870\mum$ within this region at an unprecedented sensitivity ($50-70\usk\milli\janskytab\beam^{-1}$) and angular resolution ($\sim19.2\arcsecond$); the survey is complete at $99\%$ above $0.3-0.4\usk\jansky\beam^{-1}$, allowing to detect all cold and massive clumps with $M\gtrsim1000\msun$ within $20\kpc$ \citep{Contreras+13_aa549_45,Csengeri+14_aa565_75,Urquhart+14_aap568_41}, and thus providing a full census of high-mass star-forming regions in the inner Galaxy.
        
        Taking advantage of the unbiased nature of this survey we selected a flux limited sample, the TOP100 \citep{Giannetti+14_aa570_65}, with additional infrared criteria to avoid biasing the sample towards the hottest and most luminous sources. 
        This sample includes 110 sources and consists almost entirely of clumps that have the potential of forming massive stars, following the criteria of \citet{KauffmannPillai10_apjl723_7} and \citet{Urquhart+13_mnras431_1752} \citep{Giannetti+14_aa570_65,Koenig+17_aap599_139}. The classification for the TOP100 was recently refined in the work by \koenigSedt, separating the sources into four categories, depending on their IR and radio-continuum properties as follows:
        \begin{itemize}
            \item Seventy-micron weak sources (70w; 16 objects) -- sources that are not detected at $24\mum$ and show no clear compact $70\mum$ emission or are seen in absorption at this wavelength, that should represent the earliest stages of the high-mass star-formation process;
            \item Infrared-weak clumps (IRw; 33 objects)-- clumps associated with a compact $70\mum$ object, that are not detected in emission in MIPSGAL \citep{Carey+09_pasp121_76} at $24\mum$ or that are associated with a weak IR source, with a flux below $2.6\jy$ corresponding to a B3 star at $4\kpc$ \citep{Heyer+16_aap588_29};
            \item Infrared-bright objects (IRb; 36 objects) -- objects associated with a strong IR source, with a $24\mum$ flux above $2.6\jy$. If the source is saturated in MIPSGAL we use MSX \citep{Price+01_aj121_2819} to determine the source flux;
            \item Compact \hii\ regions (\hii; 25 objects) -- IR-bright clumps also showing radio-continuum emission at $\sim5$ or $9\usk\giga\hertz$ in the CORNISH- \citep{Hoare+12_pasp124_939,Purcell+13_apjs205_1}, the RMS surveys \citep{Urquhart+08_aspc387_381,Urquhart+09_aap501_539}, or targeted observations towards methanol masers \citep{Walsh+98_mnras301_640}.
        \end{itemize}
        The association radius used for the submm, IR, and radio-continuum sources is $10\arcsec$, corresponding to the ATLASGAL HWHM and in agreement with the angular separation between high-mass star-formation indicators \citep[e.g.][]{Urquhart+13_mnras431_1752,Urquhart+13_mnras435_400}.
        Within each subsample, the sources are among the brightest ATLASGAL clumps in their class. 
        Comparing the distance and mass distributions of the different evolutionary classes, we find that no bias is present for these two parameters; furthermore, the TOP100 includes sources with widely different properties in terms of infrared- and radio-continuum emission, spanning four orders of magnitude in $L/M$ \koenigSed.{}
        These characteristics make the TOP100 a statistically significant and representative sample of high-mass clumps in the inner Galaxy, and an ideal sample for our purpose to test the evolutionary schemes described in Sect.~\ref{sec:intro}. 
        
        The present paper is part of a series investigating the process of high-mass star formation in a broad context.
        In addition to the work by \koenigSedt, which addresses the dust continuum emission and the total luminosity of the clumps, obtaining the new classification and confirming that the vast majority of the clumps in the sample have the ability to form high-mass stars, we use spectroscopic follow-up observations to characterise the properties of the molecular- and ionised gas.
        In \citet{Giannetti+14_aa570_65} we study the CO content of the clumps in the TOP100, showing that the abundance of CO is lower than expected in the cold and dense gas typical of clumps in early stages of evolution, and increases to canonical values in more evolved sources, closely mimicking the behaviours of low-mass cores. The TOP100 sample is included in the work of \citet{Kim+17_arXiv} as well, where mm-recombination lines are used to characterise the ionised gas and identify potentially young compact \hii\ regions.
        Mid-$J$ CO lines, observed with CHAMP$^{+}$ at APEX  in a region of $1^{\prime}\times1^{\prime}$ around the clumps of the TOP100, are analysed in \navareteMidJCO\ comparing their emission to warm dust, looking for evolutionary trends, and investigating how those commonly found in the low-mass regime change when considering high-mass clumps.
        Finally, the 36 sources in the first quadrant were studied by \citet{Csengeri+16_aap586_149}, focussing on the emission from the outflow tracer SiO and its excitation.

    \section{Selection of the tracers}\label{sec:selection_tracers}
    
            To reveal the progressive warm-up of the gas, we needed to rely on efficient thermometers. The selected species should be sensitive to a broad range of physical conditions (i.e. a large number of lines with widely different excitations must be observed), should be abundant, and should have different chemical properties. The latter strongly influences the emitting region of a species, substantially affecting the measured temperature. Using multiple tracers is therefore fundamental to understand the importance of feedback, to characterise the volume of gas that different species probe, and identify those more sensitive to the warm-up process, that is, those with the highest potential for the definition of an evolutionary sequence.
            
            We selected three species, methyl acetylene (\ma), acetonitrile (\an) and methanol (\mt). The first two are symmetric top molecules, among the most efficient thermometers available. 
            For each total angular momentum quantum number, $J$, they have groups of rotational transitions with different $K$ quantum numbers that span a wide range in energy above the ground state, but are close together in frequency.
            Because $K$ describes the projection of the angular momentum along the symmetry axis, the lowest energy state for each $K$-ladder is the one with $J=K$, and the energy of the level increases with increasing $K$.
            The group of transitions with $J\rightarrow J-1, \Delta K = 0$ have increasing energy for higher values of $K$ \citep[see e.g.][]{Kuiper+84_apj276_211}. 
            The dipole moment for these species is parallel to the symmetry axis, therefore radiative transitions with $\Delta K \neq 0$ are forbidden: different $K$-ladders are connected only through collisions, and this makes their ratio sensitive to the kinetic temperature \citep{Bergin+94_apj431_674}. 
            The relative intensity of $J\rightarrow J-1$ groups between different $J$ levels, on the other hand, can potentially depend on density, too. The great potential of these molecules resides, therefore, in the separation of temperature and density effects in their excitation.
        
			\medskip
			
            Methyl acetylene has a small dipole moment of $0.7804~\mathrm{D}$ \citep{MuenterLaurie66_JChPh45_855}. Several works have shown that the lower $J$ lines of this molecule are thermalised for densities of molecular hydrogen $\sim\pot{4}\cm^{-3}$ \citep{Askne+84_aap130_311,Bergin+94_apj431_674}. Because of these properties, \ma\ is commonly used as a thermometer: it is found to have an extended emitting region and to trace relatively cold gas \citep[e.g.][]{Fontani+02_aa389_603, Bisschop+07_aap465_913}. 

            \medskip

            CH$_{3}$CN has a larger dipole moment \citep[$3.922~\mathrm{D}$,][]{Gadhi+95_JChimPhys92_1984}, and is therefore harder to thermalise. Several works discuss the use of this molecule as a temperature tracer in astrophysical environments \citep[e.g.][]{Askne+84_aap130_311,Bergin+94_apj431_674}. Acetonitrile is the prototypical hot core tracer; indeed it is classified solely as a `hot' species by \citet{Bisschop+07_aap465_913} and it is one of the first choices to derive the gas properties in the hot gas therein \citep[e.g.][]{Remijan+04_apj606_917,Sanna+14_aap565_34}. Chemical models of star-forming regions, however, predict two abundance peaks for this molecule \citep{Garrod+08_apj682_283}: the early-time peak follows the evaporation of HCN from grains, from which CH$_{3}$CN is formed in the gas phase through 
            \begin{equation}
                \mathrm{HCN} + \mathrm{CH_{3}} \rightarrow \mathrm{CH_{3}CN} + \mathrm{H^{+}};
            \end{equation}
            the late-time peak, on the other hand, is due to direct evaporation of acetonitrile from the dust grains for temperatures in excess of $\sim80-100\kel$, leading to abundances $\sim2$ orders of magnitude higher than in the cold gas, explaining the efficacy of this species in tracing hot material.
                
            \medskip

            Methanol is widespread in the ISM and it is one of the basic blocks for the formation of more complex, biologically relevant molecules \citep[e.g.][]{GarrodWeaver13_ChRv113_8939}. CH$_{3}$OH is a slightly-asymmetric rotor, that can be effectively used as a probe of temperature and density in interstellar clouds \citep{Leurini+04_aap422_573,Leurini+07_aap466_215}.
            This species is formed through surface chemistry by successive hydrogenation of CO, and it is found to be one of the major constituents of ice mantles on dust grains \citep{Gibb+04_apjs151_35}.
            Its strong binding energy to the grain surface, due to hydrogen bonding, determine the co-evaporation of CH$_{3}$OH with water at $\sim110\kel$ and a strong enhancement of the abundance in hot cores. The abundance in hot gas is consistent with that observed in ices in cold protostellar envelopes, indicating that the molecules there are directly sublimated from the mantles \citep{GarrodWeaver13_ChRv113_8939}. Methanol is found in cold gas as well \citep{Smith+04_mnras350_323}, which poses a long-standing problem in understanding how this species is released into the gas phase at such low temperatures. Proposed mechanisms include, for example, spot heating, shocks, and exothermic reactions on the grain surface \citep[e.g.][and references therein]{Roberts+07_mnras382_733}.
                
            \begin{table*}
                \caption{Spectral ranges used for the fit.}\label{tab:spectral_ranges}
                \centering
                \begin{tabular}{lccc}
					\hline
					\hline
                    Species               & \an                & \ma                & \mt                \\
                    \hline
                    Spectral ranges [MHz] & 91920.0-91995.0    & 85425.0-85480.0    & 336847.3-337050.0  \\
                                          & 110305.0-110400.0  & 102510.0-102555.0  & 337150.0-337320.0  \\
                                          & 348560.0-349037.7  & 341553.0-341760.0  & 337453.4-337575.7  \\
                                          & 349080.1-349477.3  &                    & 337590.1-337761.5  \\
                                          &                    &                    & 337850.0-338055.6  \\
                                          &                    &                    & 338101.0-338832.0  \\
                    \hline
                \end{tabular}
            \end{table*}
            
            In the following we shall compare the physical properties traced by these species to search for evidence of progressive warm-up in the evolutionary sequence defined by the infrared and radio-continuum properties of the clump and its $L/M$ ratio. In addition we will characterise the volume of gas that these molecules are tracing, and we will compare the results with other tracers from previous works on the same sample: submm dust emission by \koenigSedt, NH$_{3}$ (1,1) and (2,2) inversion transitions from \citet{Wienen+12_aap544_146}, and CO isotopologues from \citet{Giannetti+14_aa570_65}.

    \section{Observations}\label{sec:obs}
    
        We observed multiple lines of \an, \ma, and \mt, with the APEX, IRAM-30\,m, and Mopra telescopes, between $85\usk\giga\hertz$ and $350\usk\giga\hertz$. Table~\ref{tab:table_obs} gives an overview of these observations. All line temperatures in this work are reported in the main-beam temperature scale, according to the efficiencies listed in the table. 

		\begin{figure}
			\includegraphics[width=\columnwidth]{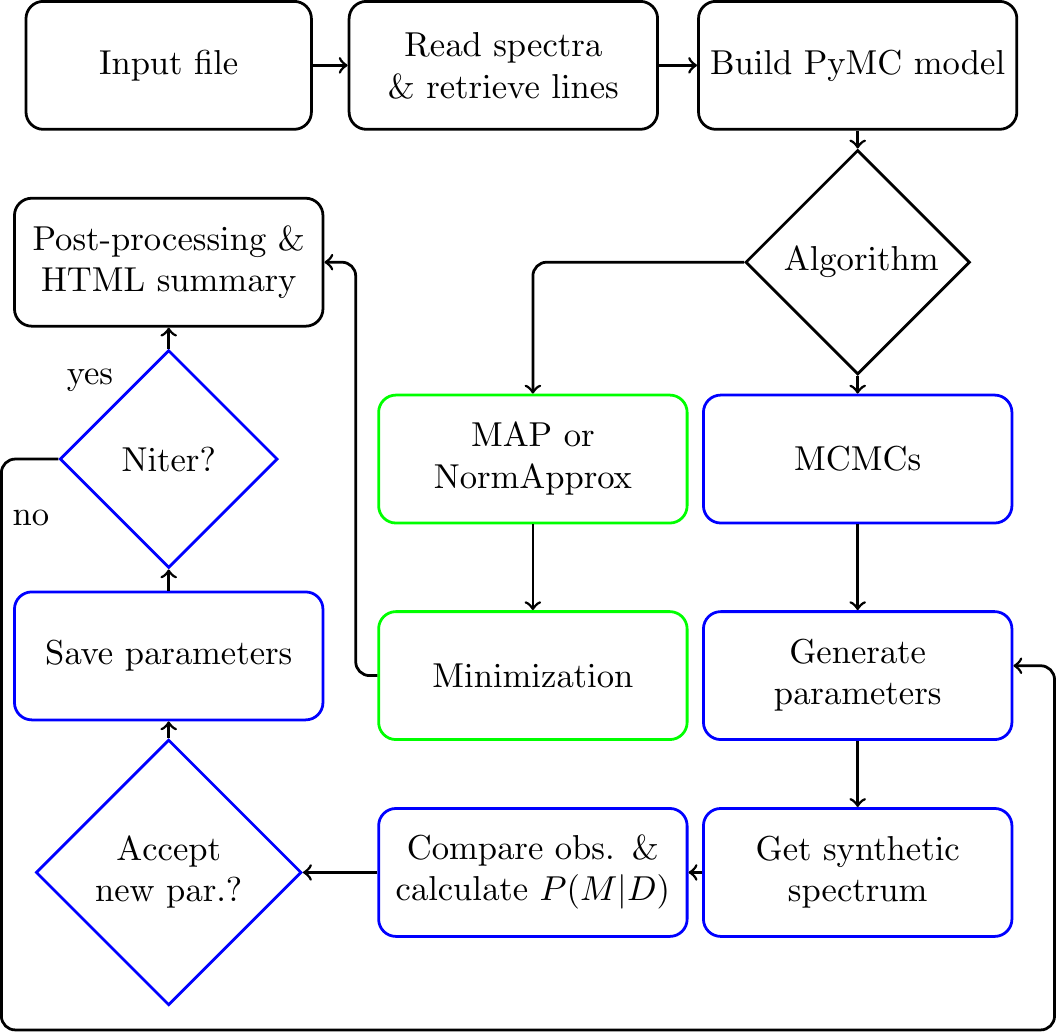}
			\caption{Simple schematic flow-chart of MCWeeds. The steps performed when using MCMCs are highlighted in blue and those for MAP or normal approximation in green. A more detailed description is given in the text.}\label{fig:mcweeds_flow_chart}
		\end{figure}
        \subsection{IRAM-30\,m}\label{iram}
            Thirty-six  of the TOP100 sources are observable from the northern sky and were observed in the 3-mm band with the IRAM-30\,m telescope.   
            These observations were previously presented by
            \citet{Csengeri+16_aap586_149} and we refer the reader to their paper for a detailed discussion. The authors report a calibration uncertainty $\lesssim10\%$. These observations were carried out between the 8th and 11th of April 2011. The velocity resolution of the lines reported in this paper is listed in Table~\ref{tab:table_obs}.

        \subsection{Mopra}\label{mopra}
            The observations were carried out with the ATNF Mopra 22\,m telescope in May 2008 as pointed observations in position switching mode. The HEMT receiver was tuned to 89.3 GHz and the MOPS spectrometer was used in broadband mode to cover the frequency range from 85.2 -- 93.4\,GHz with $\sim0.9 \kms$ velocity resolution. The beam size at these frequencies is $\sim35\arcsec$. The typical observing time per source was 15\,min. 
            The system temperature was 200\,K on average. Pointing was checked every hour with line pointings on SiO masers. Each day G327 and M17 were observed as a check in the line setup chosen for the survey. The calibration uncertainty for this dataset is $20\%$.

            The processing of the spectra resulting from the on-off observing mode, the time and polarisation averaging, and baseline subtraction were done with the ASAP package. The data was then exported to the CLASS spectral line analysis software of the GILDAS package for further processing. 
            
        \subsection{APEX}\label{apex}

            The FLASH$^+$ \citep{Klein+14_TransThzSciTech4_588} heterodyne receiver on the APEX 12\,m telescope was used to observe the ATLASGAL TOP100 sample in CH$_3$CN\,(19--18), CH$_3$CCH\,(20--19), and CH$_3$OH\,(7--6) on June 15th and 24th, August 11th, 23rd, 24th and 26th 2011.
            The observations were performed in position switching mode with reference positions offset from the central position of the sources by (600\arcsec,0\arcsec). Pointing and focus were checked on planets at the beginning of each observing session. The pointing was also regularly checked on Saturn and on hot cores (G10.62, G34.26, G327.3$-$0.6, NGC6334I and Sgr B2 (N)) during the observations. Comparison of sources observed multiple times show that the calibration uncertainty is $\lesssim15\%$ for this dataset.

			\begin{figure*}
				\centering
				\includegraphics[width=0.95\columnwidth]{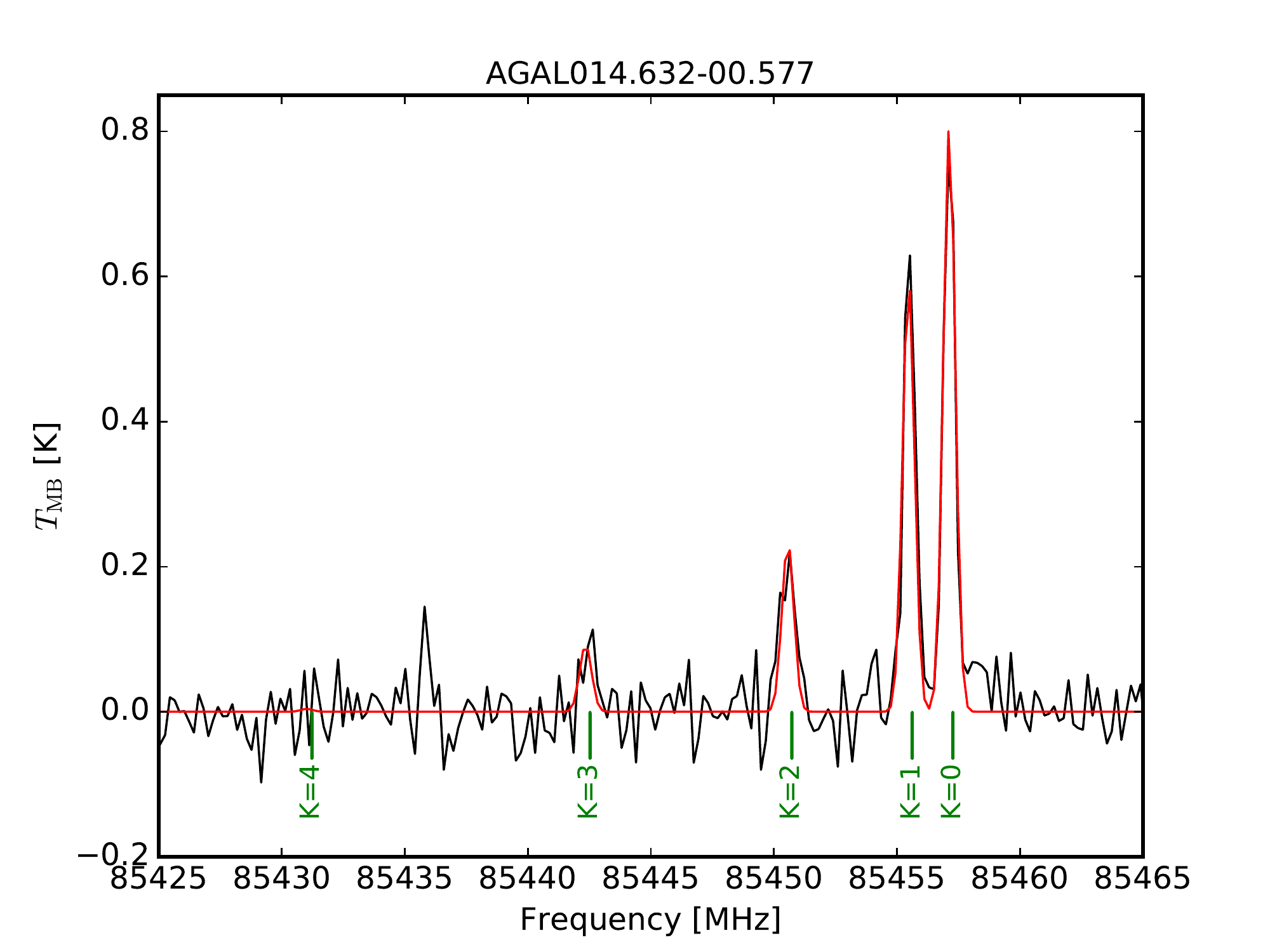}
				\includegraphics[width=0.95\columnwidth]{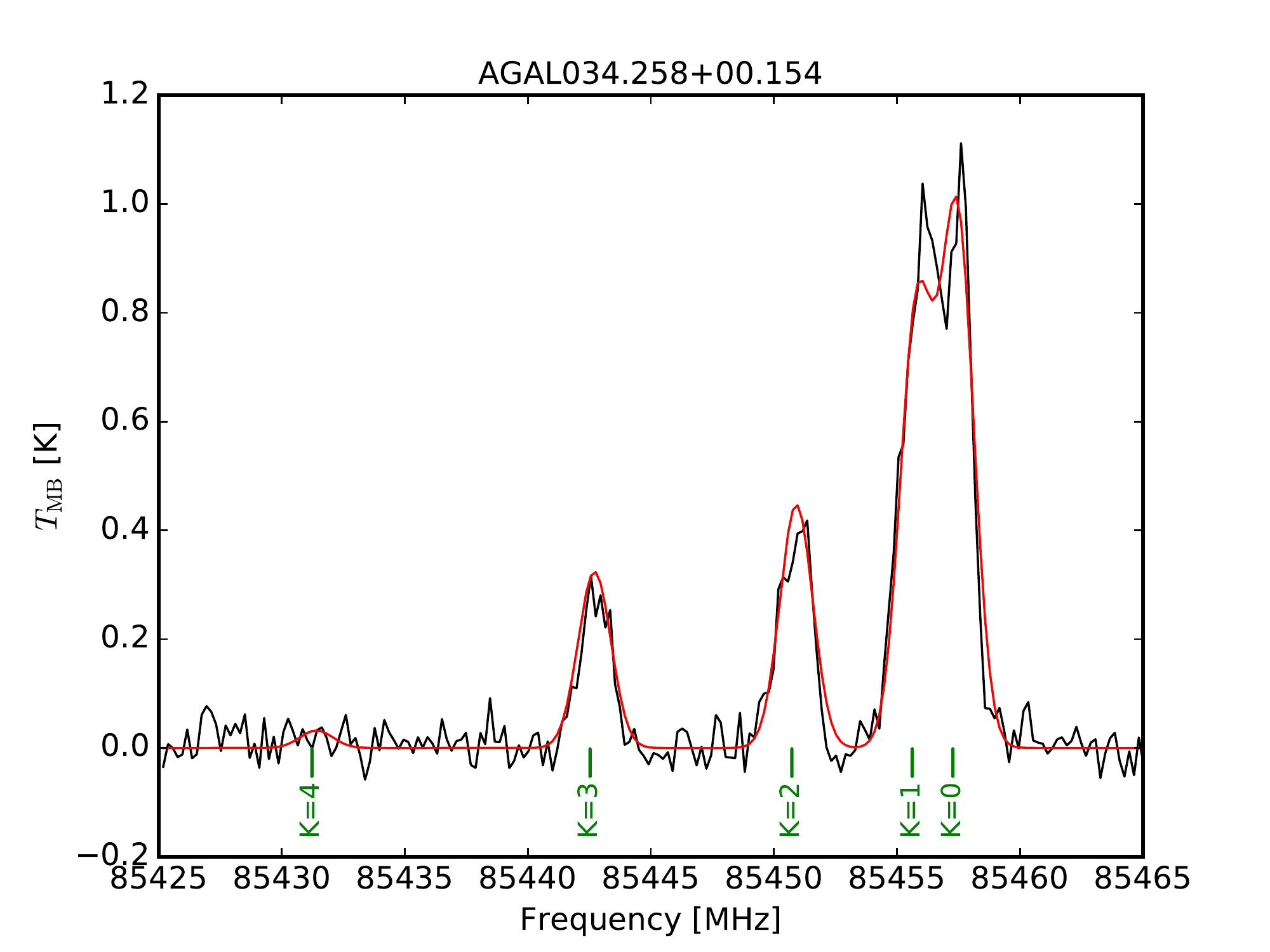}\\
				\includegraphics[width=0.95\columnwidth]{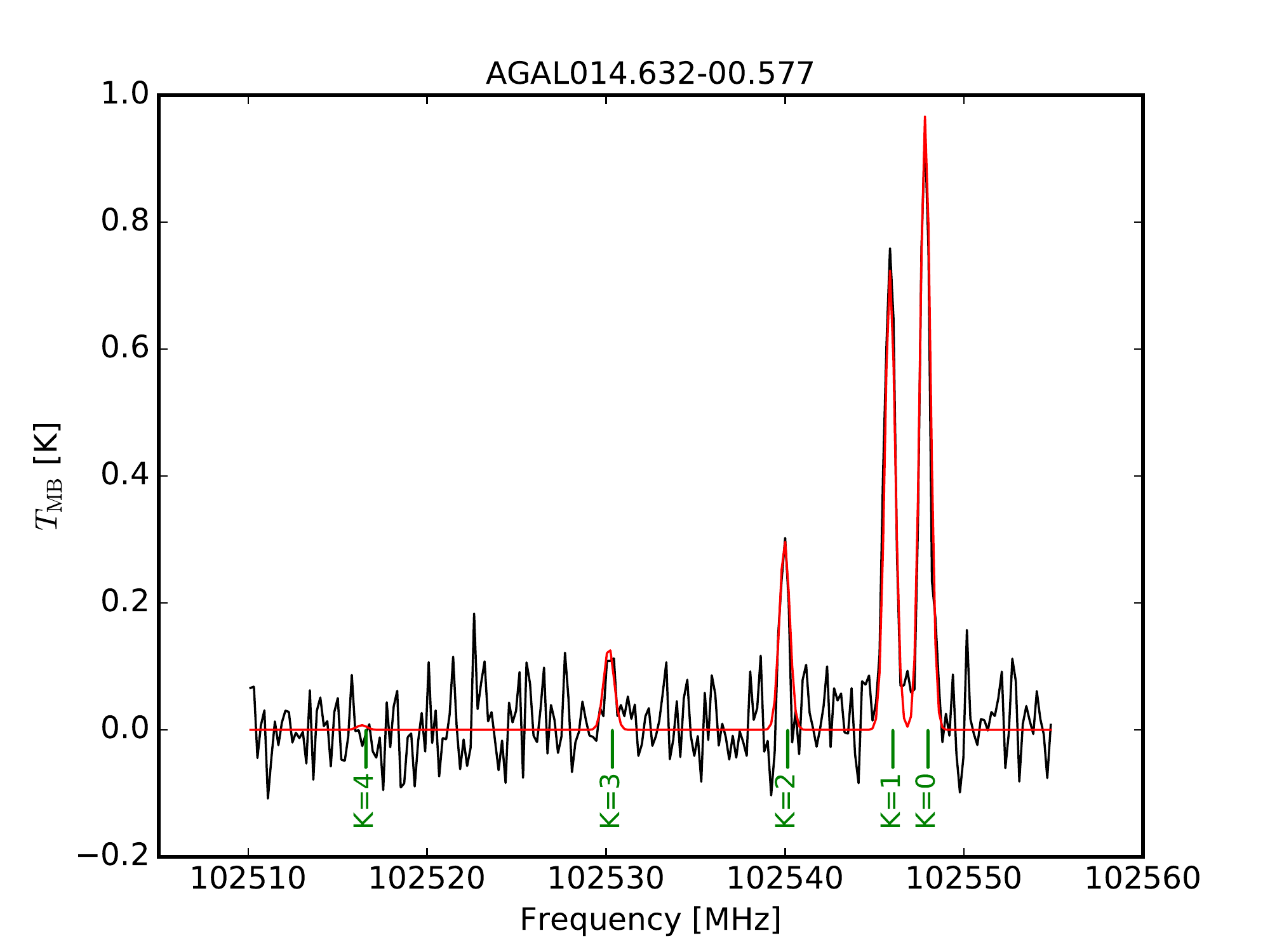}
				\includegraphics[width=0.95\columnwidth]{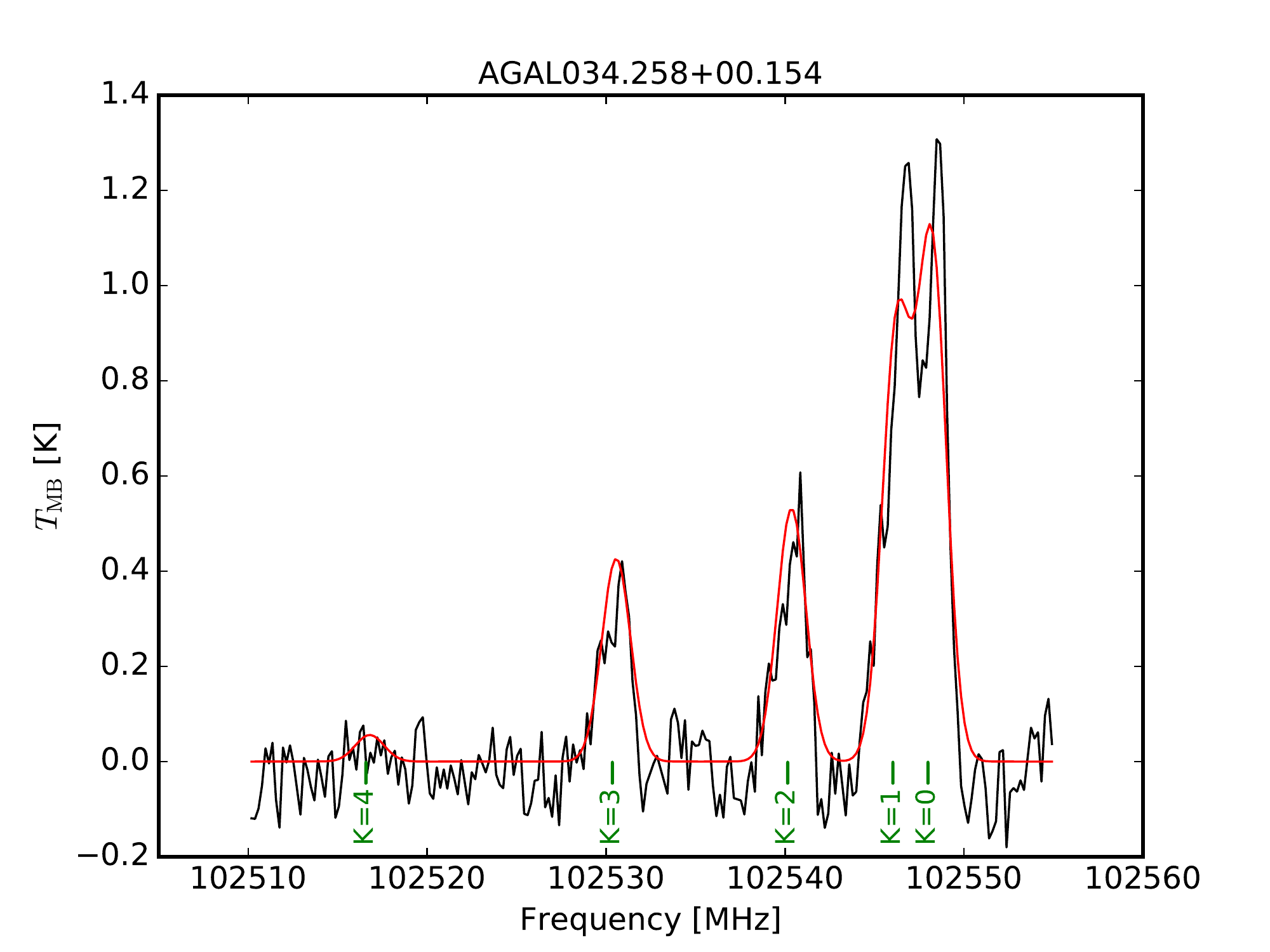}\\
				\includegraphics[width=0.95\columnwidth]{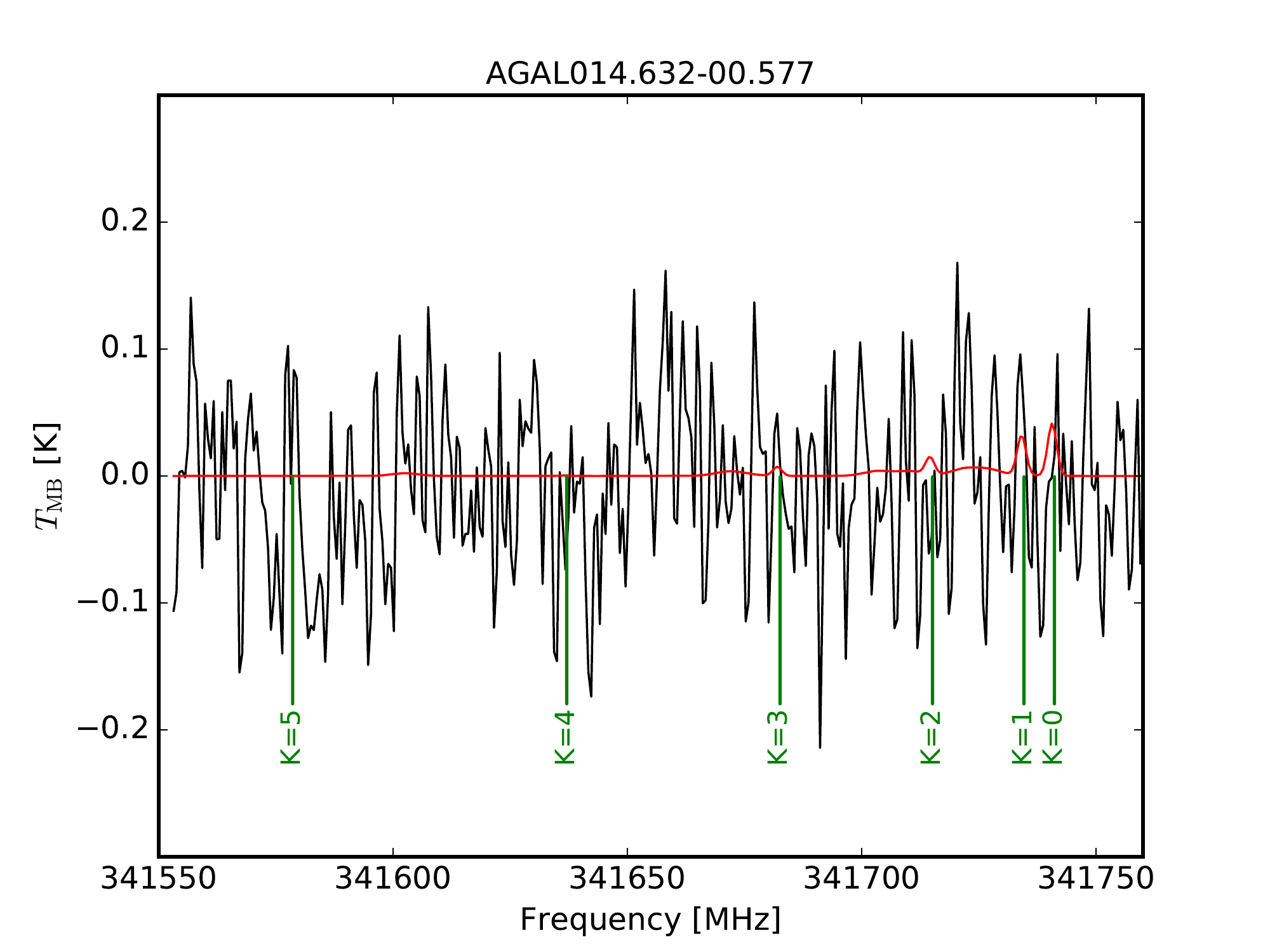}
				\includegraphics[width=0.95\columnwidth]{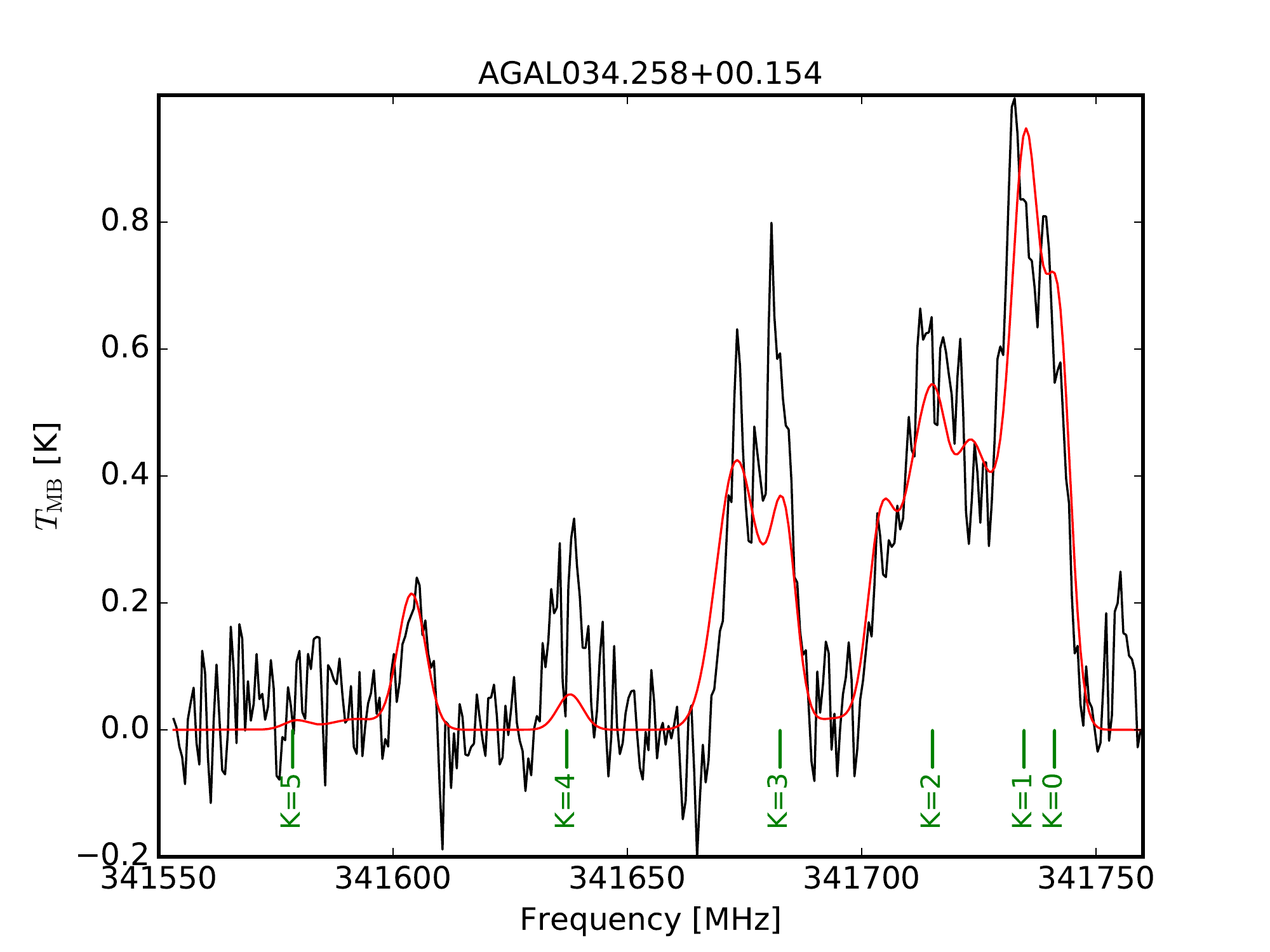}
				\caption{Example of the fit performed with MCWeeds for \ma. The $K$ components are indicated in green. The best fit model is shown in red. The source on the left is an IRw and the source on the right belong to \hii.}\label{fig:spectral_fit_ma}
			\end{figure*}

			\begin{figure*}
				\centering
				\includegraphics[width=0.95\columnwidth]{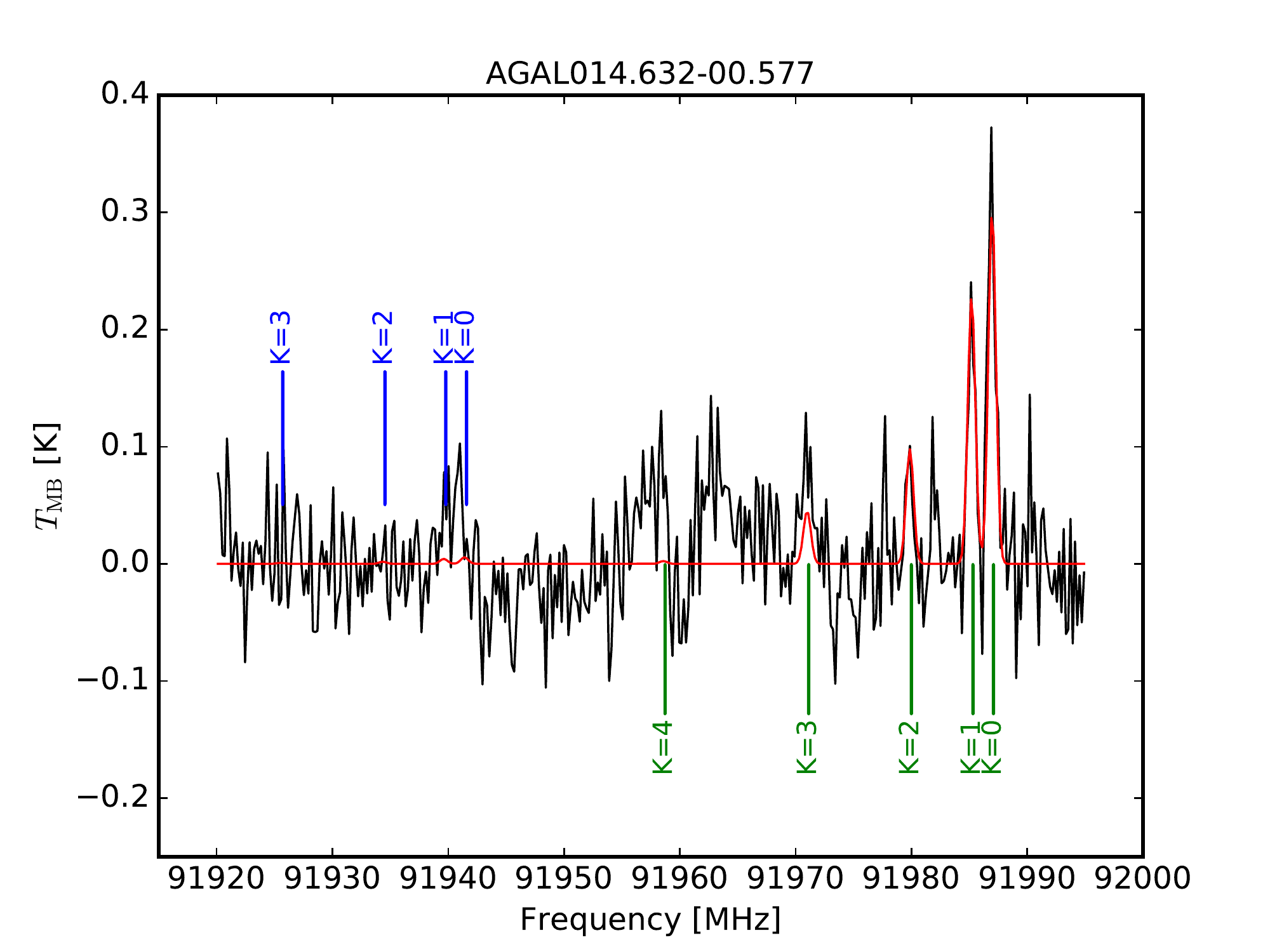}
				\includegraphics[width=0.95\columnwidth]{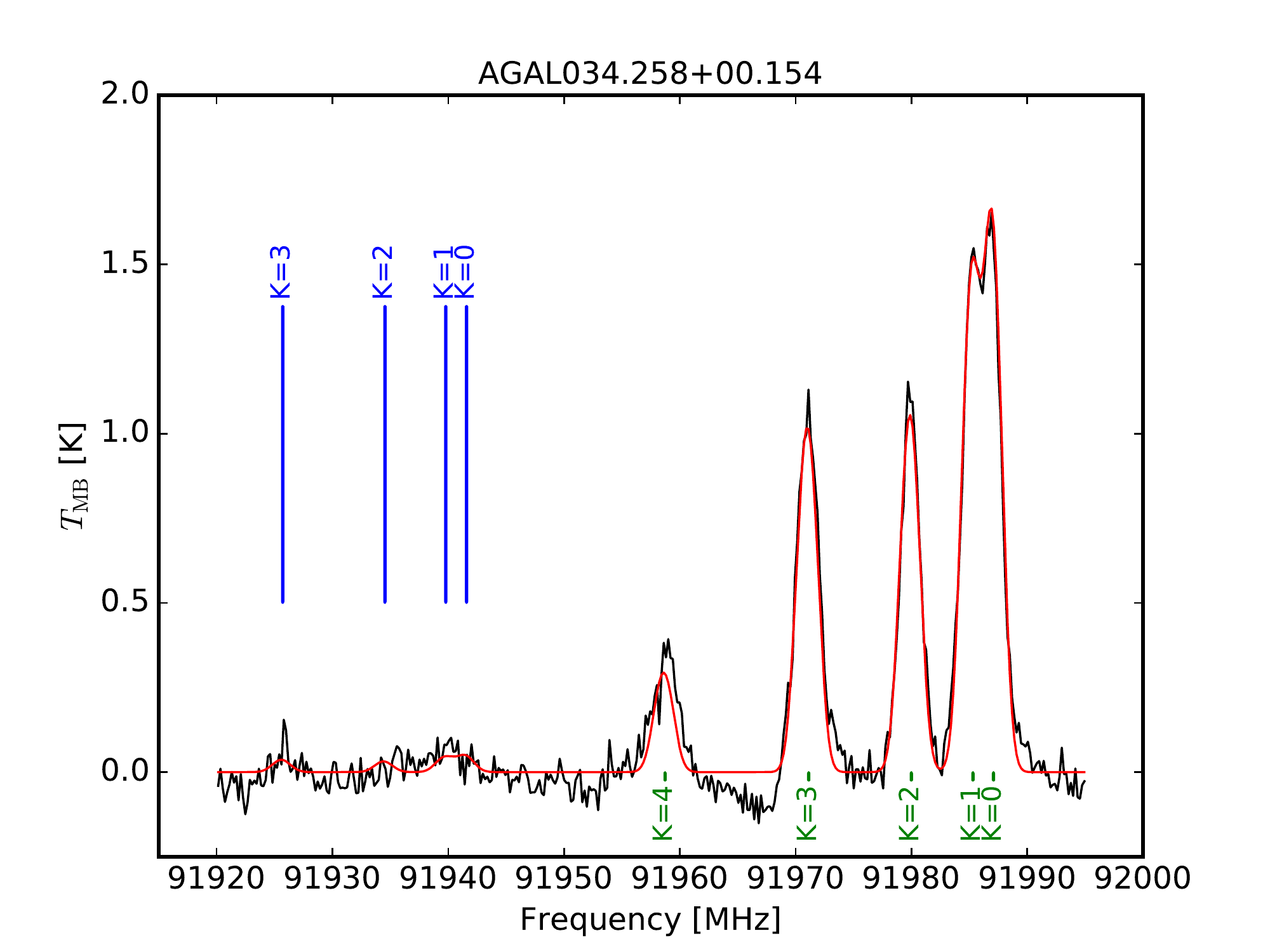}\\
				\includegraphics[width=0.95\columnwidth]{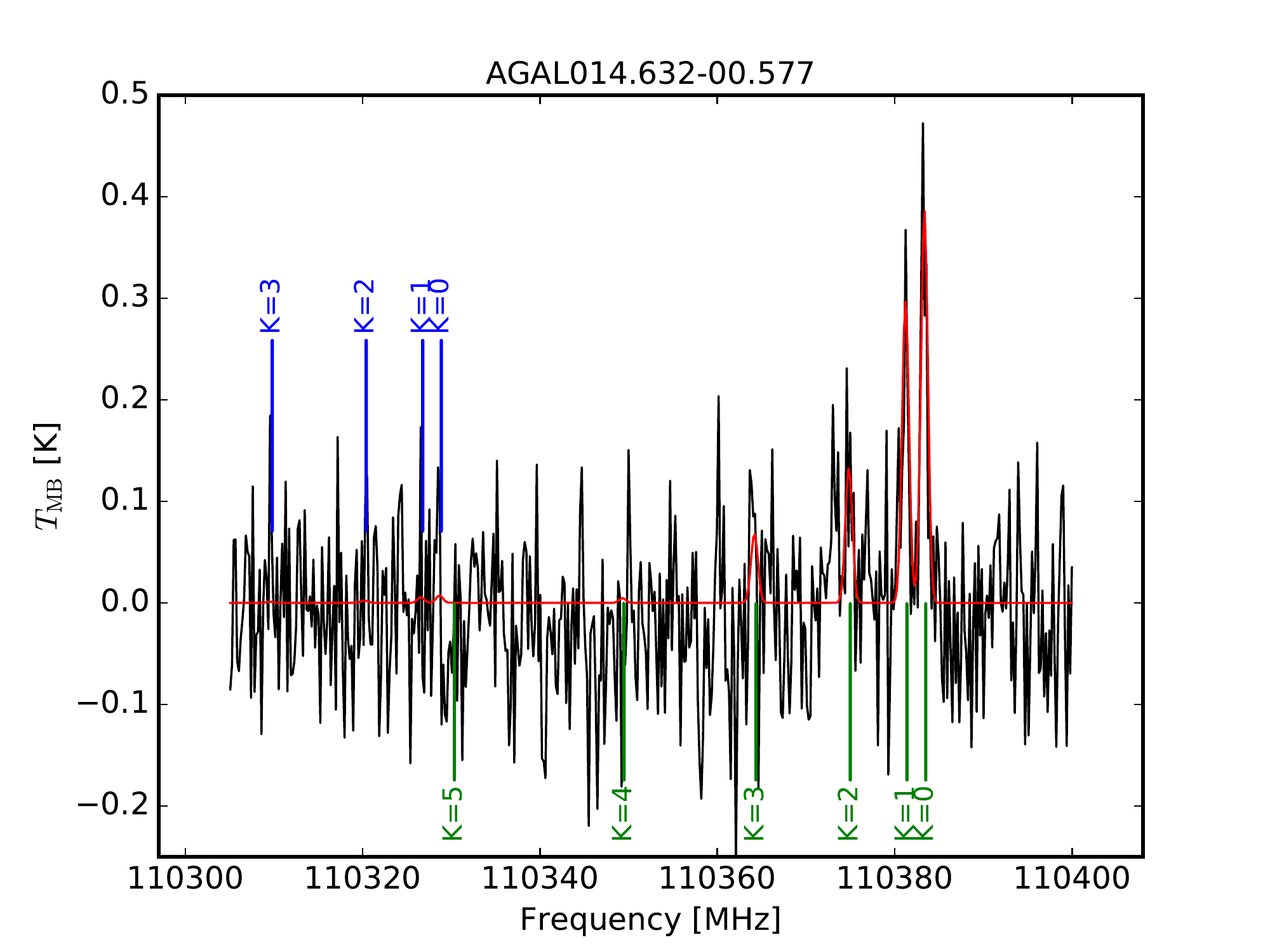}
				\includegraphics[width=0.95\columnwidth]{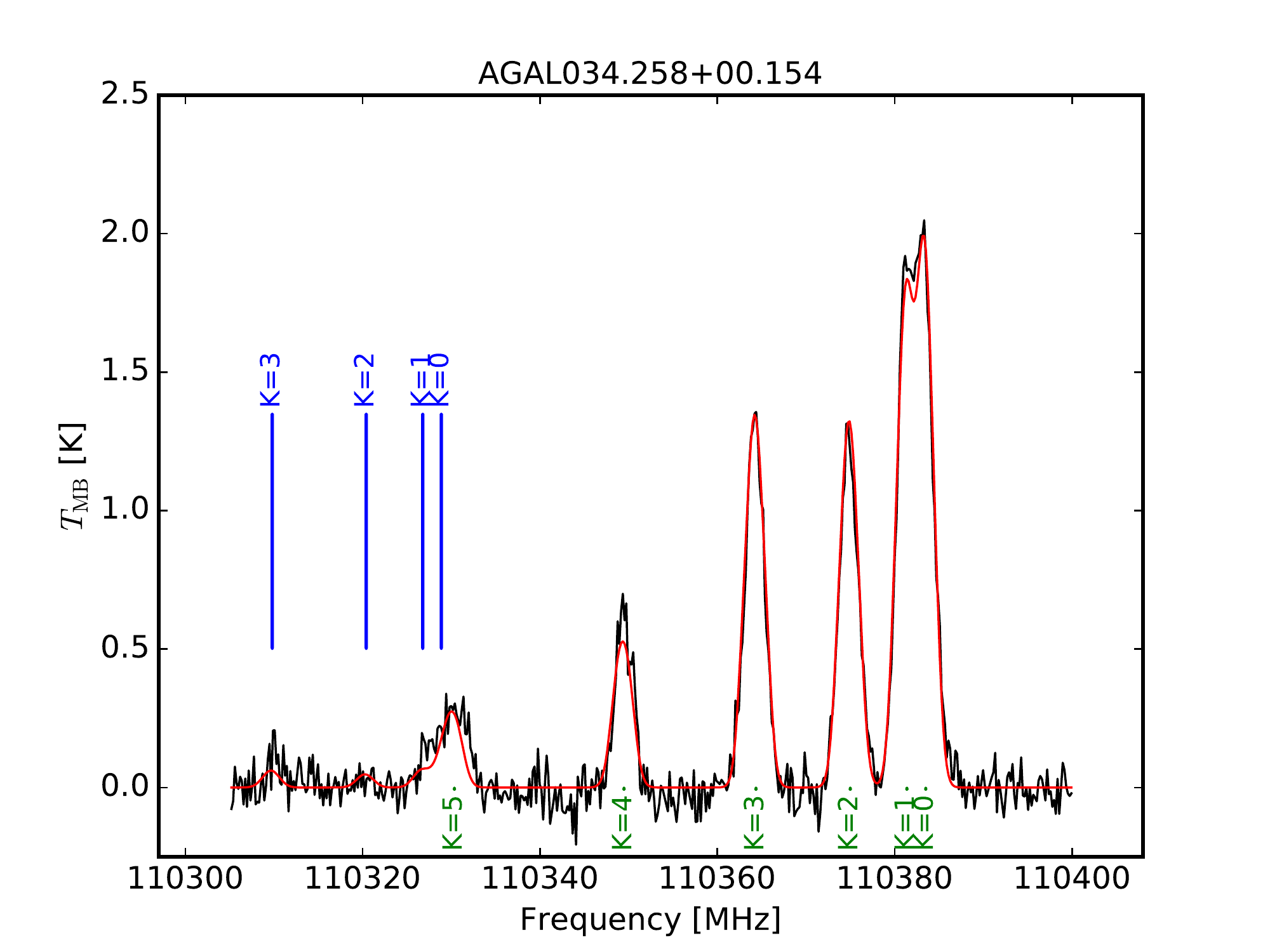}\\
				\includegraphics[width=0.95\columnwidth]{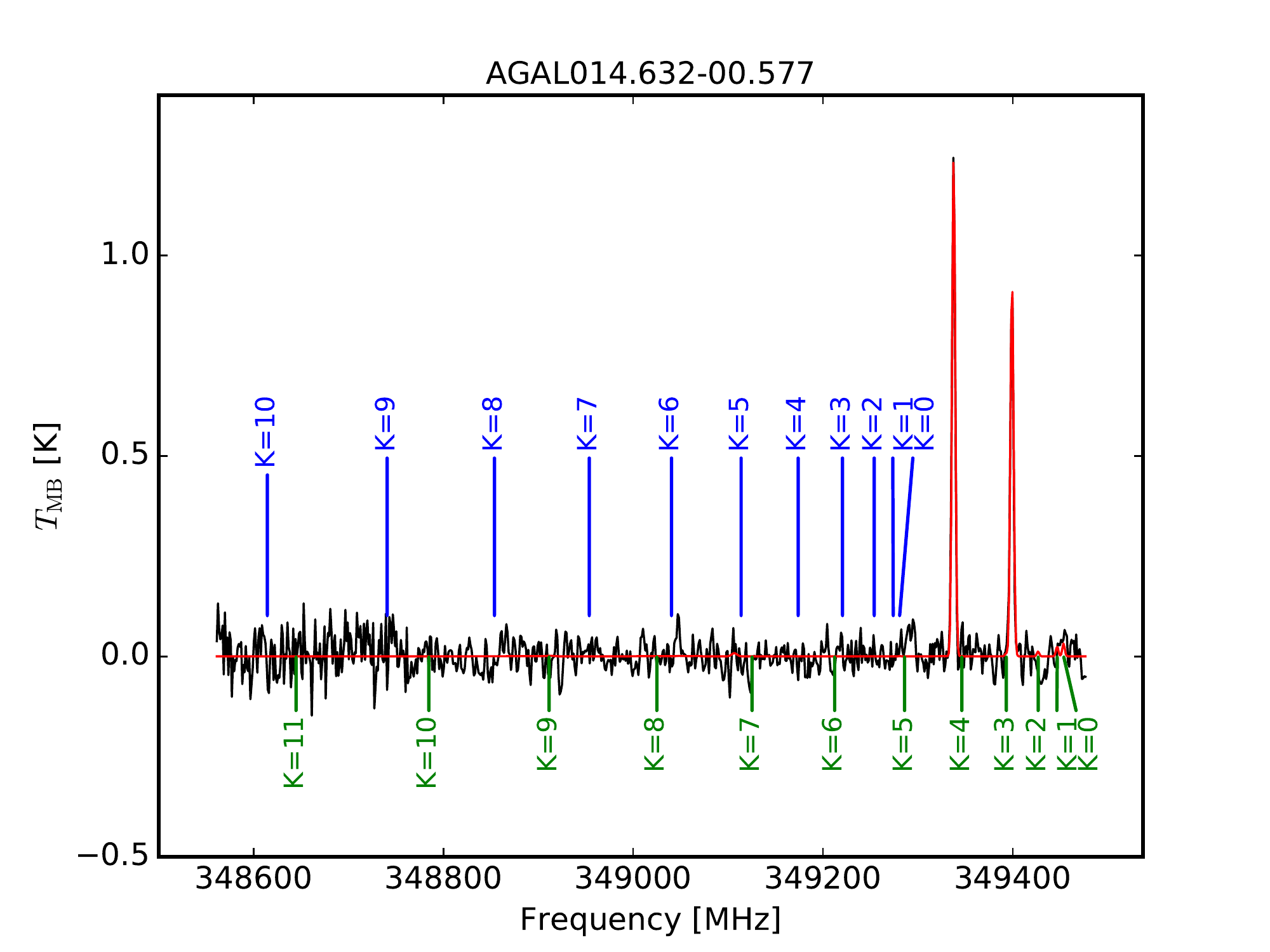}
				\includegraphics[width=0.95\columnwidth]{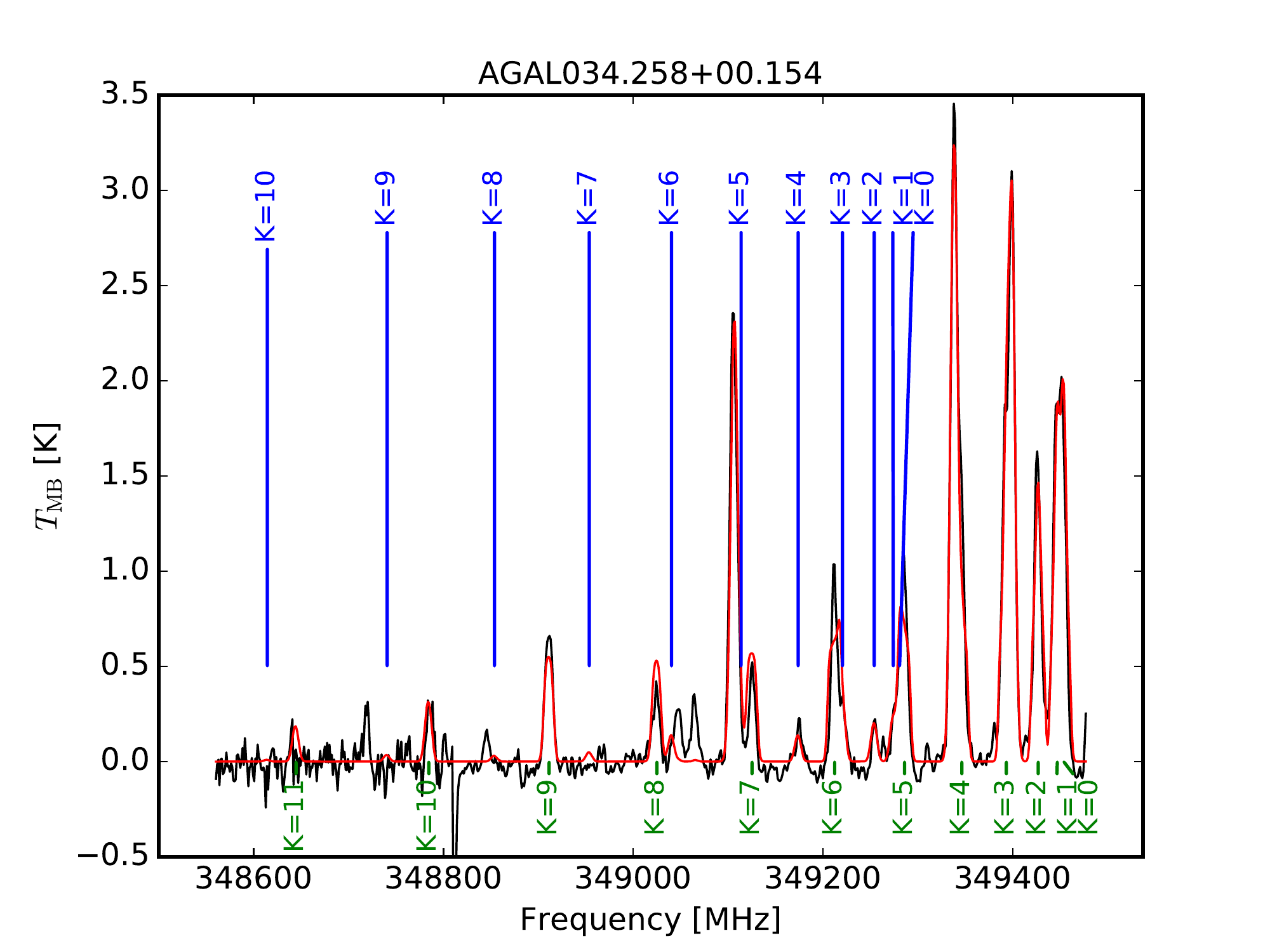}
				\caption{Example of the fit performed with MCWeeds for \an. The $K$ components of \an\ are indicated in green and those of CH$_{3}$$^{13}$CN are indicated in blue. The best fit model is shown in red. The source on the left is an IRw and the source on the right belong to \hii.}\label{fig:spectral_fit_an}
			\end{figure*}

        \section{MCWeeds}\label{sec:MCWeeds}

            To extract physical parameters from this large set of lines, a simultaneous fit was performed, including all available spectra for a given source.
            WEEDS \citep{Maret+11_aap526_47} provides a simple and fast way to generate synthetic spectra assuming LTE, but lacks automatic optimisation algorithms. Here PyMC \citep{Patil+10_jstatsoft35_1} comes into play, implementing Bayesian statistical models and fitting algorithms; therefore we developed MCWeeds, providing an external interface between these two packages to automatise the fitting procedure.
            
            MCWeeds allows to simultaneously fit the lines from an arbitrary number of species and components, combining multiple spectral ranges.
            The fitting is performed using one of the methods offered by PyMC, that is, maximum a posteriori estimates (MAP), `normal approximations' (NormApprox) \citep{gelman2003bayesian} or Monte Carlo Markov Chains (MCMCs)\footnote{A more detailed description of these methods is given in the PyMC documentation \url{https://pymc-devs.github.io/pymc/index.html}}. This is therefore a very flexible approach, giving the possibility of obtaining both a very fast fitting procedure or a very detailed one, generating probability distribution functions for each of the model parameters. MCWeeds can be accessed both from inside CLASS or from shell.
            
            A simple flow chart of the main steps performed by the programme is shown in Fig.~\ref{fig:mcweeds_flow_chart}. As indicated in the figure, an input file must be defined, containing all the information to perform the fit, including:
			the species that must be included in the fit, the spectral ranges to be considered and the corresponding data file, the fitting algorithm to be used and its parameters, and the priors or the values for all of the models parameters (i.e. column density, temperature, size of the emitting region, radial velocity and linewidth for each species, rms noise, and calibration uncertainty for each of the spectral ranges).
            MCWeeds retrieves the spectra in the specified frequency ranges from the data files, and the line properties from the JPL- \citep{Pickett+98_jqsrt60_883}, CDMS- \citep{Mueller+01_aap370_49}, or a local database, to generate the synthetic spectrum, given the model parameters. The PyMC model is automatically built 
            using the priors specified in the input file;
            the likelihood is computed as the product of the probabilities for each channel, assuming a Gaussian uncertainty, with a $\sigma$ equal to the rms noise.
            
            A detailed discussion of priors, their meaning and choice can be found in, for example, \citet{Jaynes03}. For the purposes of the present work, the definition of a prior for the fit parameters avoids the generation of unphysical parameters (e.g. a negative linewidth, or temperature) during the fit procedure, as well as giving the possibility to incorporate any previous information available in the fit (see Sect.~\ref{sec:fit_procedure}).
            
			\begin{figure*}
				\centering
				\includegraphics[width=\textwidth]{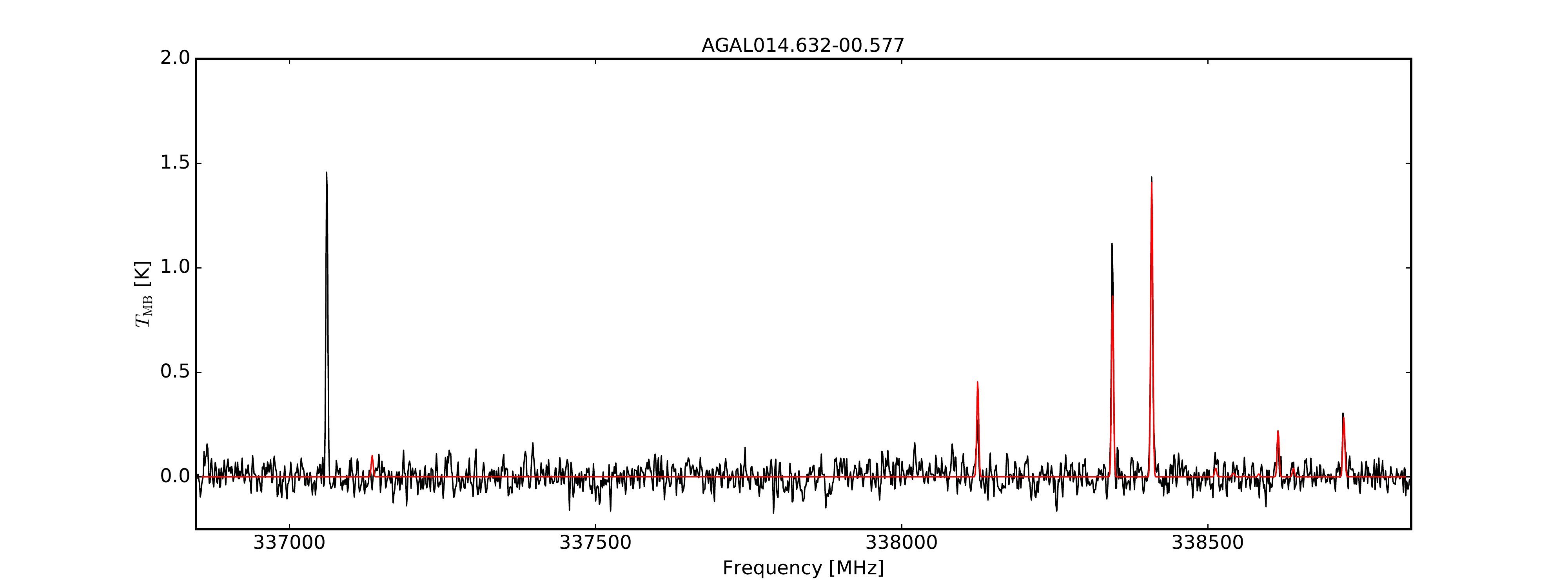}\\
				\includegraphics[width=\textwidth]{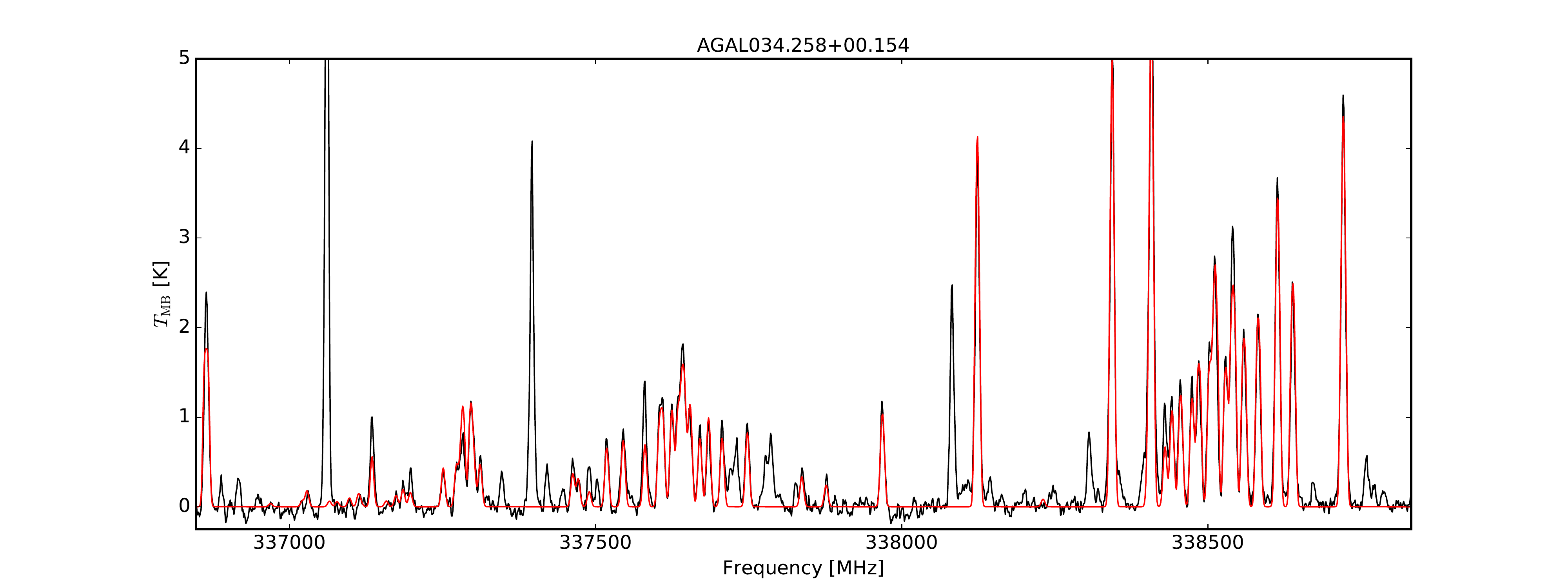}
				\caption{Example of the fit performed with MCWeeds for \mt. The best fit model is shown in red. The lines not reproduced by the model are contaminating species and are excluded from the spectral ranges used for the fit (cf. Table~\ref{tab:spectral_ranges}) The top panel show the results for an IRw source and the bottom panel for an \hii. The $\nu_{t} = 1$ band is visible between $\sim337500$ and $\sim380000\usk\mega\hertz$.}\label{fig:spectral_fit_mt}
			\end{figure*}
            The fit is performed according to the method specified in the input file, as indicated in the first decision box of Fig.~\ref{fig:mcweeds_flow_chart}. In the case of MAP or NormApprox, the optimisation is performed with one of the methods available in SciPy optimize package. For MCMCs a set of parameters is generated, which is used to derive the synthetic spectrum. The model spectrum is compared to the observations; the likelihood $P(D|M)$ (the probability of the data given the model) and the posterior $P(M|D)$ (the probability of the model given the data) are computed. Based on this, the newly-generated set of parameters is accepted or rejected \citep[see, e.g. ][]{Hastings70_Biometrika57_97}, and the procedure is repeated until the number of iterations requested is reached (last decision point).
            Finally, the results are summarised in an HTML page, containing the best fit parameters, their uncertainties, the best fit spectrum superimposed on the observations and, for MCMCs, the probability distribution function, the trace plot and the autocorrelation of each of the fit parameters. 
			\onlfig{
			\begin{figure*}
				\centering
				\includegraphics[width=0.4\columnwidth]{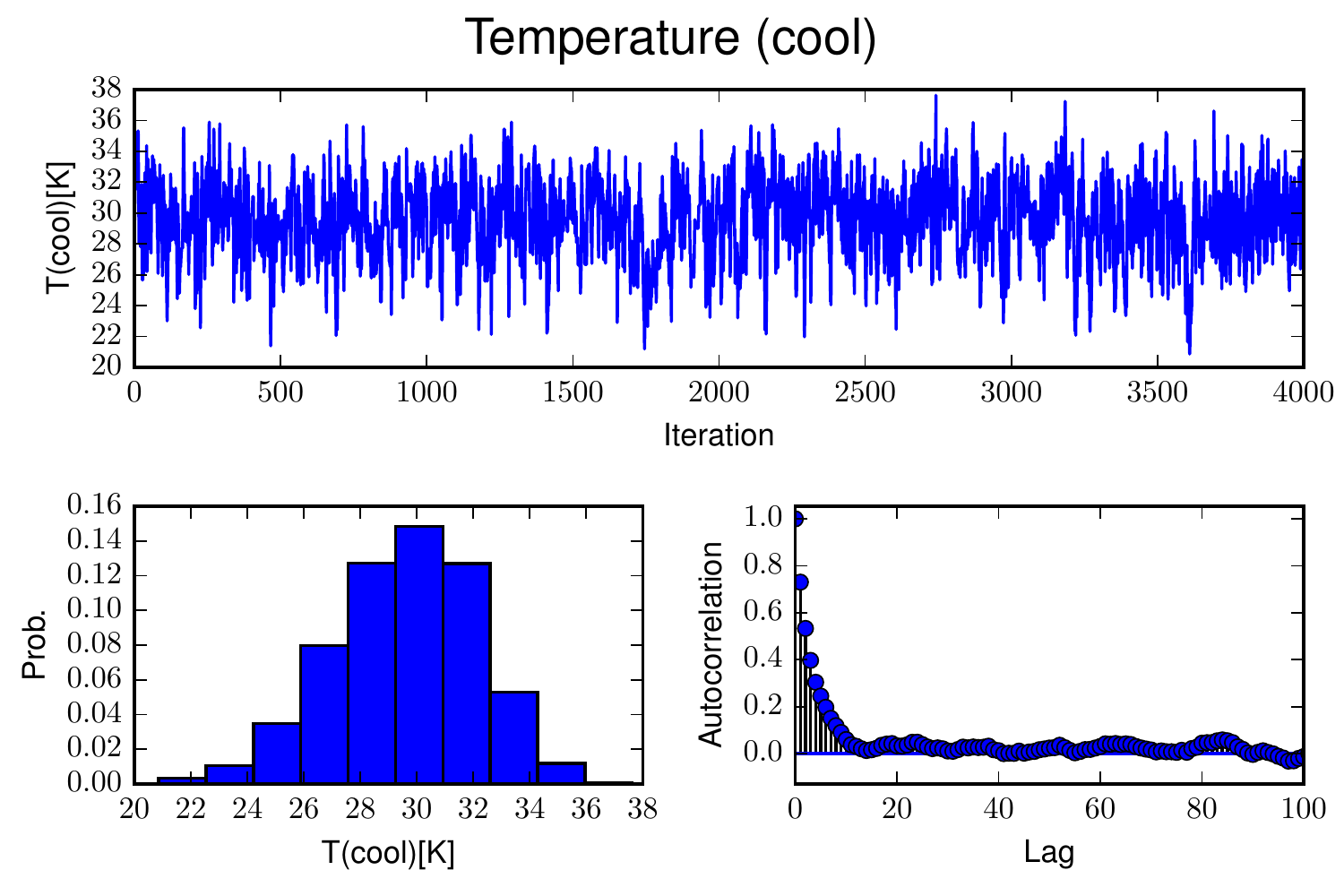}\hspace{2cm}
				\includegraphics[width=0.4\columnwidth]{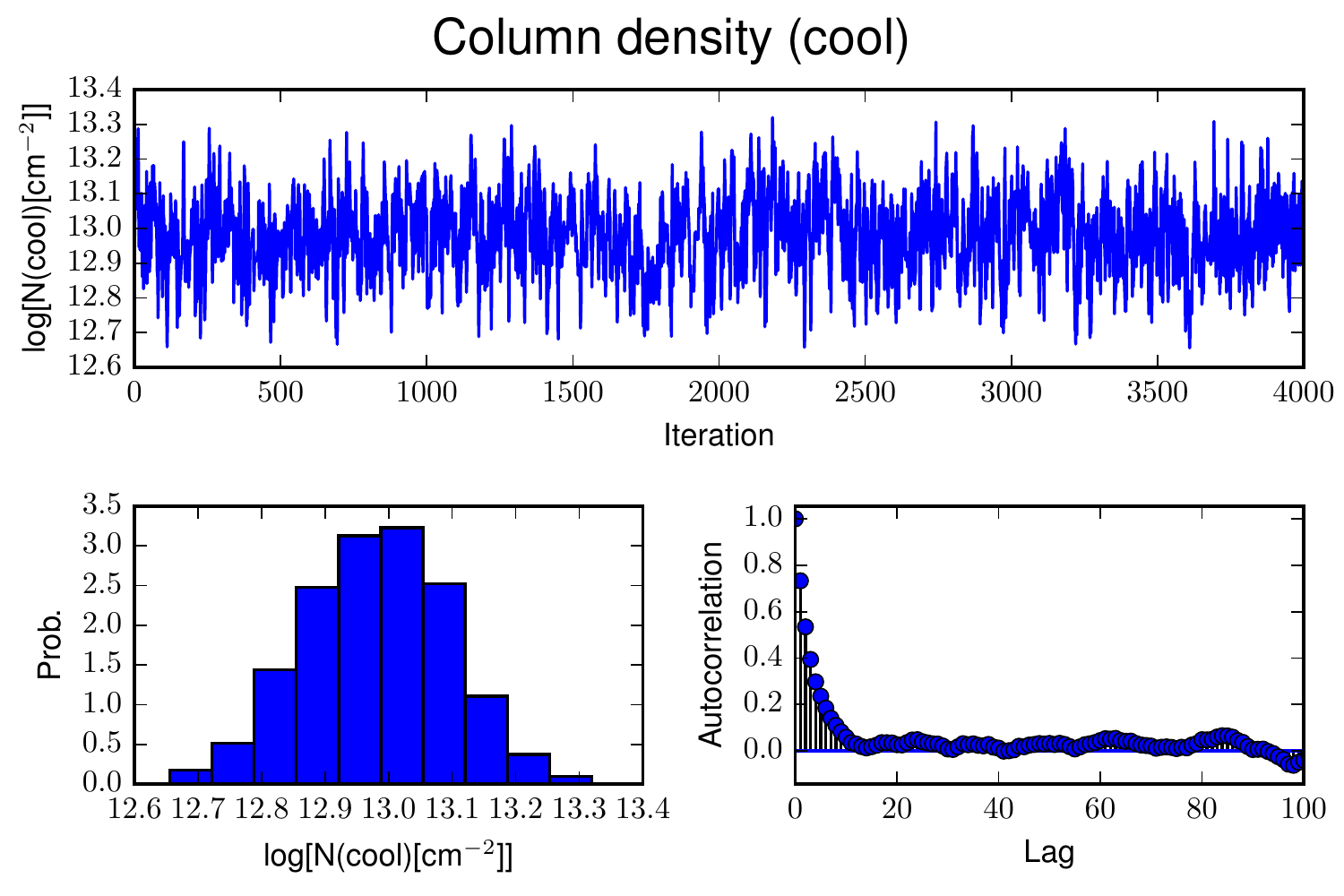}\\
				\includegraphics[width=0.4\columnwidth]{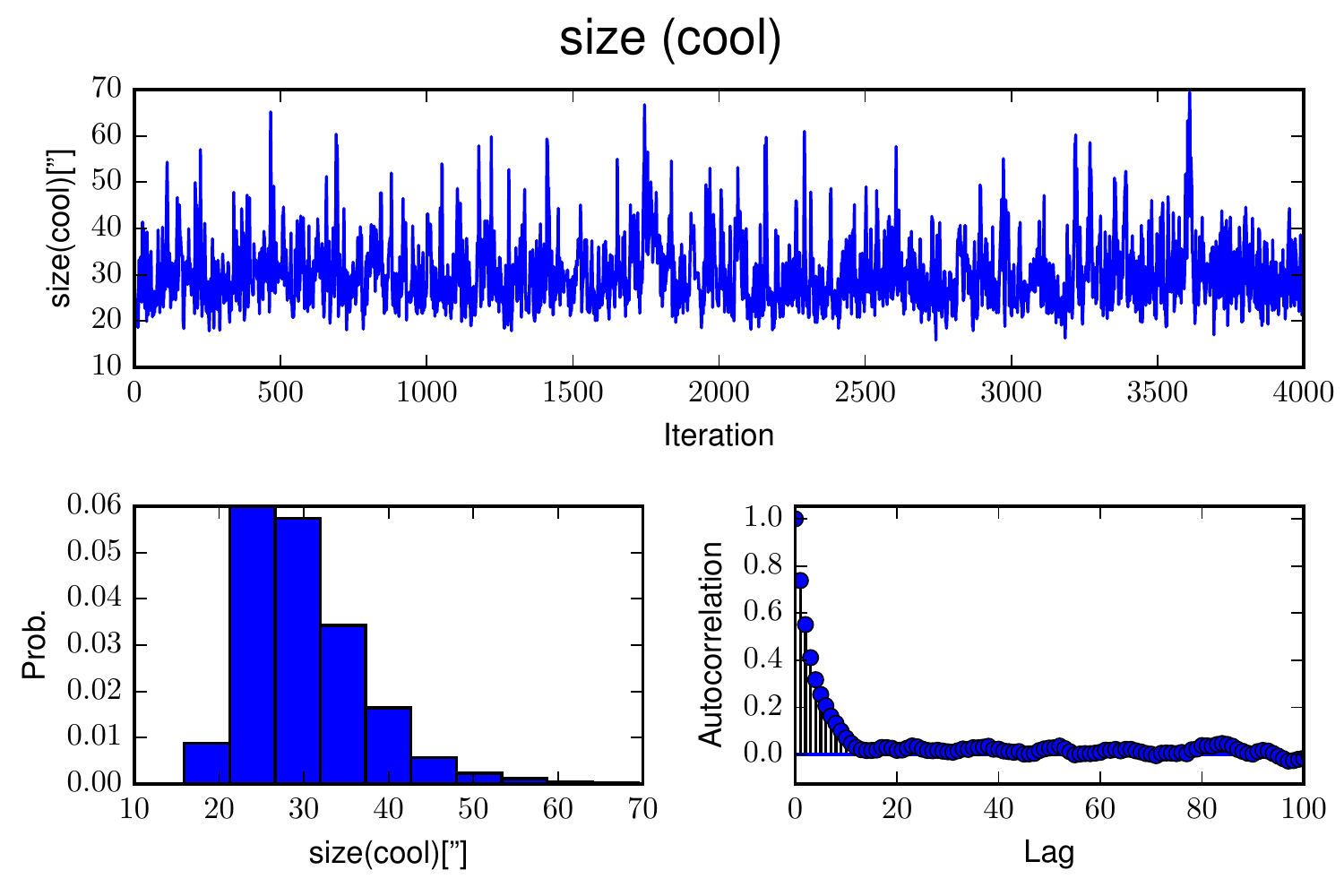}\hspace{2cm}
				\includegraphics[width=0.4\columnwidth]{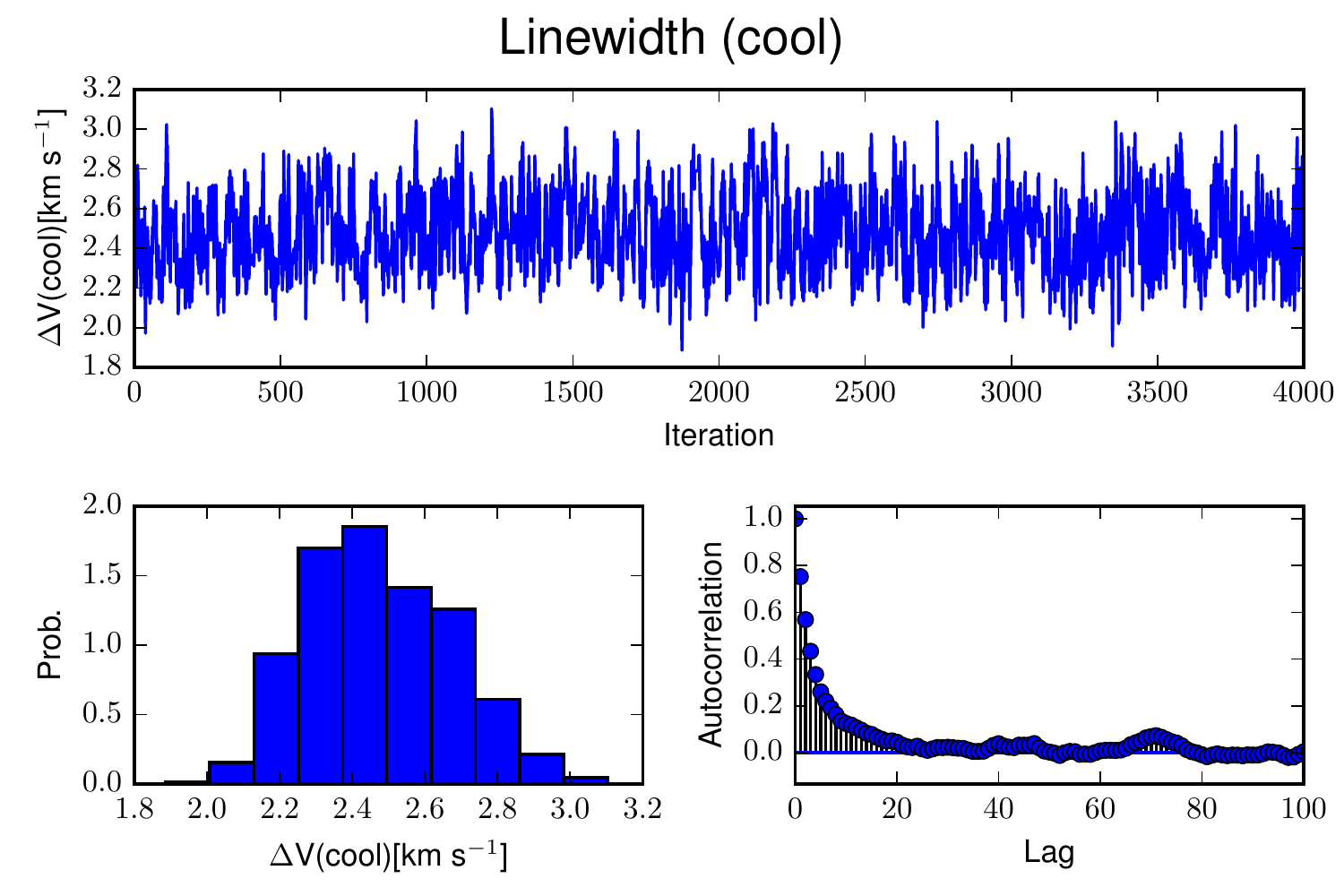}\\
				\includegraphics[width=0.4\columnwidth]{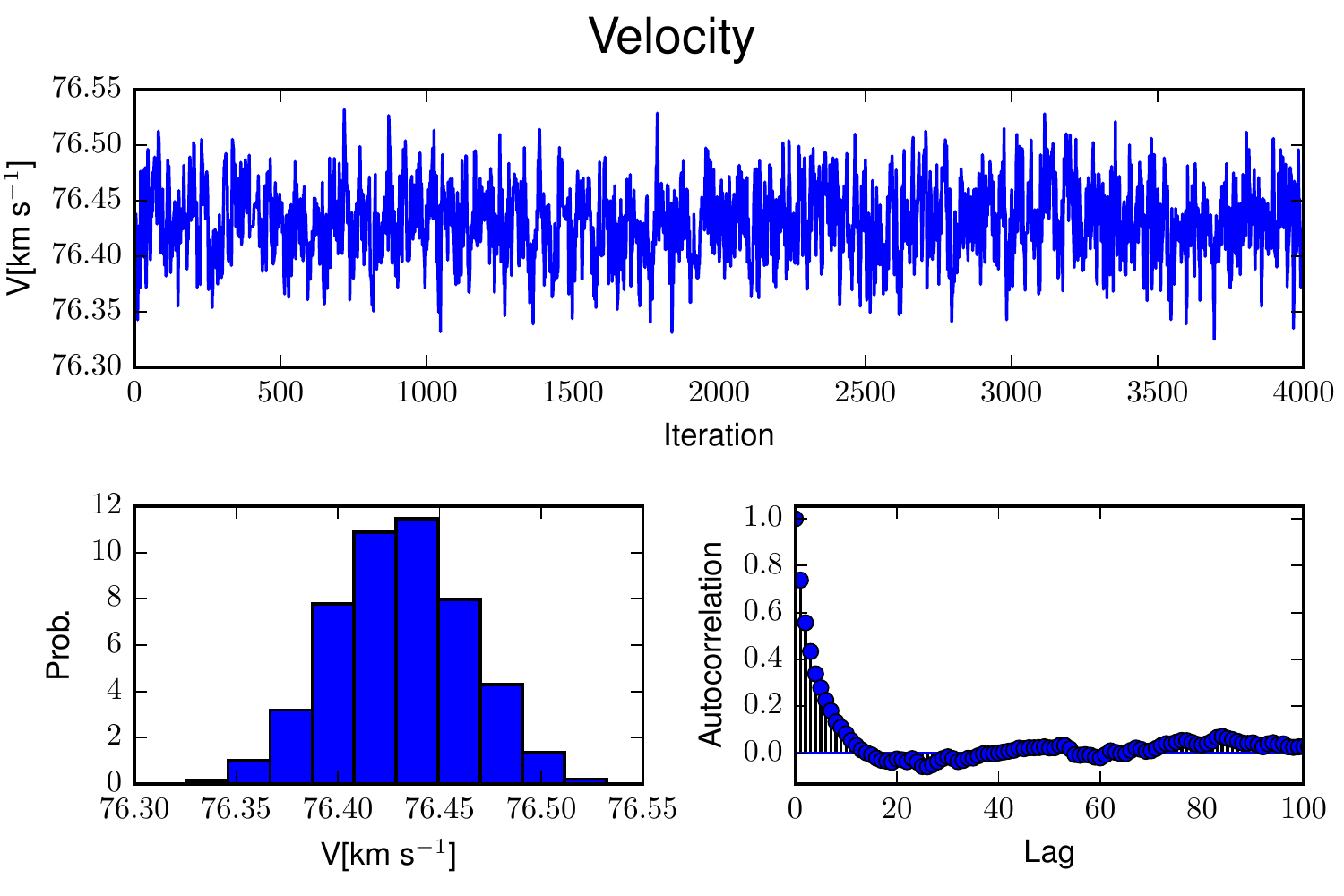}
				\caption{For each parameter, indicated at the top of each panel, we show an example of trace plot (top), probability distribution function (bottom left), and autocorrelation plot (bottom right), for AGAL014.632--00.577. The plots refer to \an.}\label{fig:mcmc_ex_g14}
			\end{figure*}}
			\onlfig{
			\begin{figure*}
				\centering
				\includegraphics[width=0.4\columnwidth]{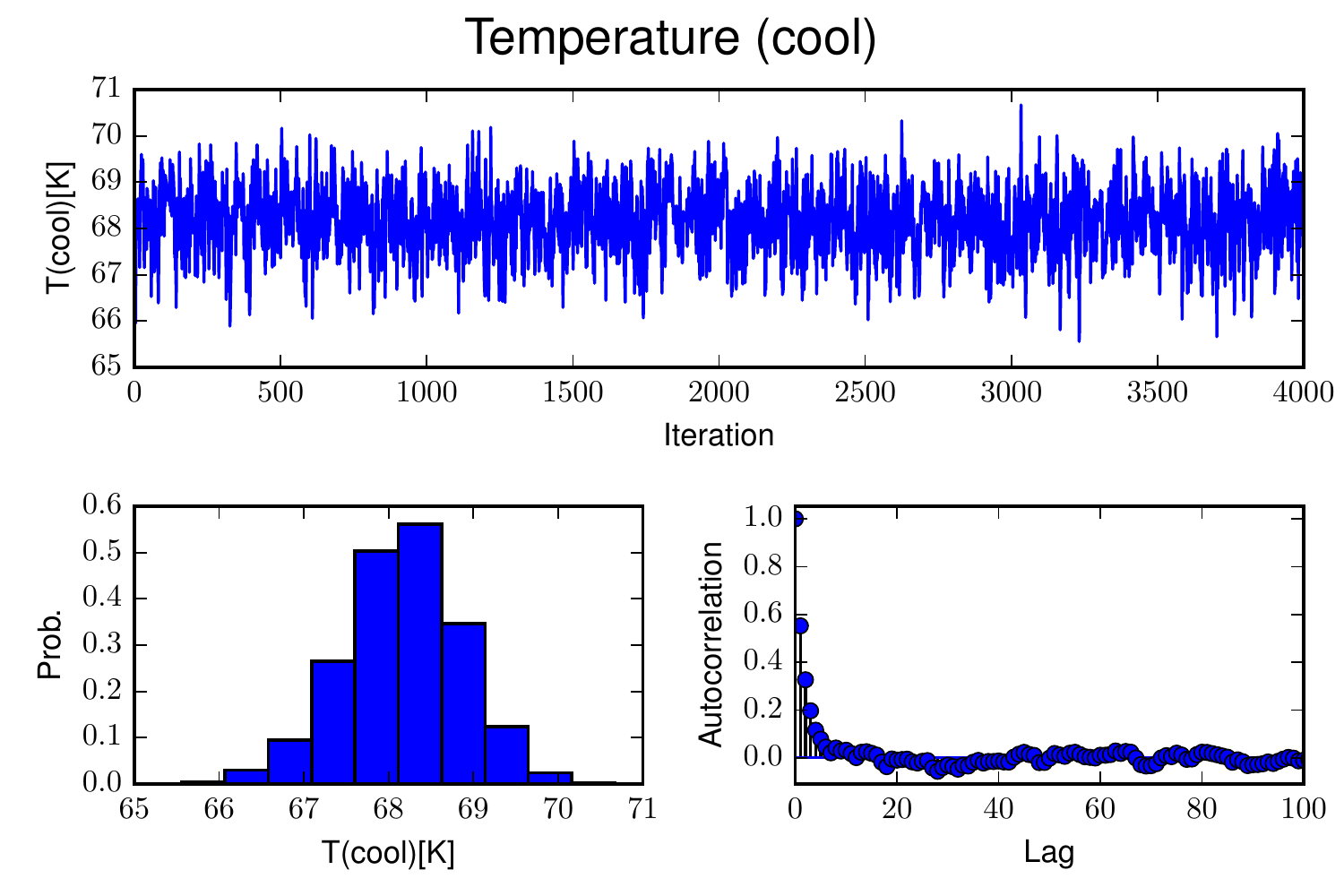}\hspace{2cm}
				\includegraphics[width=0.4\columnwidth]{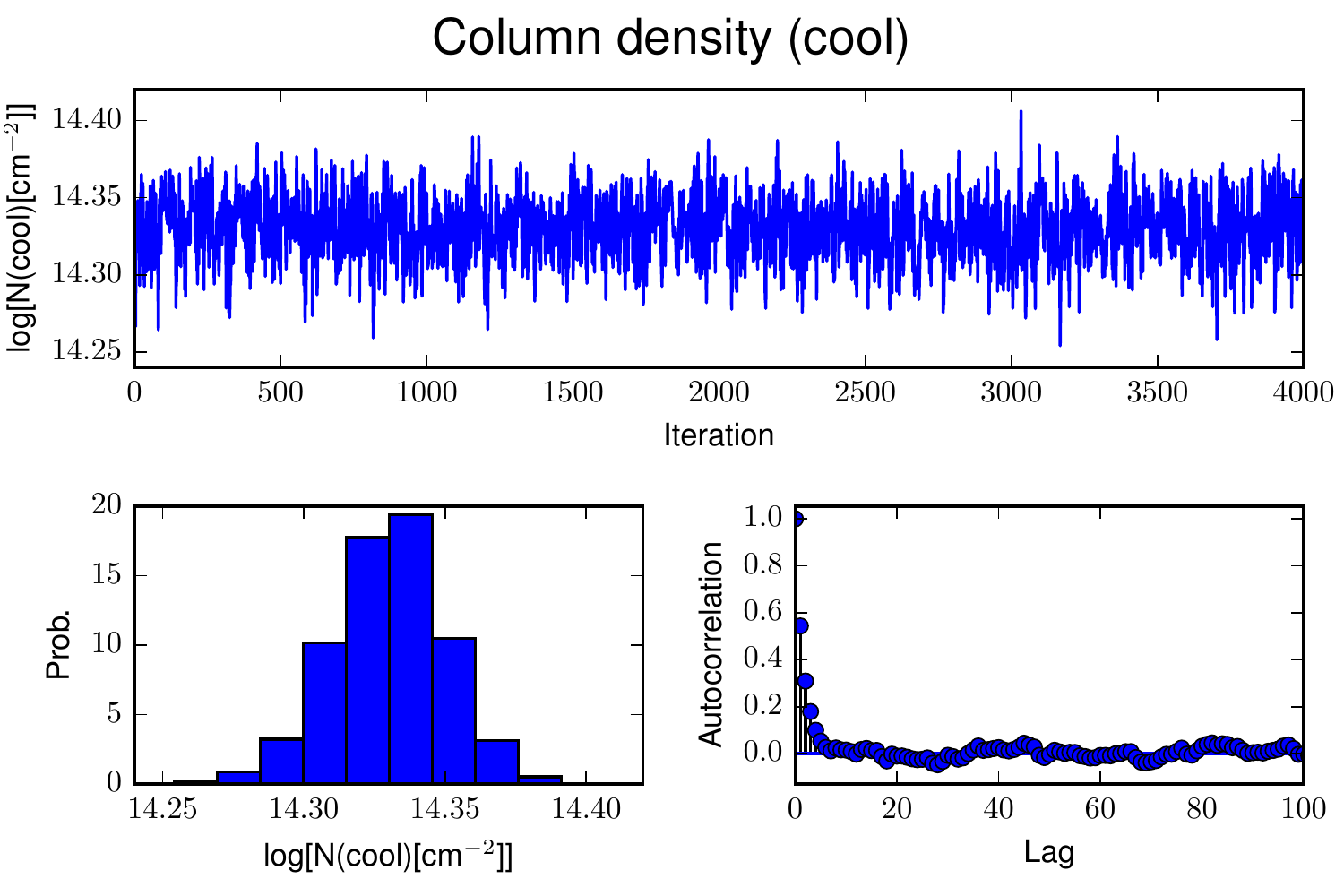}\\
				\includegraphics[width=0.4\columnwidth]{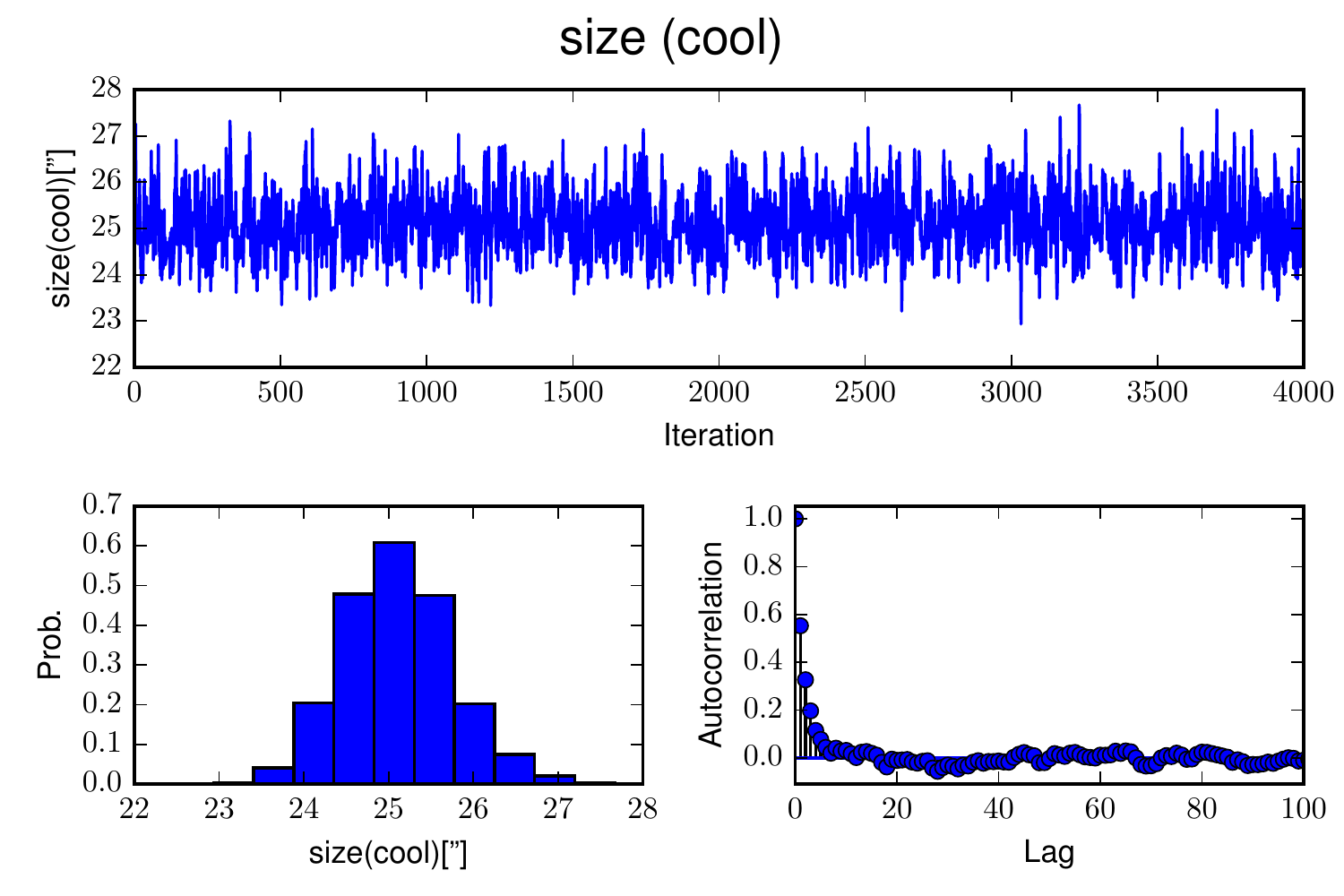}\hspace{2cm}
				\includegraphics[width=0.4\columnwidth]{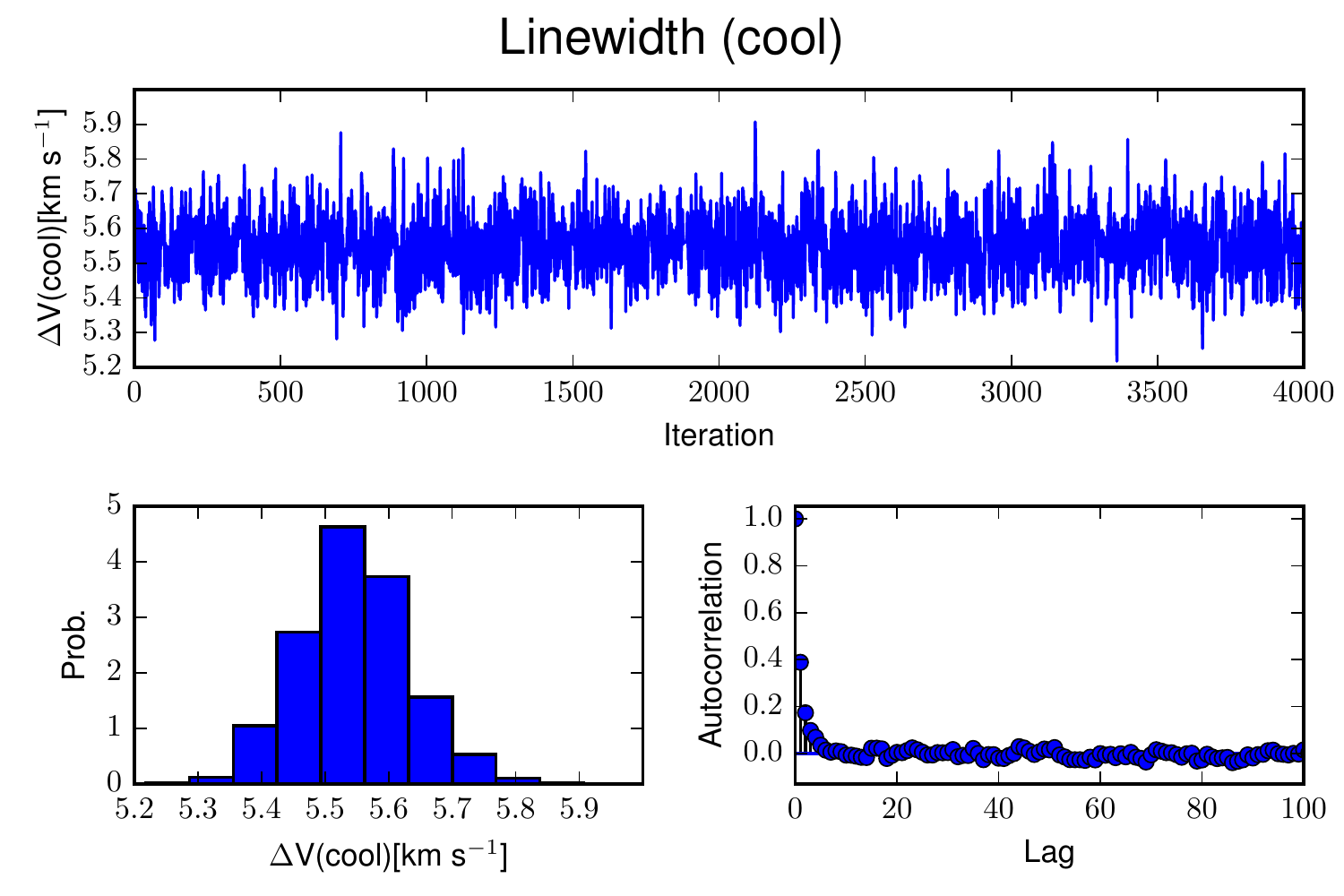}\\
				\includegraphics[width=0.4\columnwidth]{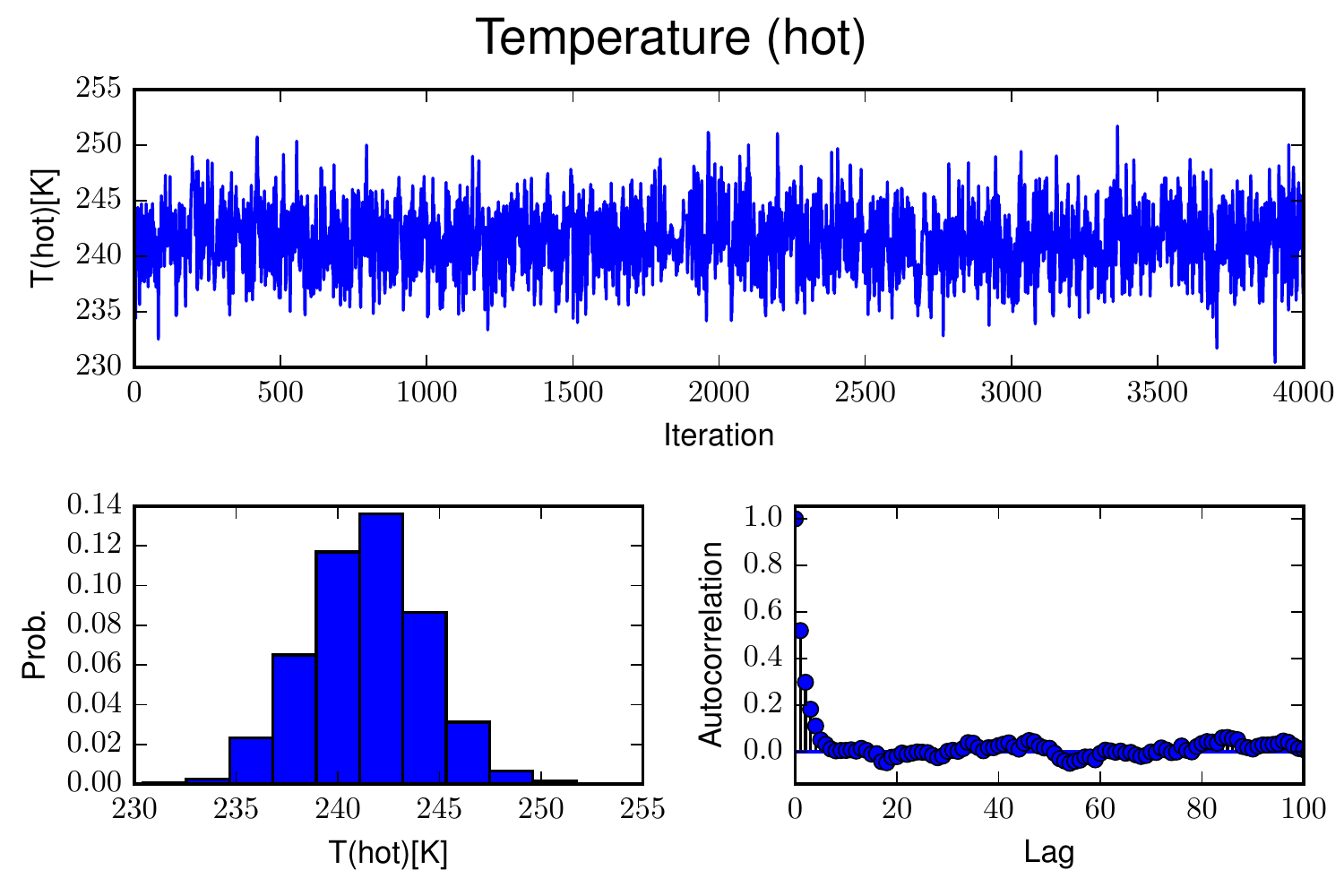}\hspace{2cm}
				\includegraphics[width=0.4\columnwidth]{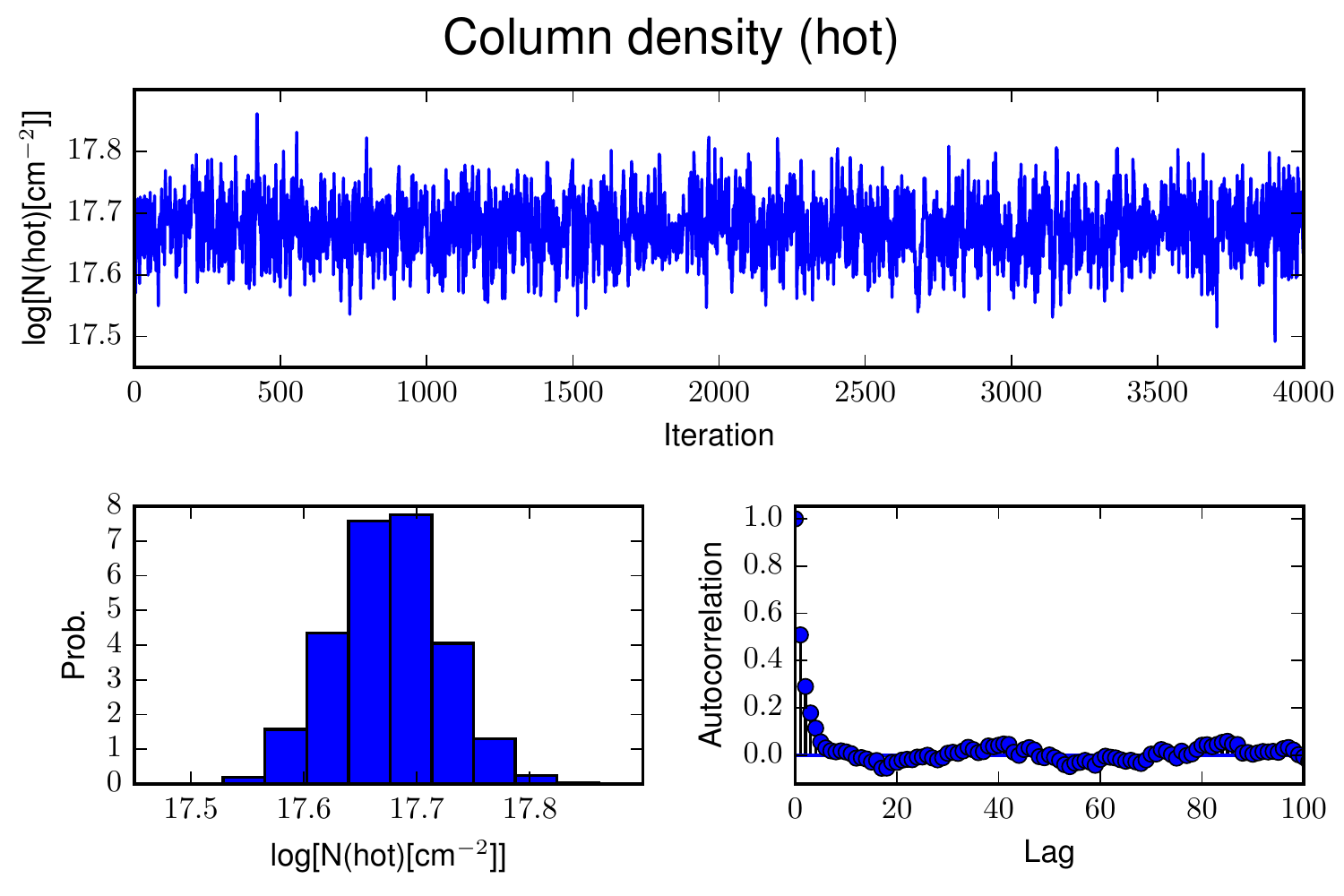}\\
				\includegraphics[width=0.4\columnwidth]{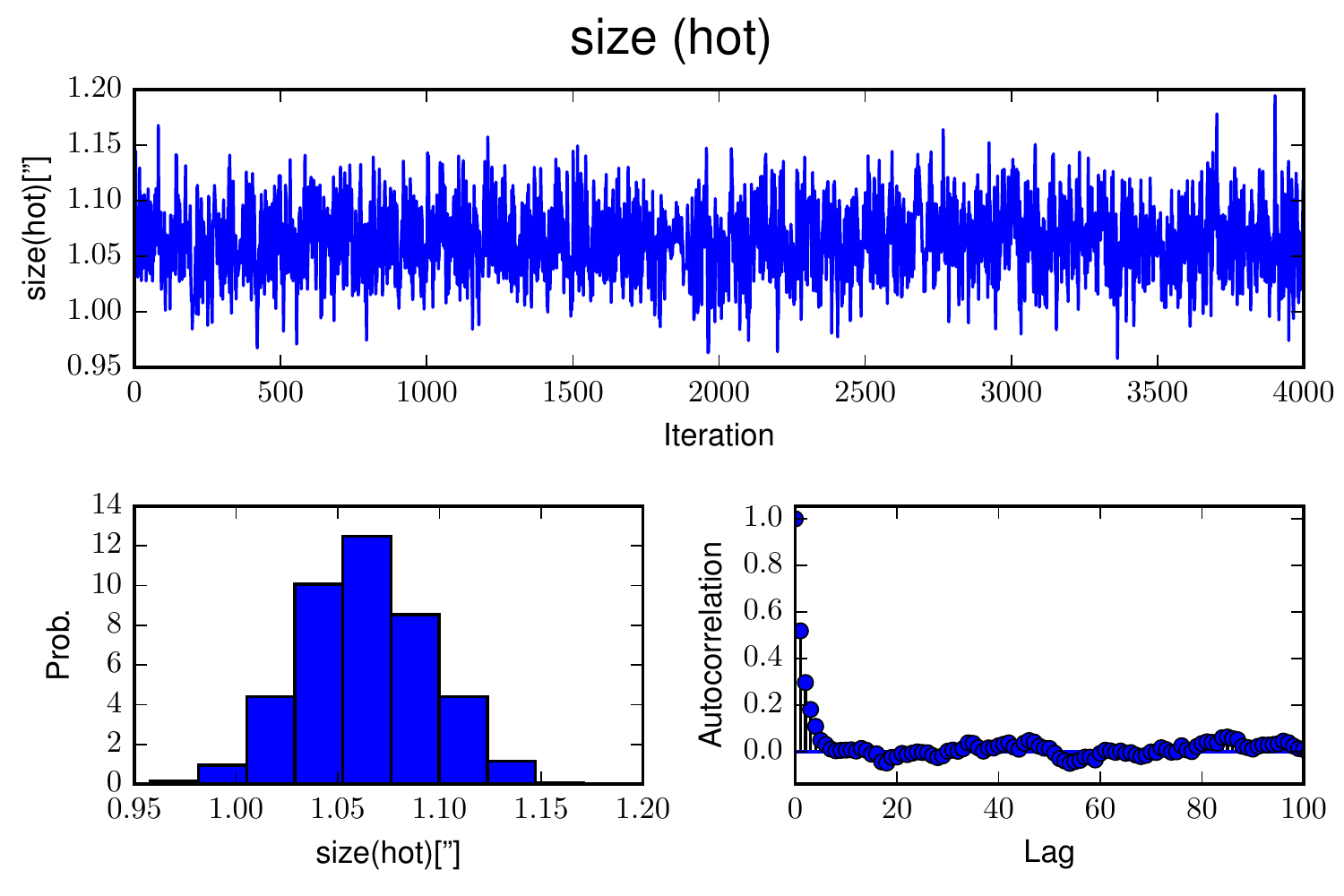}\hspace{2cm}
				\includegraphics[width=0.4\columnwidth]{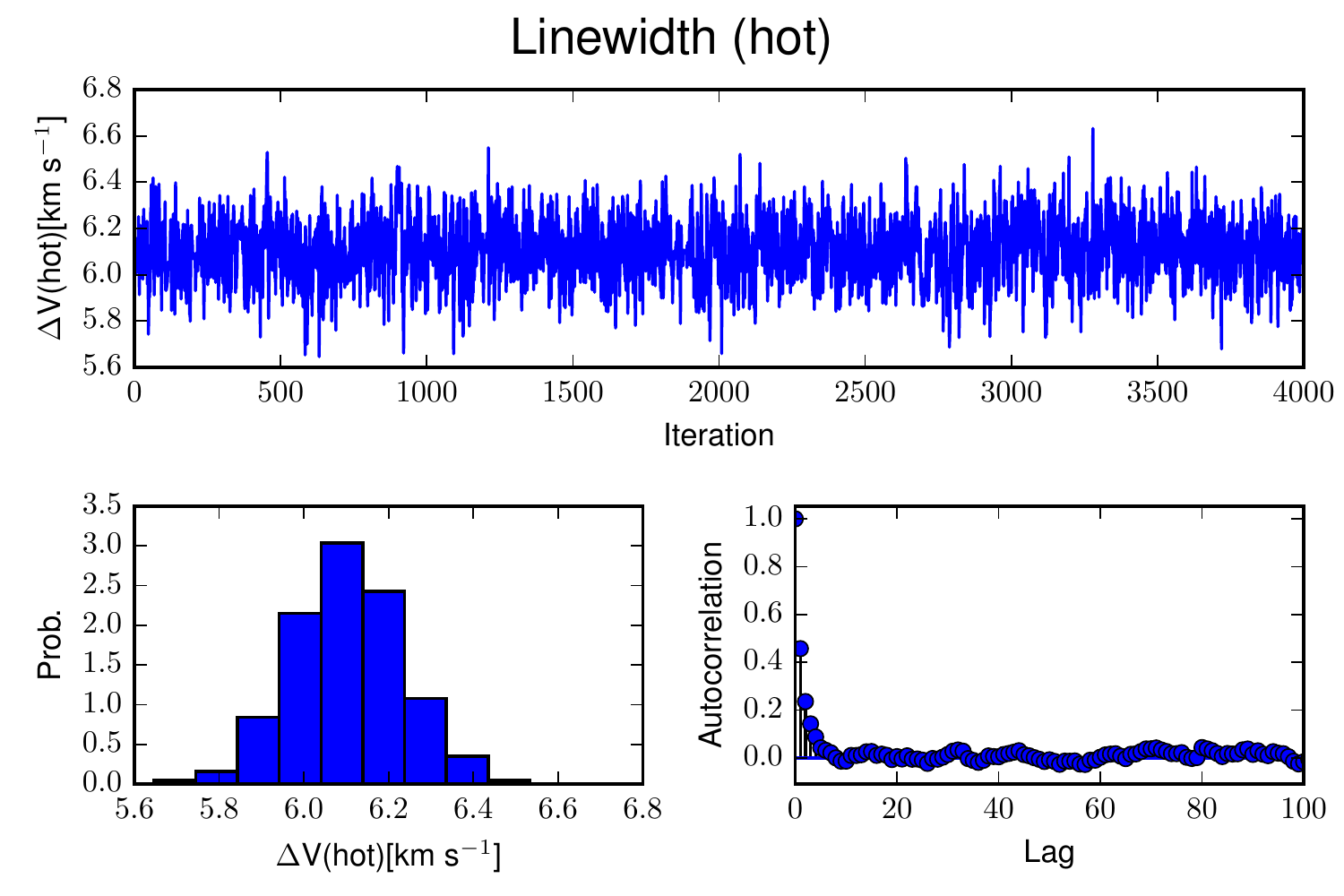}\\
				\includegraphics[width=0.4\columnwidth]{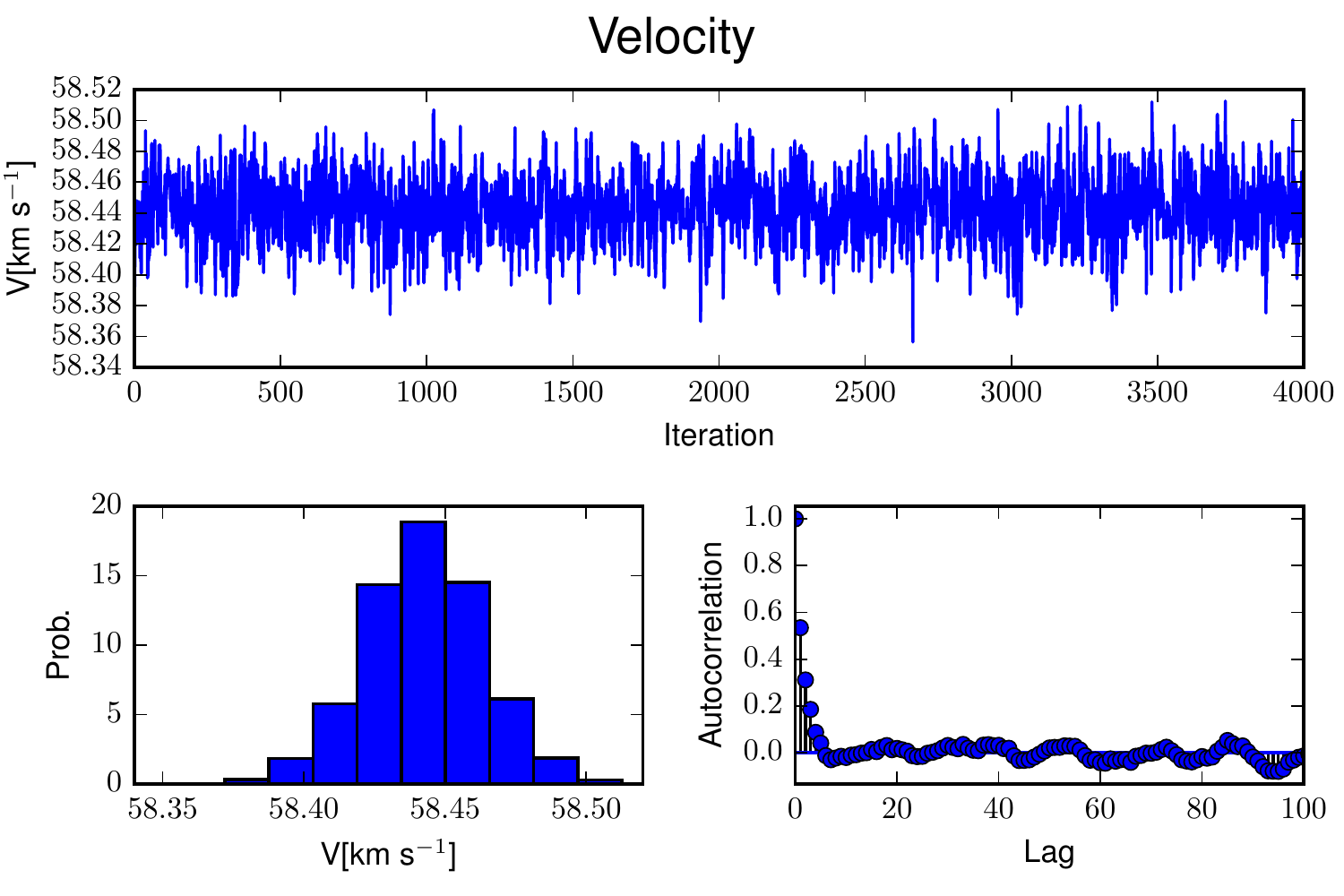}
				\caption{For each parameter, indicated at the top of each panel, we show an example of trace plot (top), probability distribution function (bottom left), and autocorrelation plot (bottom right) for AGAL034.258$+$00.154. The plots refer to \an.}\label{fig:mcmc_ex_g34}
			\end{figure*}}
            When using MCMCs additional convergence tests can be performed: 
            \begin{itemize}
	            \item the Raftery-Lewis procedure \citep{RafteryLewis95_PracticalMCMC_115} estimates the number of iterations requested to achieve convergence, as well as the burn-in period and the thinning factor (see Sect.~\ref{sec:fit_procedure}) to get independent samples;
	            \item the Geweke test \citep{Geweke92_BS4_169}, that checks whether the mean estimates converged by comparing the mean and the variance of the samples in consecutive intervals from the beginning and the end of the chains; 
	            \item and the Gelman-Rubin statistic \citep{GelmanRubin92_BS4_625}, which tests if multiple chains converge to the same distribution from random starting points, comparing the variance within- and between chains.
			\end{itemize}

			An example of the fits produced with MCWeeds in the context of this work are shown in Figs.~\ref{fig:spectral_fit_ma}, \ref{fig:spectral_fit_an}, and \ref{fig:spectral_fit_mt}, \ma, \an, and \mt, respectively. Trace plots, probability distribution functions, and autocorrelation plots are shown in Figs.~\ref{fig:mcmc_ex_g14} and \ref{fig:mcmc_ex_g34} in the online material.

        \subsection{Fit procedure and parameters}\label{sec:fit_procedure}

            The synthetic spectrum is produced according to the following equations \citep[][]{Comito+05_apjs156_127,Maret+11_aap526_47}:
            \begin{equation}
                \tau(\nu) = \frac{c^2}{8\pi\nu^2}\frac{N_{tot}}{Z(\tex)}\Sigma_i A^i g_u^i \expo{-E_u^i/(k_B \tex)} \left(\expo{h\nu_0^i/(k_B \tex)} - 1 \right) \phi^i\label{eq:weeds_tau},
            \end{equation}
            \begin{equation}
                \tb(\nu) = \eta \left[ J_\nu(\tex) - J_\nu(\tbg) \right] \left( 1 - \expo{-\tau(\nu)} \right)\label{eq:weeds_tb},
            \end{equation}
            where $\tau(\nu)$ and $\tb(\nu)$ are the line optical depth and brightness temperature at a given frequency $\nu$; $k_{B}$ is the Boltzmann constant, $h$ is the Planck constant, $N_{tot}$ is the column density of the molecule considered, $Z(\tex)$ is the rotational partition function, $A^i$ is the Einstein coefficient for spontaneous emission, $g_u^i$, $E_u^i$, $\nu_0^i$ are the upper level degeneracy and energy, and the rest frequency of line $i$, and $\phi^i$ is its profile. $\tex$ and $\tbg$ are the excitation temperature and the background temperature, respectively. We assumed that the lines have a Gaussian profile and that $\tbg=2.73\kel$.
            In Eq.~\ref{eq:weeds_tb}, $\eta$ is the beam dilution factor, calculated as:
            \begin{equation}
                \eta = \frac{\vartheta_s^2}{\vartheta_s^2 + \vartheta_{beam}^2}, \label{eq:beam_dil}
            \end{equation}
            where $\vartheta_s$ and $\vartheta_{beam}$ are the source- and beam FWHM sizes. 

            \citet{Wilner+95_apjl449_73}, on the basis of the calculations by \citet{Rowan-Robinson80_apjs44_403} and \citet{WolfireCassinelli86_apj310_207}, show that the temperature structure in a clump heated by a central massive star can be described by the power-law:
            \begin{equation}
                T = 233 \left(\frac{L}{\lsuntab} \right)^{\frac{1}{4}} \left(\frac{R}{\mathrm{AU}} \right)^{-\frac{2}{5}} [K],\label{eq:t_str}
            \end{equation}
            where $R$ is the radius of the region with a temperature $T$, around a star with bolometric luminosity $L$. We therefore inverted this expression to define the size of the emitting region as a deterministic variable, dependent on the bolometric luminosity of the source, and on the temperature generated in each step of the chain (see below). Assigning a size for the emitting region of each species and component allows us to derive $\eta$, correcting for the dilution; this correction considers the appropriate beam for each combination of frequency and telescope.
			
			\begin{figure*}
				\centering
				\includegraphics[width=\textwidth]{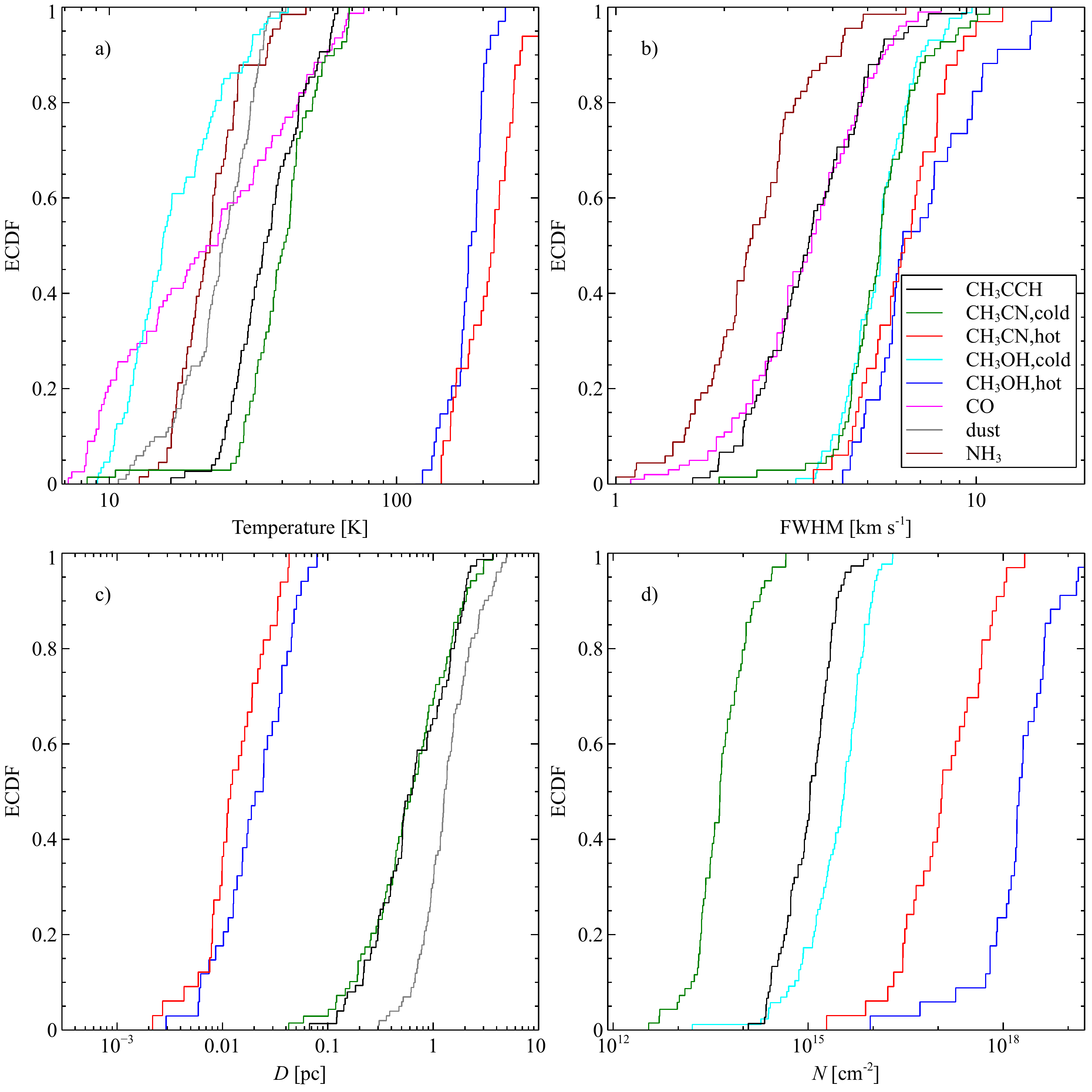}
				\caption{Empirical cumulative distribution functions for temperatures (a), line FWHMs (b), source diameter (c), and molecular column density (d). The different tracers and components are drawn with different colours, as indicated in the legend in the lower right corner of panel b). ``CO'' refers to properties obtained from C$^{17}$O and C$^{18}$O transitions from \citet{Giannetti+14_aa570_65}, and NH$_{3}$ properties are taken from \citet{Wienen+12_aap544_146}. Dust properties from \koenigSedt\ are included only in panels a) and c).}\label{fig:ecdf_T_DV_theta_N}
			\end{figure*}
            \begin{table*}
                \caption{Priors used in the models for the molecules of interest.}\label{tab:priors}
                \centering
                \begin{tabular}{lcccc}
					\hline
					\hline
					CH$_{3}$CN        & Temperature           & Column density   & Linewidth            & Velocity     \\
					                  & [K]                   & [log(cm$^{-2}$)] & [$\kmstab$]          & [$\kmstab$]  \\
					\hline
                    Prior             & Truncated normal      & Normal           & Truncated normal     & Normal       \\
                    Parameters (cool) & $\mu = 50$            & $\mu = 14$       & $\mu = 4$            & $\mu=\vlsr\tablefootmark{a}$ \\
                                      & $\sigma = 15$         & $\sigma = 1.5$   & $\sigma = 2$         & $\sigma = 1$ \\
                                      & $low = 8, high=100$   &                  & $low = 0.5, high=34$ &              \\
                    Parameters (hot)  & $\mu = 150$           & $\mu = 17$       & $\mu = 7$            & $\mu=\vlsr\tablefootmark{a}$  \\
                                      & $\sigma = 80$         & $\sigma = 2$     & $\sigma = 2$         &              \\
                                      & $low = 100, high=650$ &                  & $low = 1.5, high=37$ &              \\
					\hline
					CH$_{3}$CCH       & Temperature           & Column density   & Linewidth            & Velocity     \\
					                  & [K]                   & [log(cm$^{-2}$)] & [$\kmstab$]          & [$\kmstab$]  \\
					\hline
                    Prior             & Truncated normal      & Normal           & Truncated normal     & Normal       \\
                    Parameters        & $\mu = 50$            & $\mu = 14$       & $\mu = 4$            & $\mu=\vlsr\tablefootmark{a}$  \\
                                      & $\sigma = 15$         & $\sigma = 1.5$   & $\sigma = 2$         & $\sigma = 1$ \\
                                      & $low = 8, high=150$   &                  & $low = 0.5, high=34$ &              \\
					\hline
					CH$_{3}$OH        & Temperature           & Column density   & Linewidth            & Velocity     \\
					                  & [K]                   & [log(cm$^{-2}$)] & [$\kmstab$]          & [$\kmstab$]  \\
					\hline
                    Prior             & Truncated normal      & Normal           & Truncated normal     & Normal       \\
                    Parameters (cool) & $\mu = 20$            & $\mu = 14$       & $\mu = 4$            & $\mu=\vlsr\tablefootmark{a}$  \\
                                      & $\sigma = 30$         & $\sigma = 1.5$   & $\sigma = 3$         & $\sigma = 1$ \\
                                      & $low = 8, high=100$   &                  & $low = 0.5, high=12$ &              \\
                    Parameters (hot)  & $\mu = 200$           & $\mu = 18$       & $\mu = 7$            & $\mu=\vlsr\tablefootmark{a}$  \\
                                      & $\sigma = 50$         & $\sigma = 1.5$   & $\sigma = 3$         &              \\
                                      & $low = 100, high=700$ &                  & $low = 1, high=17$   &              \\
                    \hline
                \end{tabular}
                \tablefoot{\tablefoottext{a}{The mean for the radial velocity is set equal to the measured velocity in C$^{17}$O (3--2) from \citet{Giannetti+14_aa570_65} (if C$^{17}$O $J=3\rightarrow2$ was not detected, we use the velocity of N$_{2}$H$^{+} J=1\rightarrow0$ from the 3~mm surveys; see Urquhart et al., in prep., and \citealt{Csengeri+16_aap586_149}).}
            }
            \end{table*}

            With respect to conventional rotation diagrams, our method has several advantages: 
                firstly, the opacity is taken into account for each line, 
                secondly, all lines have the same intrinsic width, because line broadening by optical depth effects is taken into account, 
                and lastly, blending of the lines can be taken into account including the contaminating species in the routine.

            The spectral ranges used and the priors considered in the model for each molecule of interest are listed in Tables~\ref{tab:spectral_ranges} and \ref{tab:priors}, respectively. 
            We used loosely-informative priors for the parameters of the fit, choosing broad Gaussian functions centred on representative of typical values of temperatures, column densities and linewidths in regions of high-mass star formation \citep[e.g.][]{vanderTak+00_aap361_327,Fontani+02_aa389_603,Beuther+02_apj566_945}. For the source radial velocity we chose a mean ($\mu$) equal to the measured velocity in C$^{17}$O(3--2) from \citet{Giannetti+14_aa570_65} (if C$^{17}$O $J=3\rightarrow2$ was not detected, we use the velocity of N$_{2}$H$^{+} J=1\rightarrow0$ from the 3~mm surveys; see Urquhart et al., in prep., and \citealt{Csengeri+16_aap586_149}). For the linewidths and the temperatures we used truncated Gaussians (the lower and upper limits are indicated in Table~\ref{tab:priors} with ``low'' and  ``high'', respectively), in order to exclude very low, highly unlikely values for the analysed data; the upper limits are sufficiently high to have no impact on the procedure. For the hot component we increased the mean of the priors, in agreement with the idea that the material in the hot core is warmer and more turbulent; we also scaled the mean of the column density distribution using Eq.~\ref{eq:beam_dil}, assuming a size of $\sim1\arcsec$. As a final remark, in the case of temperature, we made use of the prior to keep the cool and the hot components distinct, by defining non-overlapping domains, separated at $100\kel$.
            Multiple priors were tested, to ensure that the results are not significantly influenced by this choice; the fit has very strong observational constraints that dominate the information.
            Values of the rotational partition functions used, evaluated for a set of temperatures, are listed in Table~\ref{tab:part_funct}. 
            The values of the partition function $Z$ for the temperatures in our modelling were determined by interpolation, using a linear fit of the measured values in a $log(T)$ vs. $log(Z)$ space.
            
            In order to obtain the full posterior distribution, and thus the most accurate estimates for the parameters, as well as their uncertainties, the spectra were fitted using MCMCs with an adaptive Metropolis-Hastings sampler as the fitting algorithm \citep{Haario+01_Bernoulli7_223}. The number of iterations for \ma\ and \mt\ was set to 210000 for each source, the burn-in \citep[number of iterations to be discarded at the beginning of the chain, when the distribution is not yet representative of the stationary one;][]{RafteryLewis95_PracticalMCMC_115} was chosen to be $50000$, and the thinning factor \citep[number of iterations to be discarded between one sampling and the next, to minimise correlation between successive points; ][]{RafteryLewis95_PracticalMCMC_115} was set to 20 for \ma\ and to 50 for \mt. The adaptive sampling was delayed by 20000 iterations \citep[i.e. the number of iterations performed before starting to tune the standard deviation of the proposal distribution;][]{Haario+01_Bernoulli7_223}. For \an\ we used longer chains with 260000 iterations, a burn-in of 60000 and a delay of 30000; we used a thinning factor of 50 for this species as well.
            \begin{table*}
                \caption{Measurements of the rotational partition function for the species of interest.}\label{tab:part_funct}
                \centering
                \begin{tabular}{lrrrrrrr}
                    \hline
                    \hline
                    Temperatures [K]                 & 9.375  & 18.75   & 37.5   & 75      & 150     & 225     &  300    \\
                    \hline 
                    Q(\an)                           & 64.1   & 164.3   & 449.1  & 1267.7  & 3807.2  & 8044.5  & 14682.5 \\
                    Q(CH$_{3}$$^{13}$CN)             & 64.1   & 164.4   & 449.3  & 1265.6  & 3577.7  & 6573.6  & 10122.8 \\
                    Q(\mt)                           & 19.5   & 68.7    & 230.3  & 731.1   & 2437.8  & 5267.4  & 9473.3  \\
                    Q(\ma)                           & 33.4   & 88.3    & 241.3  & 679.7   & 1920.9  & 3524.5  & 5428.8  \\
                    \hline
                \end{tabular}
                \tablefoot{The partition functions were obtained from the JPL database.}
            \end{table*}
            For each molecule and iteration, a set of column density, velocity, linewidth and temperature was generated. According to Eq.~\ref{eq:t_str}, from the generated temperature and bolometric luminosity of the source \citep[taken from][]{Koenig+17_aap599_139}, a size of the emitting region was computed and translated to arcseconds using the distances listed in Table~\ref{tab:source_properties}. For isotopologues, all parameters were the same, except for column density, which was reduced by a factor equal to the relative isotopic abundance. 
            In order to take into account the uncertainty in the relative calibration of different telescopes and wavelengths, we used different calibration factors for the separate $J$ multiplets of \ma\ and \an. This allowed the programme to rescale the spectra by a normally-distributed factor with mean one and $\sigma=0.07$, according to Sect.~\ref{sec:obs}; this distribution was truncated at a maximum of $30\%$ in either direction.
            
            In order to account for contamination and for strong lines in the spectra, CH$_{3}$OCHO, C$_{2}$H$_{5}$CN, $^{33}$SO$_{2}$, and c-C$_{3}$H were included in the model of methyl acetylene (these lines are relevant only for the $J=20\rightarrow19$ multiplet in bright sources, like the one shown in Fig.~\ref{fig:spectral_fit_ma}, right); CCH, CH$_{3}$OCHO, and \mt\ are considered for that of acetonitrile. 
            All methyl acetylene lines, from $(5-4)$ to $(20-19)$, are well described by a single temperature component; on the other hand, \an\ and \mt\ need a second temperature component for sources detected in the highest excitation lines, as discussed in Sect.~\ref{sec:fit_results}. The radial velocity of both temperature components is assumed to be the same, whereas all other parameters are independent. To take into account the variation of the $^{12}$C/$^{13}$C isotopic ratio with galactocentric distance, we use the relation
            \begin{equation}
				\frac{[^{12}\mathrm{C}]}{[^{13}\mathrm{C}]} = 6.1 D_{GC} [\mathrm{kpc}] + 14.3\label{eq:c_isotopic_ratio}
			\end{equation}
			from \citet{Giannetti+14_aa570_65}, and we fix the carbon isotopic abundance for each source to the appropriate value obtained with this expression.

        \section{Detection rates and fit results}\label{sec:fit_results}

            In the following we consider only sources with a `good detection', that is, those sources detected with a $S/N > 3$ for at least two transitions. 
            Detection rates for individual molecules are listed in Table~\ref{tab:det_rates_total}, and the number of observed sources per species is indicated. 
            The table also shows how the detection rate depends on the evolutionary class for the TOP100, increasing with time, from $30-40\%$ for sources quiescent at 70 microns, up to $95-100\%$ for \hii\ regions.             
            In addition, for the analysis we discard all the sources with a $95\%$ highest probability density (HPD) interval exceeding $20\kel$ for cool components and $60\kel$ for hot components. This removes sources with excessive uncertainties in the temperatures. Tests with model spectra show that in these cases the best-fit temperature is overestimated.
 
            \begin{table}
                \caption{Detection rates for the different species considered in this work.}\label{tab:det_rates_total}
                \centering
                \begin{tabular}{lccccc}
                    \hline
                    \hline
                    \an\                     & Obs. sources  & cool          & 90\% CI\tablefootmark{a} & hot           & 90\% CI     \\
                    \hline                                                                                                            
                    Total $[\%]$             & 99            & 71            & 63-78                    & 34            & 27-42       \\       
                    \hii\ $[\%]$             & 22            & 100           & 94-100                   & 77            & 61-89       \\       
                    IRb $[\%]$               & 33            & 76            & 62-86                    & 36            & 24-51       \\       
                    IRw $[\%]$               & 31            & 61            & 47-75                    & 16            & 8-29        \\       
                    70w $[\%]$               & 13            & 31            & 11-58                    & 0             & 0-13        \\       
                    \hline
					\mt\                     &               &               &                          &              &              \\
					\hline                   
                    Total $[\%]$             & 100           & 88            & 82-93                    & 35           & 28-43        \\
                    \hii\ $[\%]$             & 22            & 100           & 94-100                   & 82           & 66-92        \\
                    IRb $[\%]$               & 34            & 94            & 85-98                    & 32           & 21-46        \\
                    IRw $[\%]$               & 30            & 93            & 83-98                    & 20           & 10-34        \\
                    70w $[\%]$               & 14            & 43            & 23-64                    & 0            & 0-9          \\
                    \hline
					\ma\                     &               &               &                          &              &              \\
					\hline                   
                    Total $[\%]$             & 99            & 78            & 70-84                    & \dots        & \dots        \\
                    \hii\ $[\%]$             & 22            & 95            & 84-99                    & \dots        & \dots        \\
                    IRb $[\%]$               & 33            & 85            & 73-93                    & \dots        & \dots        \\
                    IRw $[\%]$               & 31            & 74            & 60-85                    & \dots        & \dots        \\
                    70w $[\%]$               & 13            & 38            & 19-61                    & \dots        & \dots        \\
                    \hline
                \end{tabular}
                \tablefoot{The columns indicate the number of observed sources, the detection rates for the cool and hot components with their respective $90\%$ credible interval. \tablefoottext{a}{The credible interval is computed using the Jeffreys prior interval for the binomial distribution \citep[cf.][]{Brown+01_statsci2_16}.}}
            \end{table}

            \begin{table}
                \caption{Properties obtained with the different species considered in this work.}\label{tab:typical_properties}
                \addtolength{\tabcolsep}{-1pt}
                \centering
                \begin{tabular}{lcccc}
					\hline                                                              
                    \hline
                    \an\                     & cool          & 90\% CI                  & hot           & 90\% CI     \\
                    \hline
					$T$[K]                   & 40.2          & 27.9-67.1                & 218           & 143-294     \\     
					$N[\cmtab^{-2}]$         & 4.4e13        & (1.0-25.0)e13            & 1.2e17        & (0.3-10.9)e17\\       
					$\Delta V[\kmstab]$      & 5.4           & 3.9-8.9                  & 6.6           & 4.2-9.7     \\     
					$\chi$                   & 4.6e$-$10     & (1.8-11.7)e$-$10         & 1.6e$-$7      & (0.2-8.7)e$-$7\\     
                    \hline
                    \mt\                     &               &                          &               &             \\
                    \hline
					$T$[K]                   & 15.3          & 10.1-31.5                & 180           & 134-217     \\
					$N   [\cmtab^{-2}]$      & 3.6e15        & (0.4-11.1)e15            & 1.8e18        & (0.1-13.7)e18\\
					$\Delta V[\kmstab]$      & 5.4           & 3.7-8.2                  & 6.2           & 4.5-14.2    \\            
					$\chi$                   & 2.3e$-$8      & (0.4-10.6)e$-$8          & 2.6e$-$6      & (0.3-11.8)e$-$6\\                   
                    \hline
                    \ma\                     &               &                          &               &             \\
                    \hline
					$T$[K]                   & 34.5          & 24.0-59.1                & \dots         & \dots       \\
					$N   [\cmtab^{-2}]$      & 1.1e15        & (0.2-3.6)e15             & \dots         & \dots       \\
					$\Delta V[\kmstab]$      & 3.4           & 1.9-6.4                  & \dots         & \dots       \\
					$\chi$                   & 1.2e$-$8      & (0.5-2.8)e$-$8           & \dots         & \dots       \\
                    \hline                                                              
                \end{tabular}
                \tablefoot{The columns indicate the median properties and their $90\%$ credible interval for the cool and hot components.}
                \addtolength{\tabcolsep}{1pt}
            \end{table}
                
                The extensive set of observations that we have for the TOP100 now permits us a consistent comparison of properties of the gas as traced by different molecules.
                Examples of the fit obtained with MCWeeds for \ma, \an, and \mt\ are shown in Figs.~\ref{fig:spectral_fit_ma}, \ref{fig:spectral_fit_an}, and \ref{fig:spectral_fit_mt}, respectively. The complete results of the fitting procedure are reported in Tables~\ref{tab:fit_res_ma} to \ref{tab:fit_res_mt_hot}; figures are incorporated in the ATLASGAL database, and can be accessed at \url{atlasgal.mpifr-bonn.mpg.de/top100}. In addition to the fit images, the trace of the chain, the histograms of the distributions, and the autocorrelation plots are available for each of the parameters of interest. Figures~\ref{fig:mcmc_ex_g14} and \ref{fig:mcmc_ex_g34} show examples of these plots for two sources in the TOP100.

                In \citet{Giannetti+14_aa570_65} several sources in the TOP100 show multiple velocity components along the line-of-sight in low-$J$ transitions of CO isotopologues. Because of the line series of the methyl acetylene and acetonitrile, and the close-by trasitions in the methanol band, multiple velocity components may complicate the derivation of temperatures and column densities.
                In contrast to what is found with CO isotopologues, the species considered in this work have relatively simple spectra, characterised by a single velocity component. The only source clearly showing two components is AGAL043.166+00.011, which was excluded from the following analysis.
                The difference in velocity among different tracers is within $1\kms$, in line with the spectral resolution of the data. This, in combination with the fact that only a single velocity component is observed for the more complex species considered in this work, makes it easy to identify the main contributor to the dust continuum emission observed in ATLASGAL and HiGAL \citep{Molinari+10_pasp122_314} and analysed in \koenigSedt.
                
                Temperatures obtained from the data of different species can be systematically different, but they are always correlated, as discussed in more detail in Sect.~\ref{sec:discussion}. 
                The properties of the emitting material for the different tracers are summarised in Table~\ref{tab:typical_properties} and illustrated in Fig.~\ref{fig:ecdf_T_DV_theta_N}, which shows the empirical cumulative distribution function (ECDF, built with all good detections, using the best-fit results) of: a) temperature, b) line FWHM, c) size of the emitting region and d) column density for all available species.
                In addition, thanks to the association of a spatial scale for the emitting region of different species and temperature components, we could correct for the beam filling factor, deriving source-averaged abundances relative to H$_{2}$ (see Fig.~\ref{fig:ECDF_abundances} and Table~\ref{tab:typical_properties}). Opacity effects were also accounted for directly in the models (cf. Eq.~\ref{eq:weeds_tau} and \ref{eq:weeds_tb}), and by simultaneously fitting the $^{13}$C-substituted isotopologue of acetonitrile. In deriving the abundances we considered two cases, comparing the size of the emitting region ($\vartheta_{s}$), as obtained from the fit, with the beam size ($\vartheta_{b}$): $\vartheta_{b} \gtrsim \vartheta_{s}$ and $\vartheta_{b} \ll \vartheta_{s}$.
                In the former case, we computed the abundance as the ratio:
                \begin{equation}
                    \chi = \frac{\overline{n_{mol}}}{\overline{n_{H_{2}}}},\label{eq:abundance_source_av}
                \end{equation}
                where $\overline{n_{mol}} = N_{mol}/(2R)$ and $\overline{n_{H_{2}}} = M(r<R)/((4\pi/3)R^{3})$. Here $M(r<R)$ indicates the mass within a radius R derived from the fit, obtained by scaling the total mass from the SED fit assuming that $n_{H_{2}} \propto r^{-1.5}$, consistent with the measurements by \citet{Beuther+02_apj566_945}.
                Assuming a different value for the power-law slope has an impact on the measured abundances; the effect of varying the exponent between $-1$ and $-2$ is discussed in the next sections. 
                On the other hand, if the beam size is much smaller than the size of the emitting region, we derive the abundance in the usual way, as the ratio of column densities:
                \begin{equation}
                    \chi = \frac{N_{mol}}{N_{H_{2}}};\label{eq:abundance_beam_av}
                \end{equation}
                the column density of molecular hydrogen is taken from \koenigSedt.
                Below we briefly summarise the fit results for each molecule.

            \subsection{Methyl acetylene}\label{sec:results_MA}
            
                The simultaneous fit of the (5--4), (6--5) and (20--19) series indicates that CH$_{3}$CCH emission can be satisfactorily reproduced with a single temperature component.
                Figure~\ref{fig:ecdf_T_DV_theta_N} shows that the temperatures derived using methyl acetylene range between $20\kel$ and $60\kel$, associated with narrow linewidths, consistent with those observed for C$^{17}$O(3--2). The fit shows that the emitting region for CH$_{3}$CCH is of the order of $0.5\pc$ (Fig.~\ref{fig:ecdf_T_DV_theta_N}), or $25\arcsec$ at $4\kpc$. Together with the necessity of a single temperature component to fit lines with a wide range of excitations and the non-detection of high-excitation lines, this indicates that this species is tracing the bulk of the dense gas in the clump, and it is not significantly enhanced at high temperatures. 
                                
\begin{figure}
                    \centering
                    \includegraphics[width=\columnwidth]{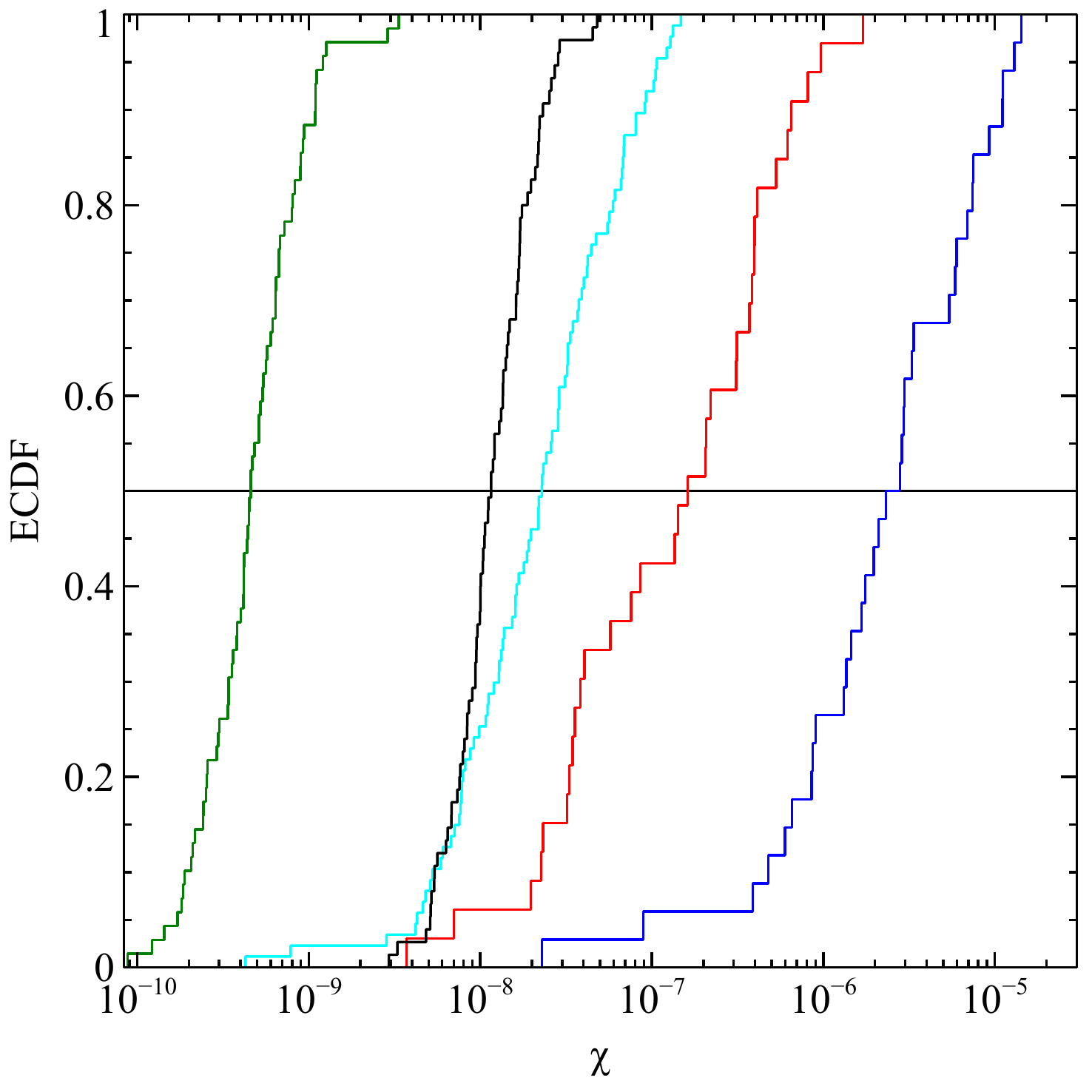}
                    \caption{Empirical cumulative distribution function for the abundance relative to molecular hydrogen.
                     \ma, and the hot and cool components of \an\ and \mt\ are drawn with different colours, as indicated in panel b) of Fig.~\ref{fig:ecdf_T_DV_theta_N}.
                     }\label{fig:ECDF_abundances}
                \end{figure}

                The minor systematic adjustments to the calibration factors (see Sect.~\ref{sec:fit_procedure}) of the spectra for different $J$ transitions lends support to the claim that this molecule is not too far from LTE, as discussed in \citet{Bergin+94_apj431_674}. In fact, if it were subthermally excited, high-$J$ transitions would be systematically fainter.
                The column densities of CH$_{3}$CCH are found to be in the range $2.3\times\pot{14}-3.6\times\pot{15}\cm^{-2}$ (central $90\%$ credible interval, hereafter CI; all ranges refer to this interval), corresponding to abundances with respect to H$_{2}$ between $5\times\pot{-9}$ and $2.8\times\pot{-8}$ (Fig.~\ref{fig:ECDF_abundances} and Table~\ref{tab:typical_properties}), calculated with Eq.~\ref{eq:abundance_source_av}. Varying the power-law index for the density profile by $\pm0.5$ changes the abundance by approximately $\sim20-30\%$ up or down, depending on the value of the exponent.
                Linewidths are narrow, distributed between $1.9-6.4\kms$ (Fig.~\ref{fig:ecdf_T_DV_theta_N}), with a median of $\sim 3.5\kms$.

            \subsection{Acetonitrile}\label{sec:results_AN}
            
                Combining the (5--4) and (6--5) transitions with the (19--18), we find that two temperature components are necessary to reproduce all observed lines for the 34 sources detected in the line series observed as part of the FLASH$^+$ survey. On the one hand, warm gas dominates at $3\mm$, with temperatures virtually always in the range $\sim30\kel-80\kel$ (hereafter the cool component), and column densities between $1.0\times\pot{13}-2.6\times\pot{14}\cm^{-2}$. This component also contributes significantly to the $K=0,1$ of the (19--18), but it is not sufficient to excite the higher-$K$ lines. The linewidths characterising acetonitrile are broader than those of C$^{17}$O and methyl acetylene, with a median of $\sim5.4\kms$ ($3.9-8.9\kms$ $90\%$ CI), showing that the material traced is more turbulent.
                The emitting region of this component is of the order of $\sim0.4\pc$, or $\sim20\arcsec$ at $4\kpc$. According to Eq.~\ref{eq:abundance_source_av} the median abundance of CH$_{3}$CN in the warm gas is $\sim4.6\times\pot{-10}$. The variation of the density profile imply variations of a factor $1.5$ for the abundance.
                When properly accounting for opacity effects and contamination by hot material, we find that acetonitrile is not exclusively a hot core tracer, but is abundant also at lower temperature, extending over dense regions in the clump.
                On the other hand, hot gas is responsible for the emission in high-excitation lines. To adequately reproduce the (19--18) multiplet, temperatures of $\sim200\kel$ are needed (Fig.~\ref{fig:spectral_fit_an}, right); the emitting region is of the order of $0.01\pc$, broadly in line with typical properties of hot cores \citep[e.g.][]{Cesaroni05_IAUS227_59,Sanna+14_aap565_34}.
                The very small emitting region implies very high column densities, with a median value of $\sim\pot{17}\cm^{-2}$ (due to the small filling factor), which translates to a median abundance of $1.6\sim\pot{-7}$. If the power-law index for the density profile changes by $\pm0.5$, the abundance is scaled by approximately a factor of $\sim10$ in each direction.
                The lines are sligthly broader than for the cool component, with a median of $\sim6.6\kms$.

            \subsection{Methanol}\label{sec:results_MT}
            
                The FLASH$^+$ survey covers the $J_{k}$ (7--6) band, both from the torsional ground state ($\nu_{t}=0$) and the first torsionally-excited state ($\nu_{t}=1$). The significant difference in excitation of the lines in the two torsionally excited states makes our observations sensitive to both hot and cold gas; in fact, a temperature in excess of $150 \kel$ is needed to reproduce the lines of the first torsionally-excited state for the band at $337\,\giga\hertz$.
                
                Methanol is the species with the highest absolute detection rate, with only a very weak dependence on the evolutionary class (see Table~\ref{tab:det_rates_total}). A total of 35 sources are detected also in the $\nu_{t}=1$ state, which mainly belong to the most evolved sources. Similarly to CH$_{3}$CN, in these sources the full set of lines is well reproduced only when including a second, hot component in the fit.
                CH$_{3}$OH traces the lowest temperatures among all the considered tracers, between $10-30\kel$. 
                Methanol is, as mentioned in Sect.~\ref{sec:selection_tracers}, a slightly-asymmetric rotor and the line ratios for this species are sensitive to both temperature and density \citep{Leurini+04_aap422_573,Leurini+07_aap466_215}. Therefore, the excitation temperature measured under the assumption of LTE may not be representative of the emitting gas kinetic temperature, and represents a lower limit for this quantity. \citet{Leurini+07_aap466_215} perform non-LTE modelling of this molecule in a sample of high-mass star-forming regions, showing that kinetic temperatures of the emitting gas range between $17$ and $36\kel$; the effect of non-thermalisation on the measured temperature is therefore non-negligible, but likely small. A direct comparison between LVG and LTE modelling for the same source is carried out by \citet{Leurini+04_aap422_573}, finding similar column densities, but a lower temperature ($10$ vs. $17\kel$) assuming LTE.
                In our model of a spherical clump with a power-law gradient in temperature, the emitting region is always more extended than the beam for the cool component. The column densities are found to range between $4.0\times\pot{14}\cm^{-2}$ and $1.1\times\pot{16}\cm^{-2}$, which are translated to a median abundance of $\sim2.3\times\pot{-8}$ (Fig.~\ref{fig:ECDF_abundances}), again demonstrating that methanol is very abundant also in cold gas. 
                The temperatures traced by torsionally-excited lines on the other hand, are only marginally lower than those determined with acetonitrile, with a median of $\sim180\kel$, coming from a region a few hundredths of a parsec in size. The properties of the gas traced by these transitions are in good agreement with interferometric studies \citep[e.g.][]{Leurini+07_aap475_925,Beuther07_aap468_1045}. 
                CH$_{3}$OH, as expected, is more abundant in hot gas on average by a factor $\sim100$, due to the evaporation from grains (Fig.~\ref{fig:ECDF_abundances}). If the density power-law exponent is changed between $-1$ and $-2$, the abundance in the hot gas is modified by a factor $\sim5$ up and down.
                The linewidths of methanol transitions are consistent with those of CH$_{3}$CN in cold gas, although the temperatures derived are closer to C$^{17}$O, and marginally larger in hot gas.

				\begin{figure}
					\centering
					\includegraphics[width=\columnwidth]{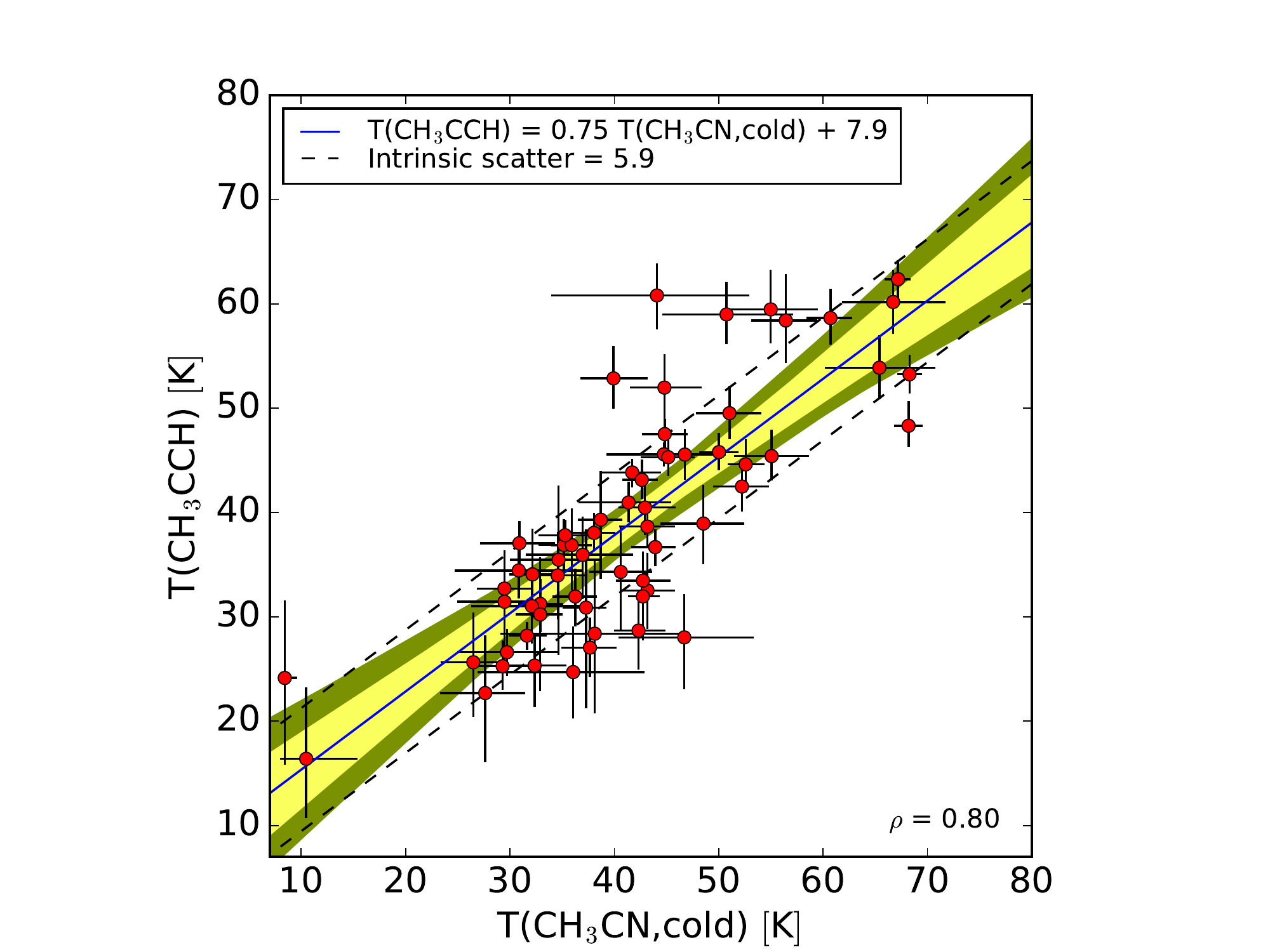}  
					\caption{Correlation between the temperature as measured by CH$_{3}$CN and CH$_{3}$CCH. All the other combinations between the molecules analysed are included in the online material in Fig.~\ref{fig:temperature_correlations_apdx}. The best fit is shown as a blue solid line. The uncertainties due to the line parameters are indicated by the shaded region - dark- and light yellow for $68\%$ and $95\%$ HPD intervals, respectively. The intrinsic scatter in the relation is indicated by the dashed black lines. At the bottom right of the panels the Spearman correlation coefficient is indicated.}\label{fig:temperature_correlations}
				\end{figure}
				\onlfig{
				\begin{figure*}
					\centering
					\includegraphics[width=0.33\textwidth]{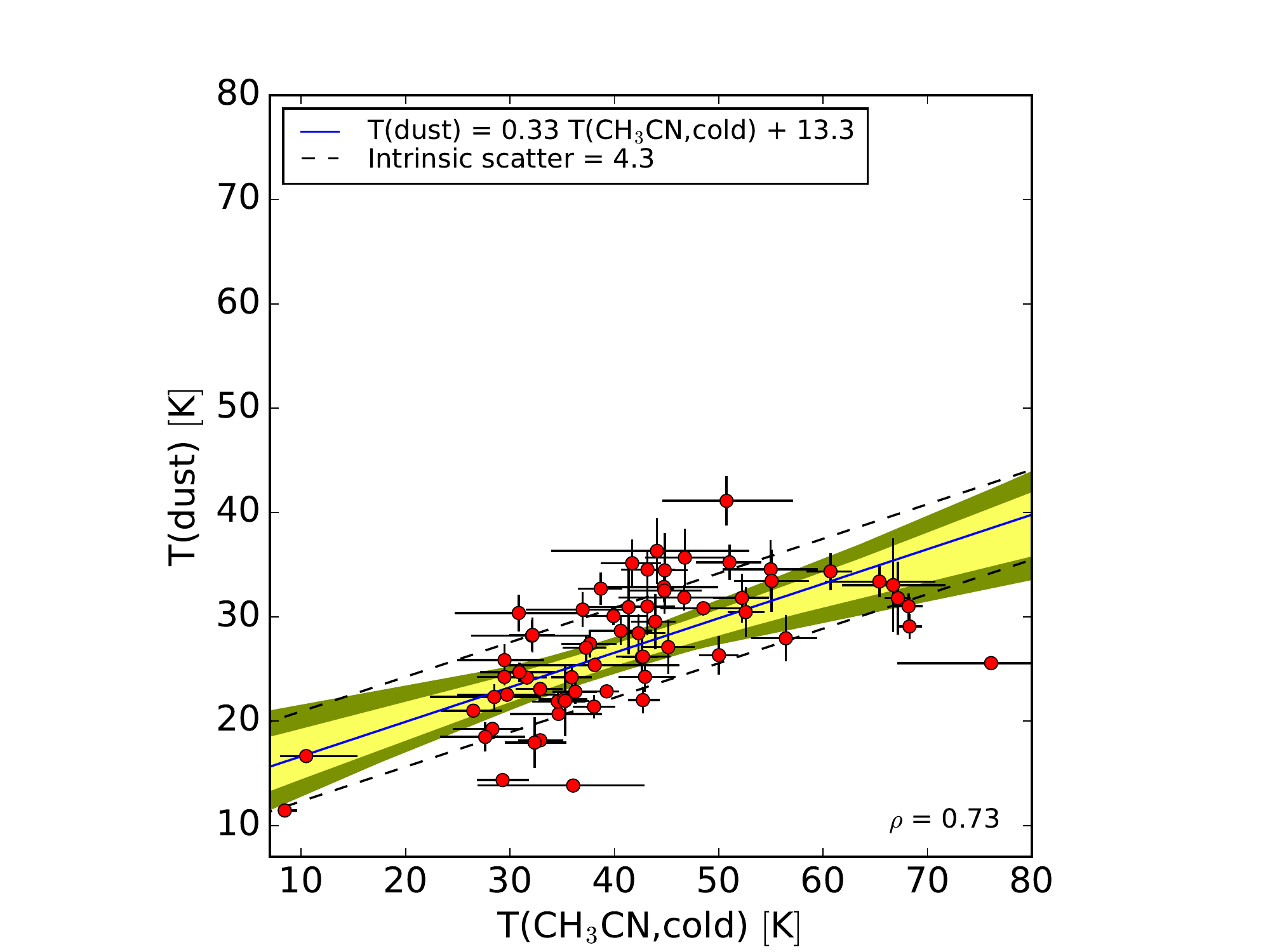}  
					\includegraphics[width=0.33\textwidth]{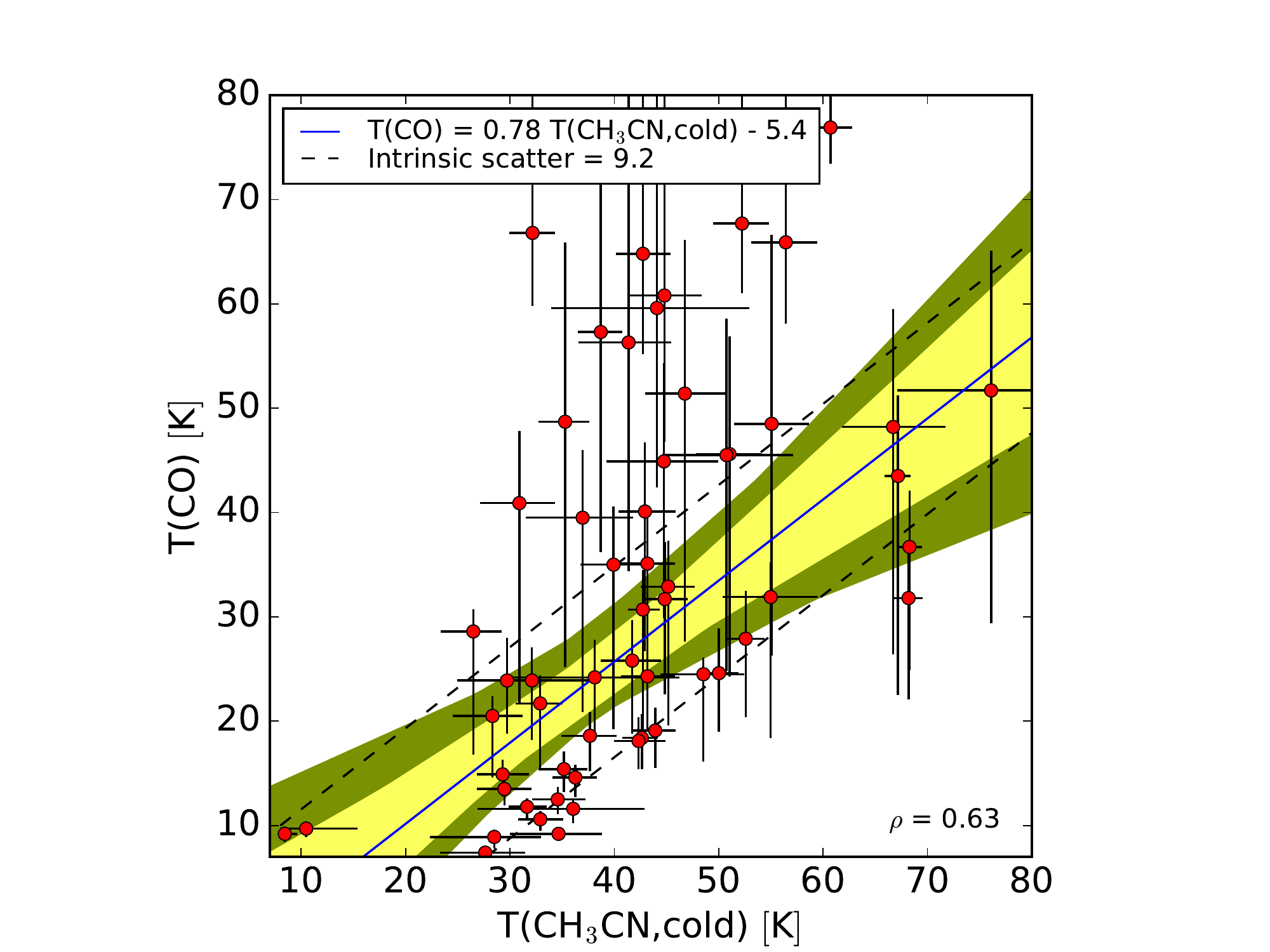}    
					\includegraphics[width=0.33\textwidth]{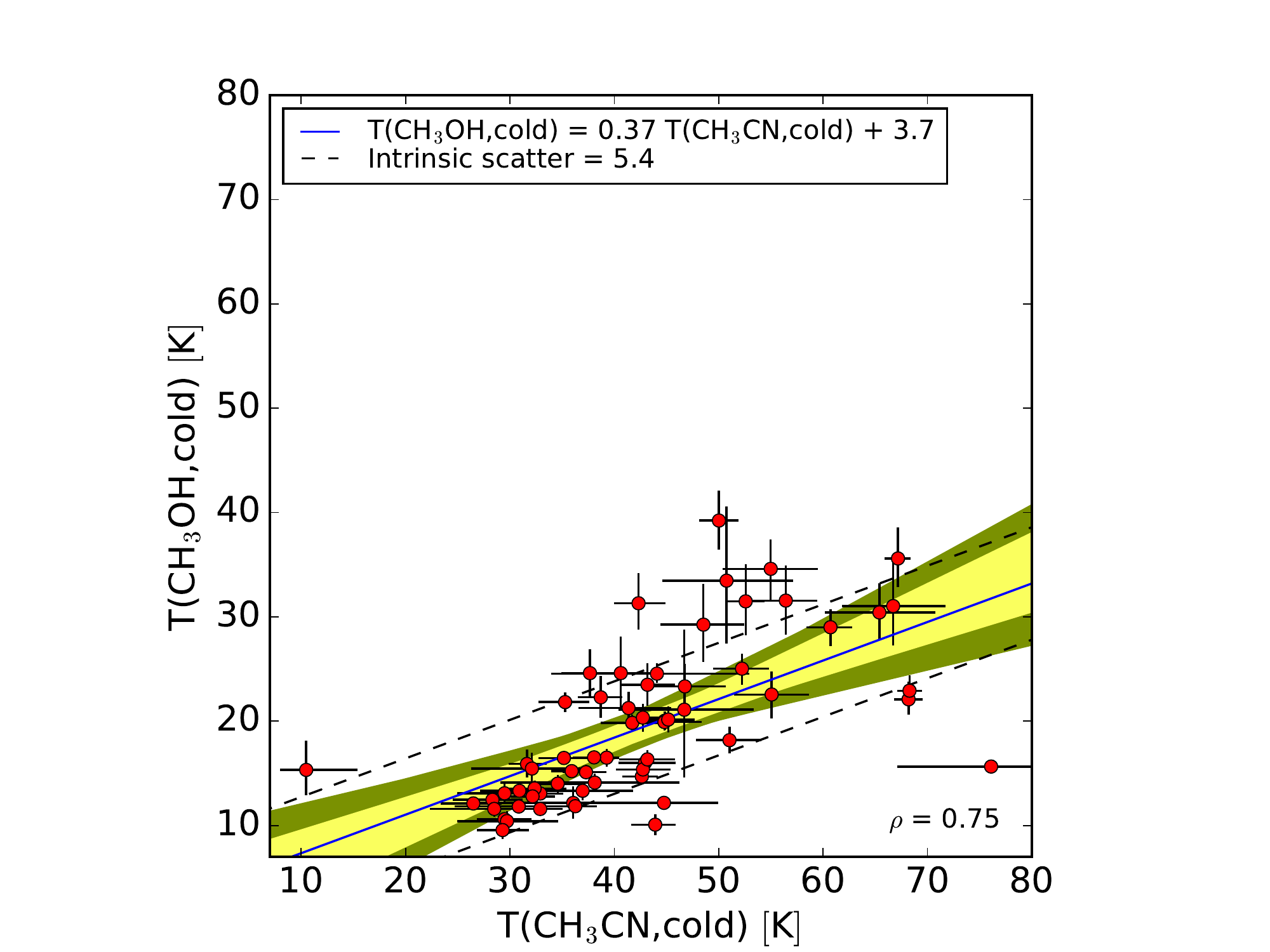}    \\
					\includegraphics[width=0.33\textwidth]{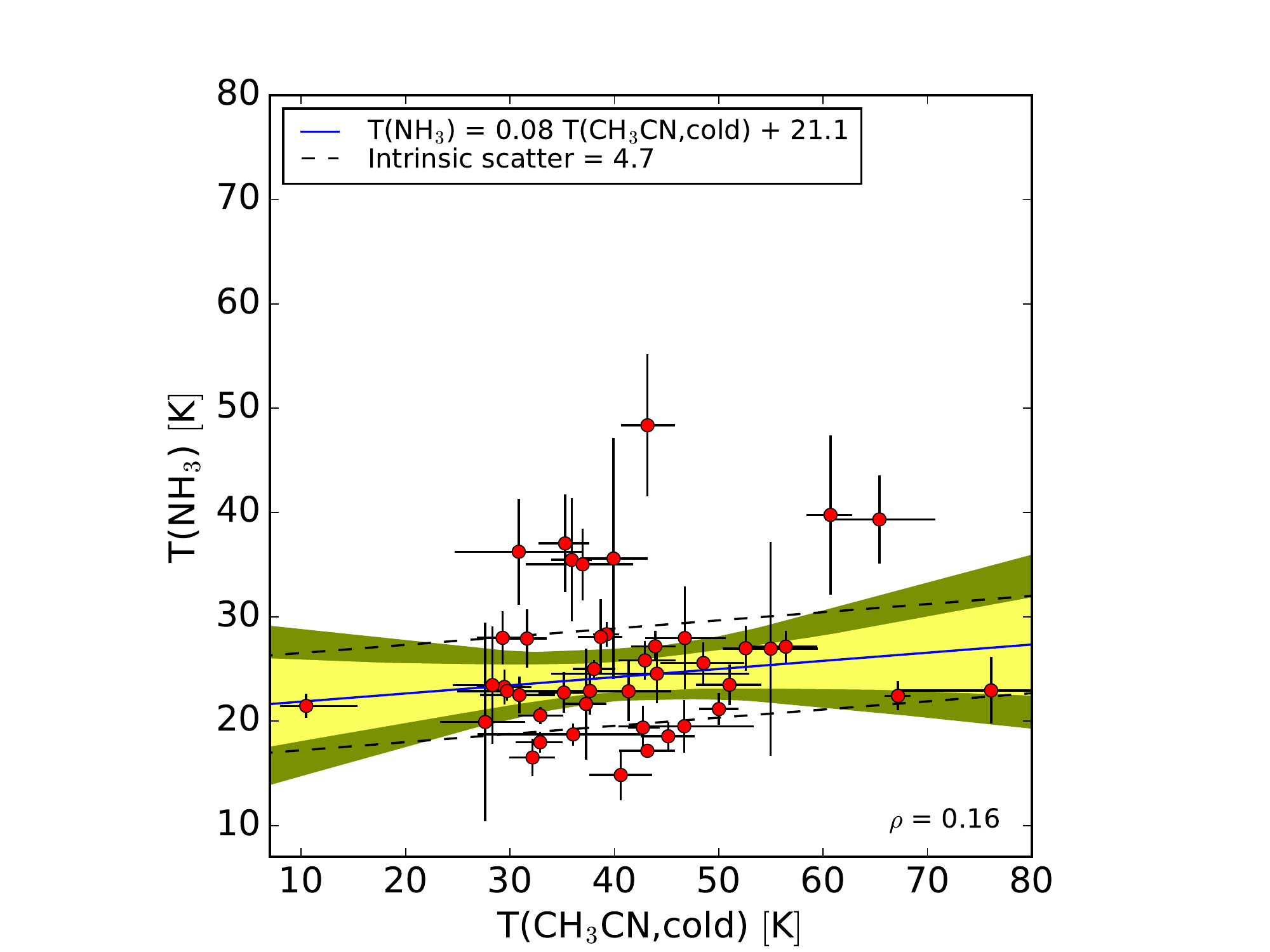}   
					\includegraphics[width=0.33\textwidth]{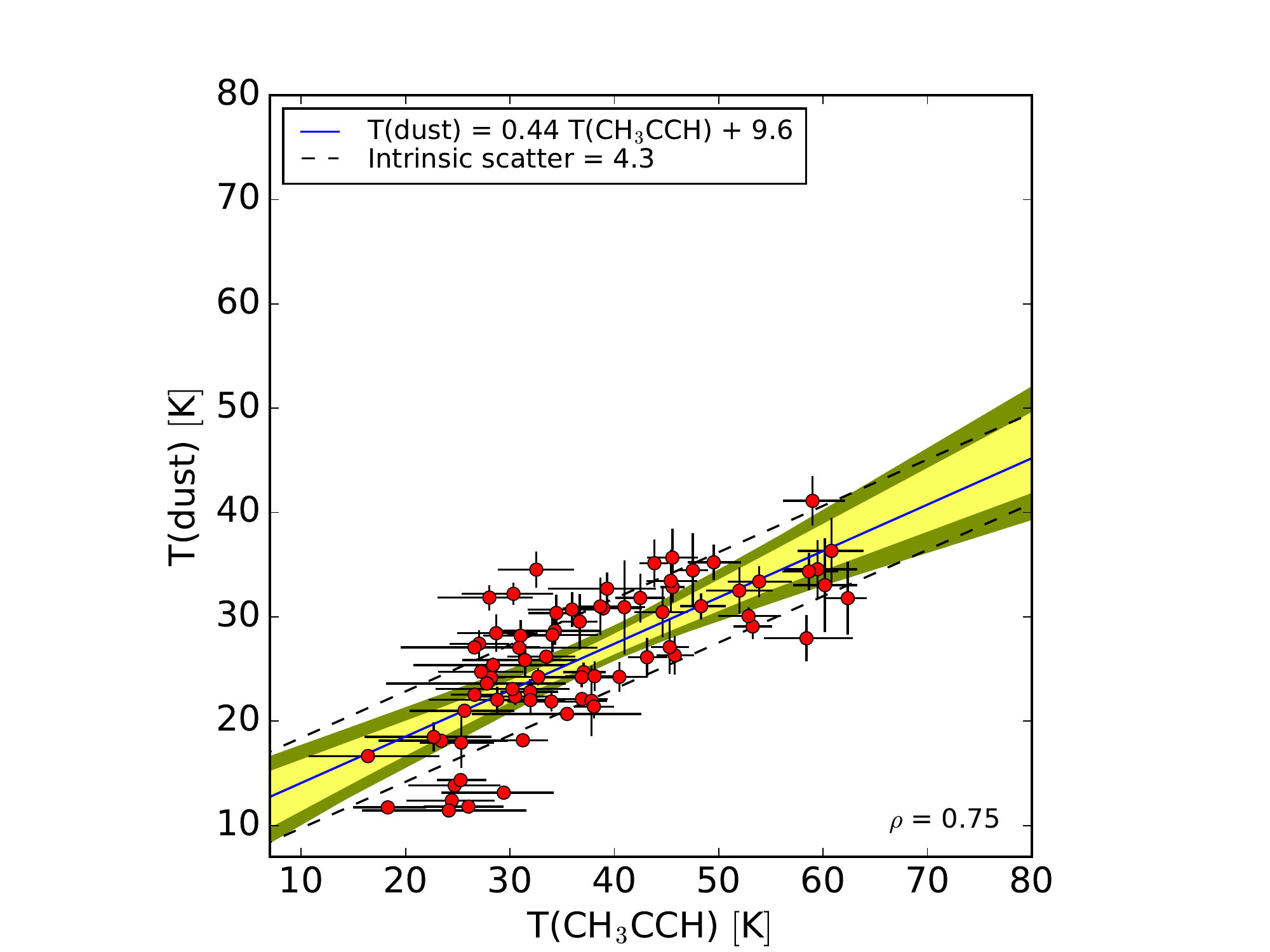}
					\includegraphics[width=0.33\textwidth]{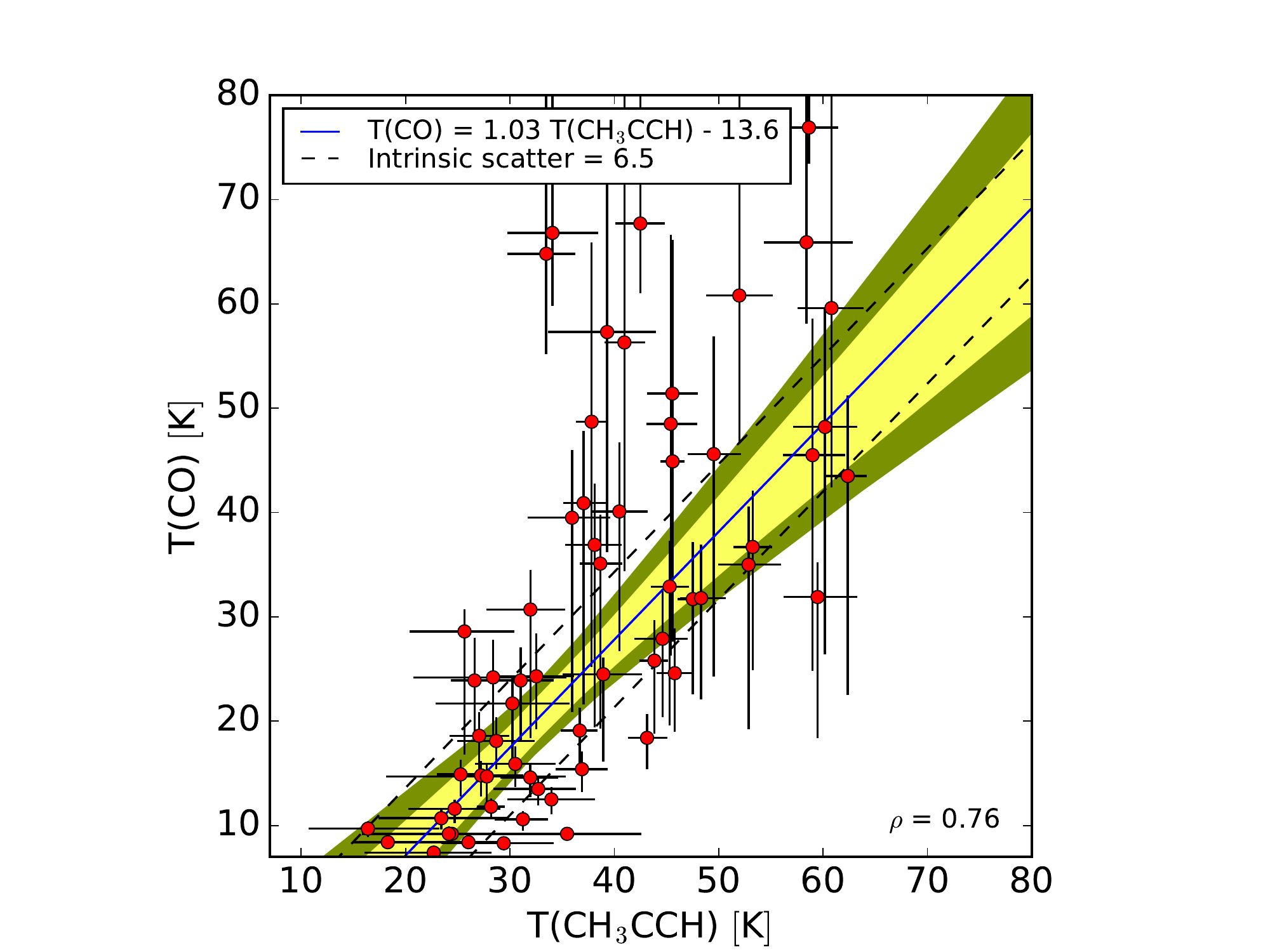} \\
					\includegraphics[width=0.33\textwidth]{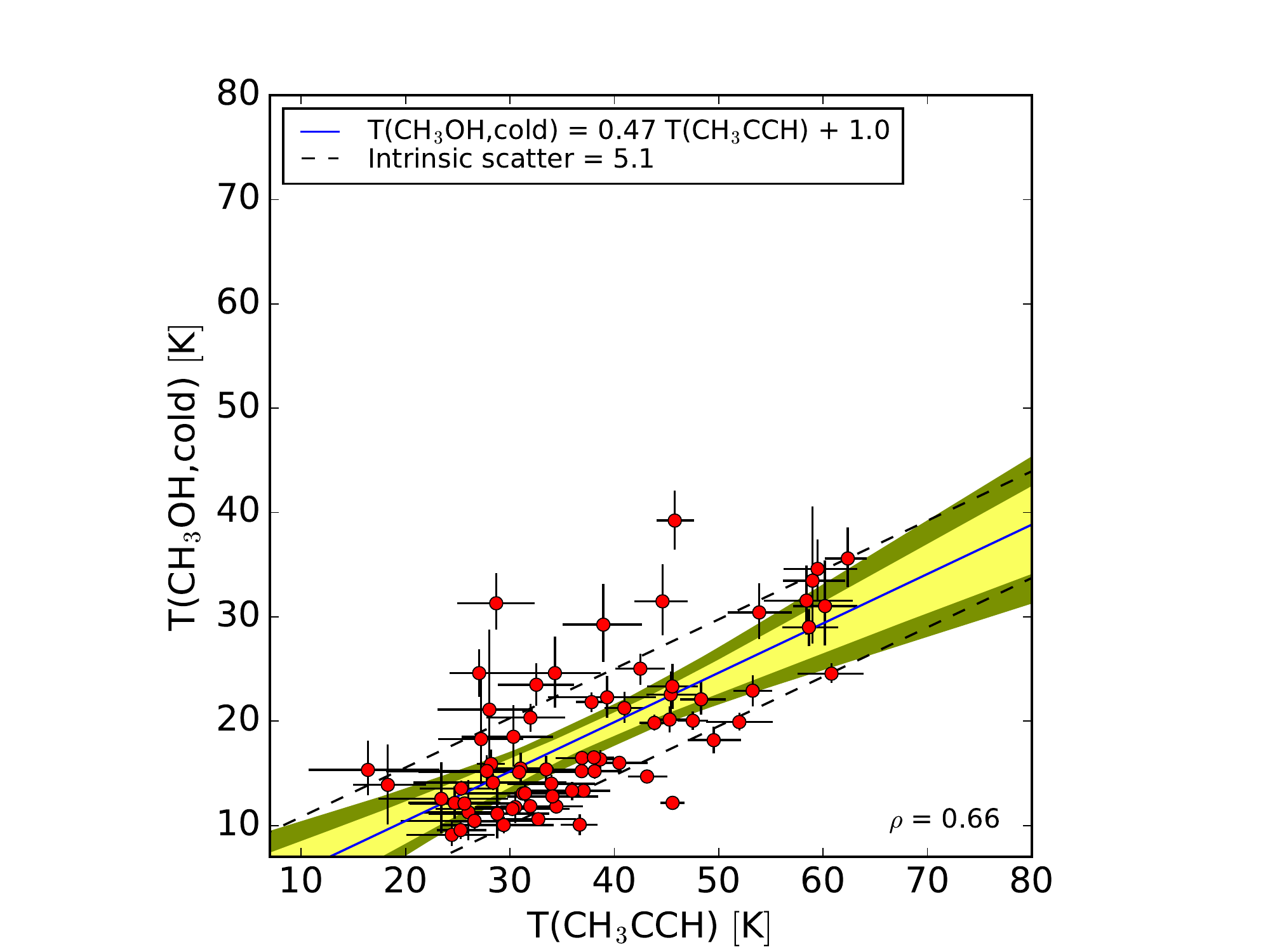} 
					\includegraphics[width=0.33\textwidth]{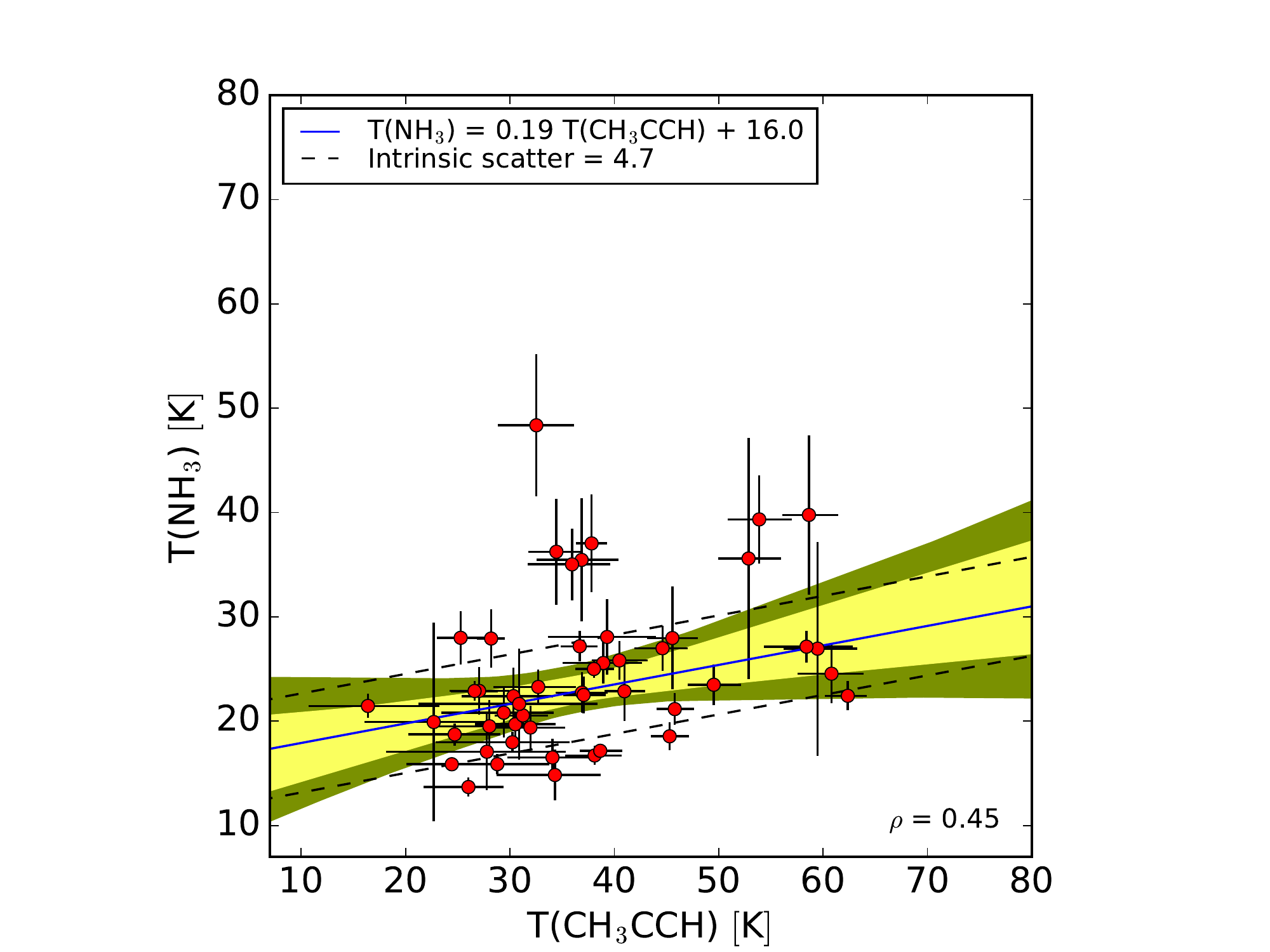}    
					\includegraphics[width=0.33\textwidth]{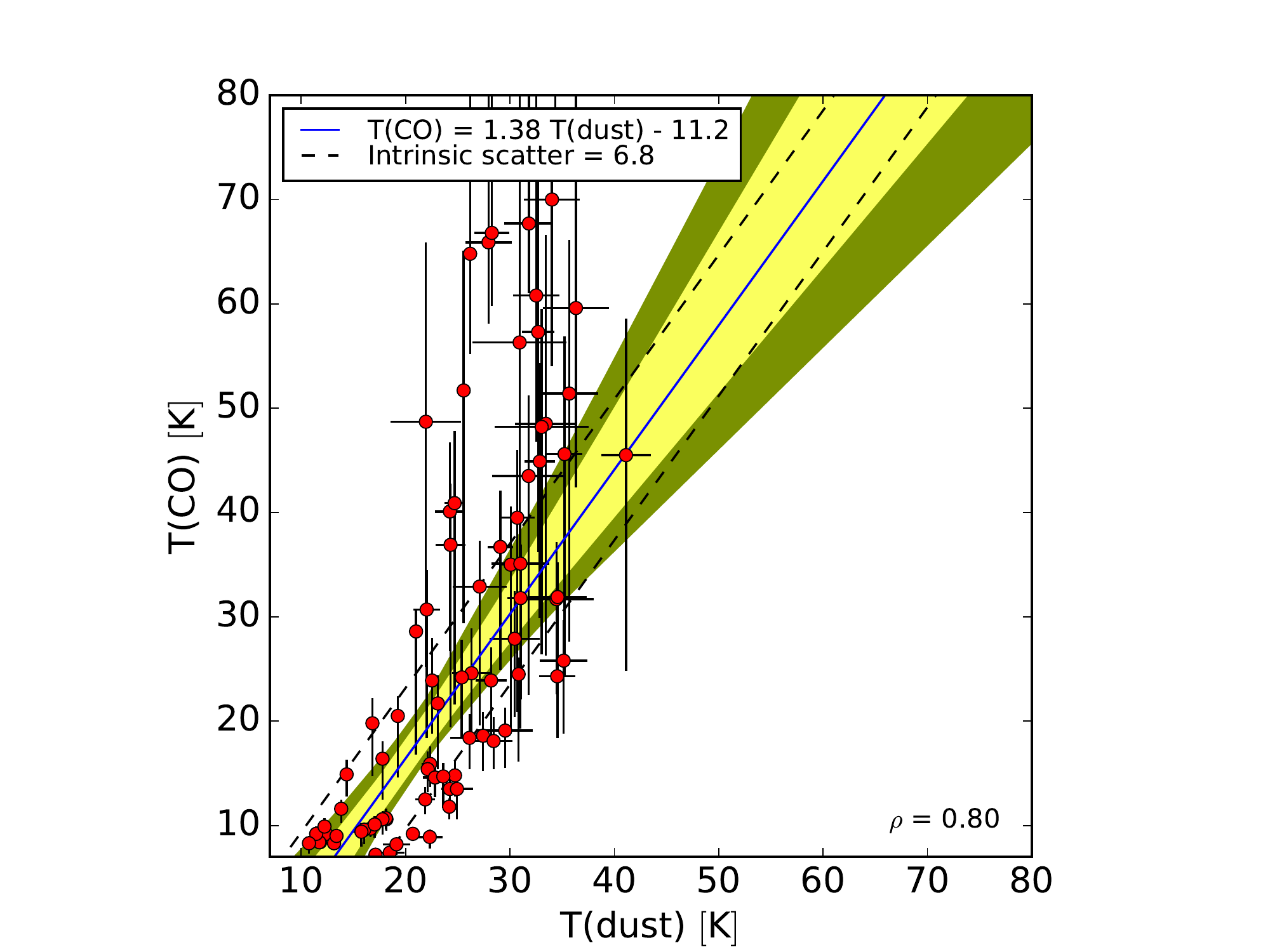}   \\
					\includegraphics[width=0.33\textwidth]{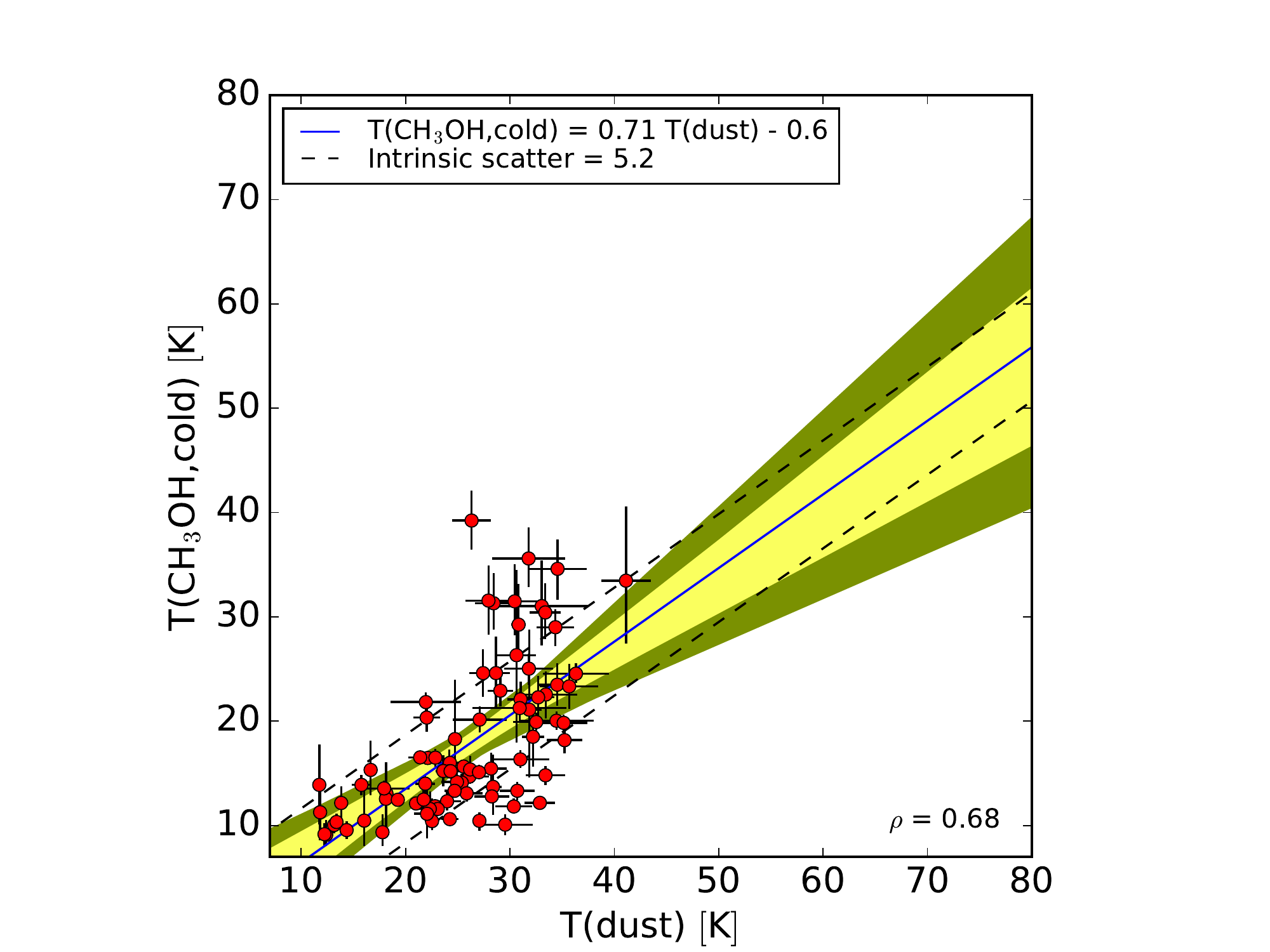}   
					\includegraphics[width=0.33\textwidth]{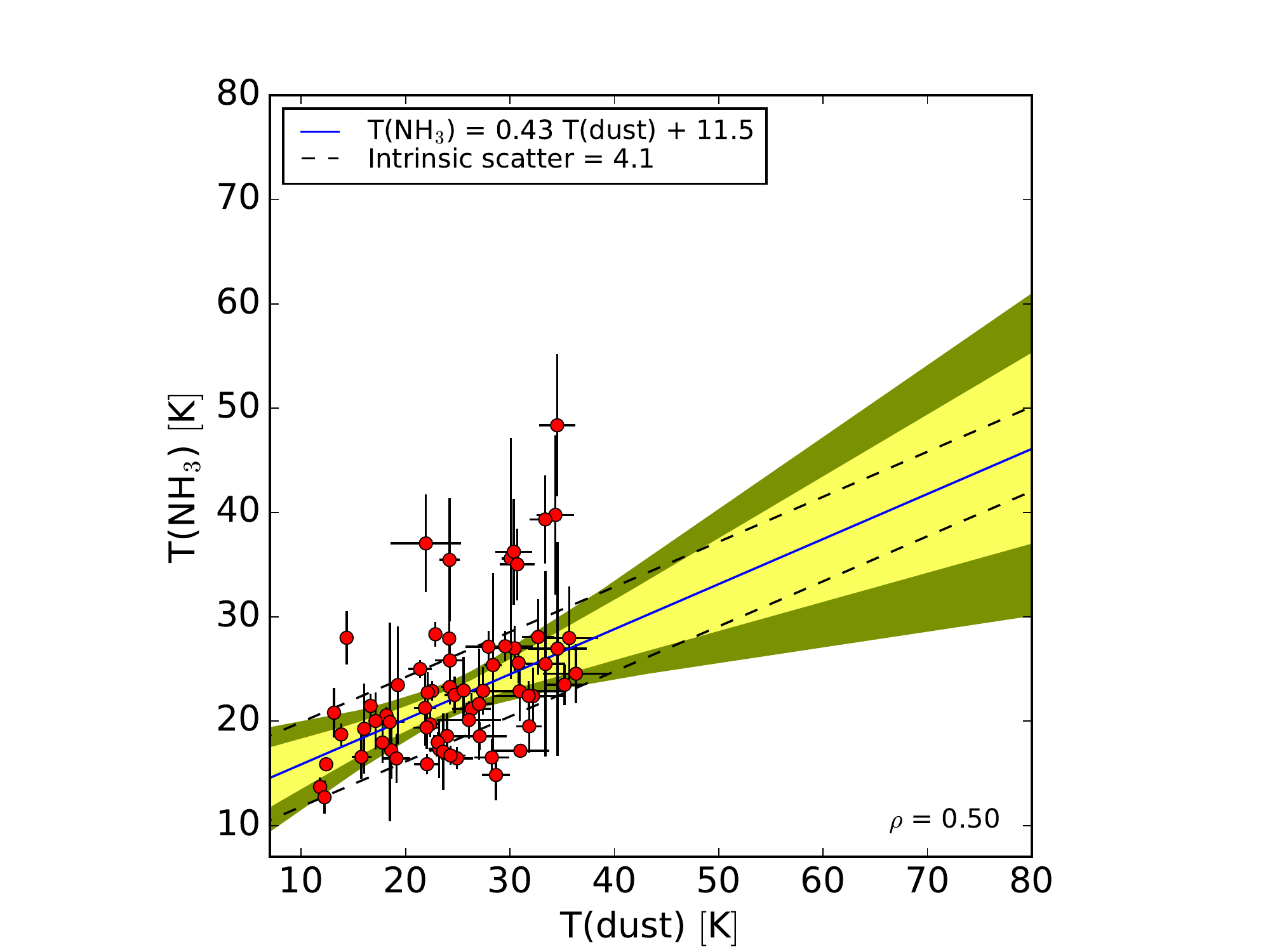}  
					\includegraphics[width=0.33\textwidth]{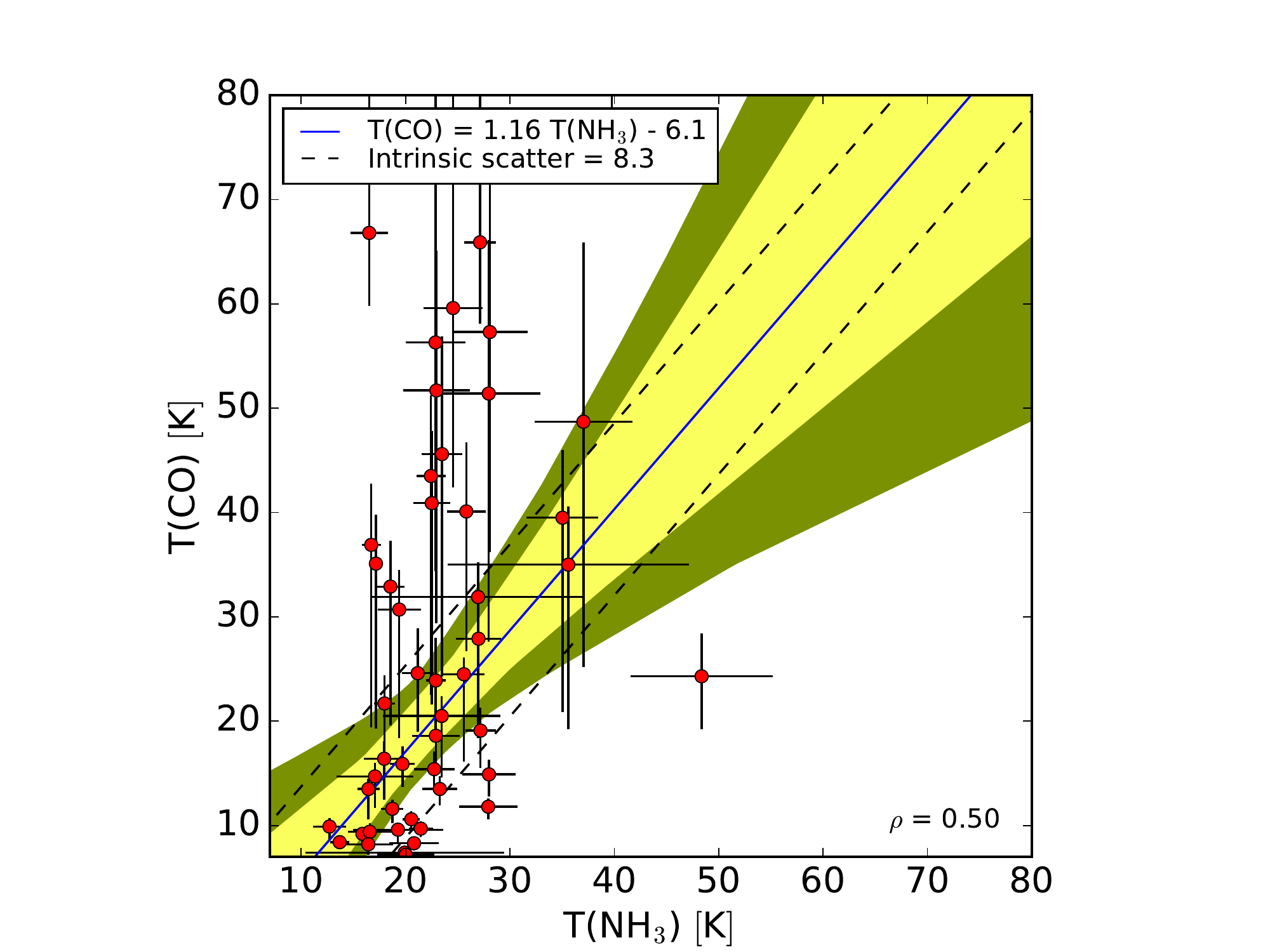}    \\
					\includegraphics[width=0.33\textwidth]{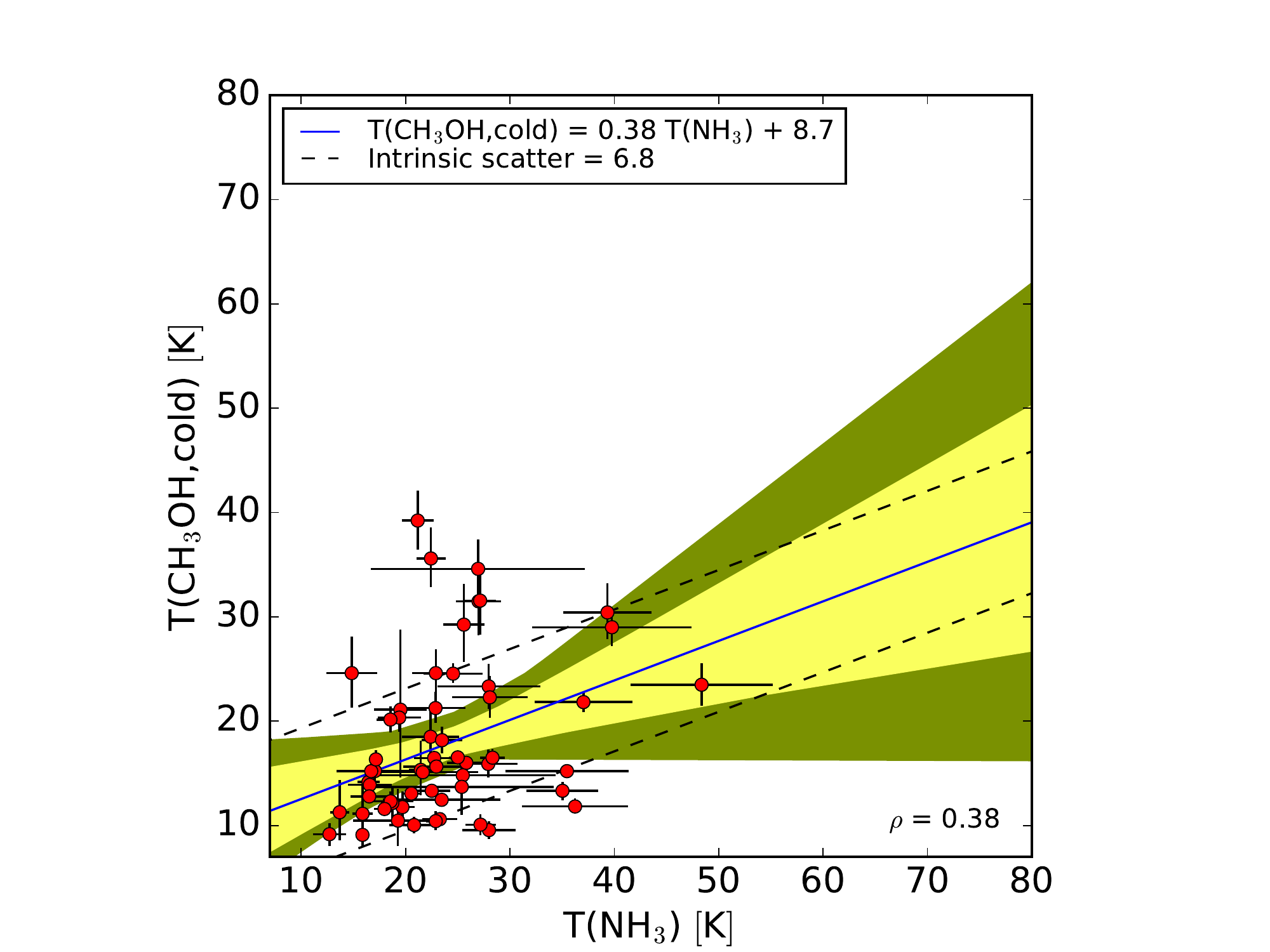}    
					\includegraphics[width=0.33\textwidth]{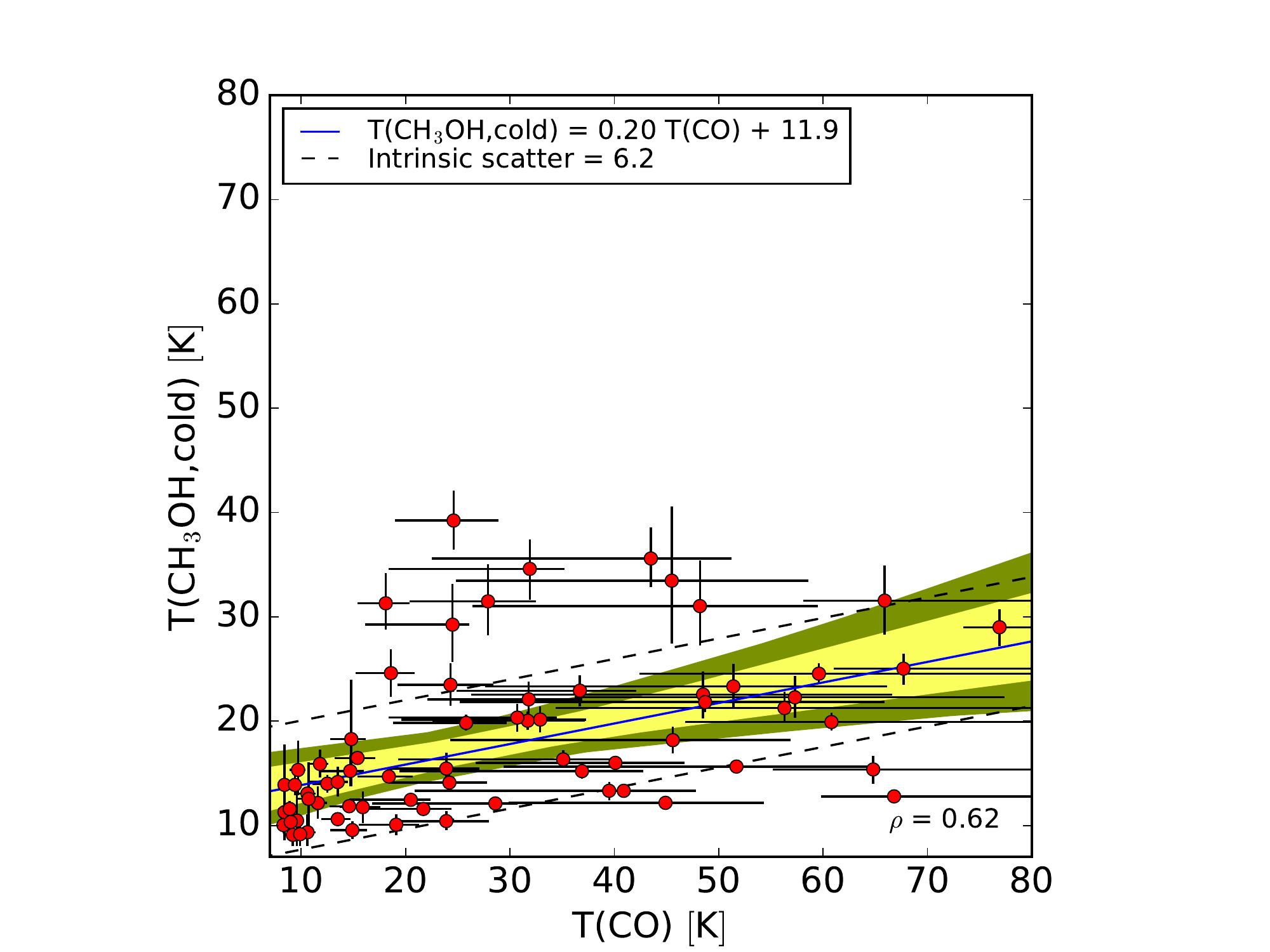}  
					\includegraphics[width=0.33\textwidth]{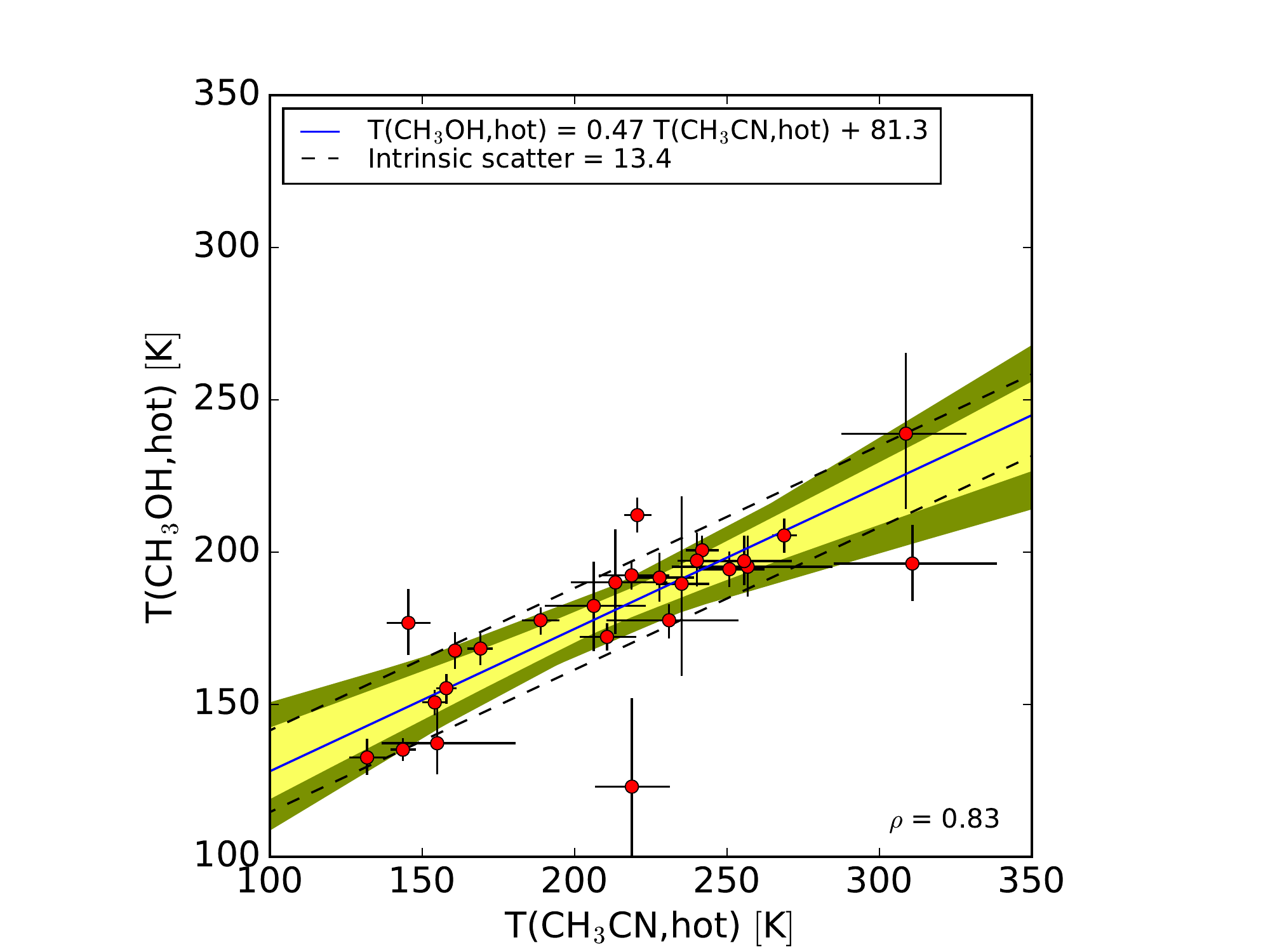}   
					\caption{Correlations between temperatures as measured by different tracers. In each panel we include the best fit result as a blue solid line. The uncertainties due to the line parameters are indicated by the shaded region - dark- and light yellow for $68\%$ and $95\%$, respectively. The intrinsic scatter in the relation is indicated by the dashed black lines. At the bottom right of the panels the Spearman correlation coefficient is shown.}\label{fig:temperature_correlations_apdx}
				\end{figure*}}

    \section{Discussion}\label{sec:discussion}

            In Sect.~\ref{sec:fit_results} we found that all temperatures from the different molecules are correlated. Figures~\ref{fig:temperature_correlations} and \ref{fig:temperature_correlations_apdx} summarise these results. The tightest correlations are found between methyl acetylene, acetonitrile and dust, which confirms that they are good tracers of temperature in the envelope. C$^{17}$O is also well correlated with these tracers, but shows a much larger scatter and uncertainties.
            The dust temperature for the sources in the TOP100 was derived and discussed in detail in \koenigSedt. Comparing their values with those obtained using the other probes available, we find that dust is tracing systematically lower temperatures than CH$_{3}$CN and CH$_{3}$CCH; this can be explained as a combination of the different regions over which the emission is integrated (only the dense, central region around the dust emission peak for molecular lines and the entire clump for dust), and the fit of the warm gas emission in the SED with a second component \citep[see][]{Koenig+17_aap599_139}.
            That a higher temperature is derived from CH$_{3}$CN compared to CH$_{3}$CCH indicates that the abundance of the former tracer is markedly more sensitive to the star-formation activity, which is also supported by the larger measured linewidths.
            The methanol results show weaker trends with respect to \an\ and \ma, suggesting that it is not as good as these species as a tracer of warm-up. It is characterised by a marginal change in the measured temperature along the evolutionary sequence for the cool component (cf. Table~\ref{tab:med_temps}). On the other hand, the temperature of the hot component is well correlated with that measured with \an.
            Ammonia does not stand out as an optimal tracer for warm-up when using only the low-excitation $J_{K}=(1,1)$ and $(2,2)$ radio inversion lines, despite being commonly used as a thermometer. The median temperature traced by NH$_{3}$ varies of only a few degrees Kelvin along the entire evolutionary sequence, as observed in previous works \citep[e.g.][and Table~\ref{tab:med_temps}]{Wienen+12_aap544_146,Giannetti+13_aa556_16}. This molecule has the lowest correlation coefficients with the other tracers, in part because it was observed with a much larger beam \citep[see][]{Wienen+12_aap544_146}.
            The temperature from NH$_3$ best correlates with $T_d$, which is consistent with the ammonia tracing an extended region (see also Sect.~\ref{sec:em_reg_chem_mods}), similar to that used to compute the SEDs. However, in a $T_d \, \mathrm{vs.} \, T_\mathrm{NH_{3}}$ the best-fit relation shows that, with respect to dust, NH$_{3}$ traces hotter material in very cold sources and colder gas in warmer, more evolved clumps, reducing the differences between different classes.
            
		\subsection{Temperature distribution from CO isotopologues: separation between IR-weak and IR-bright sources}\label{sec:CO_T_distr}
		
			The four evolutionary classes defined for the TOP100 by \koenigSedt, and used throughout this work, separate IR-bright and IR-weak sources on the basis of their $21\mum$ fluxes, using a threshold of $2.6\jy$, which roughly divides sources that need a hot component to fit the complete SED from those that do not. 
			
\begin{figure}
				\centering
				\includegraphics[width=0.66\columnwidth]{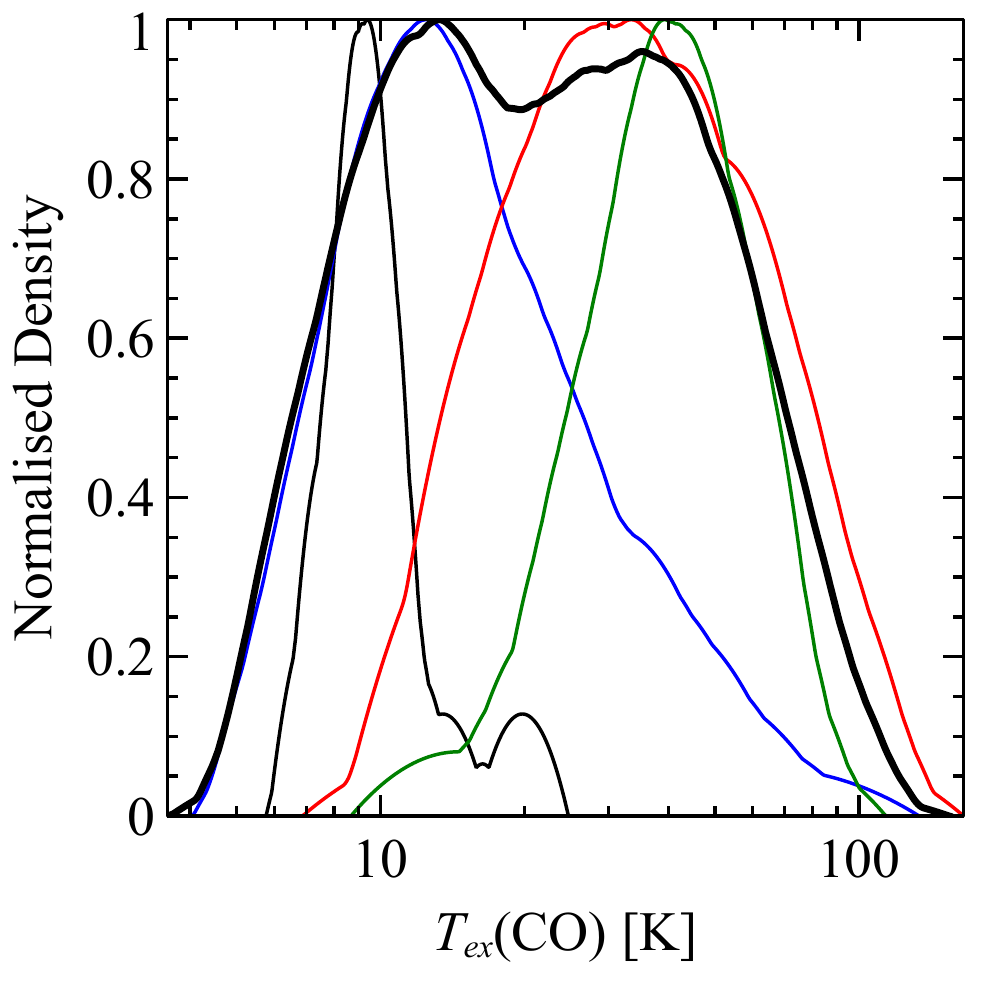}  
				\caption{Kernel density estimate of the probability density function, normalised to its maximum for displaying purposes, for the excitation temperature as traced by CO isotopologues. The thin lines represent the different classes: black - 70w, blue - IRw, red - IRb, and green - \hii. The thick black line refer to the entire TOP100 sample.}\label{fig:CO_T_distr}
			\end{figure}

			A kernel density estimation \citep{Rosenblatt56_AnnMathStat27_832,Parzen62_AnnMathStat33_1065} for the CO isotopologues excitation temperatures from \citet{Giannetti+14_aa570_65} shows a minimum at $\sim20\kel$ (Fig.~\ref{fig:CO_T_distr}), corresponding to the CO evaporation temperature, confirming that, once CO is evaporated back into the gas phase, the gas traced is hotter, and more closely affected by the feedback from young stellar objects.
			The CO temperature also nicely separates IRw and IRb: these two classes are thus not only different in terms of temperature, but also represent objects in which CO is mostly locked onto dust grains in dense gas for the former class, and in which it is mostly in the gas phase for the latter. This gives a clear physical meaning to the threshold chosen by \koenigSedt\ to distinguish these two classes. Therefore, the four classes defined for the TOP100 all represent physically distinct evolutionary stages.
			\begin{figure*}
				\centering
				\includegraphics[width=\textwidth]{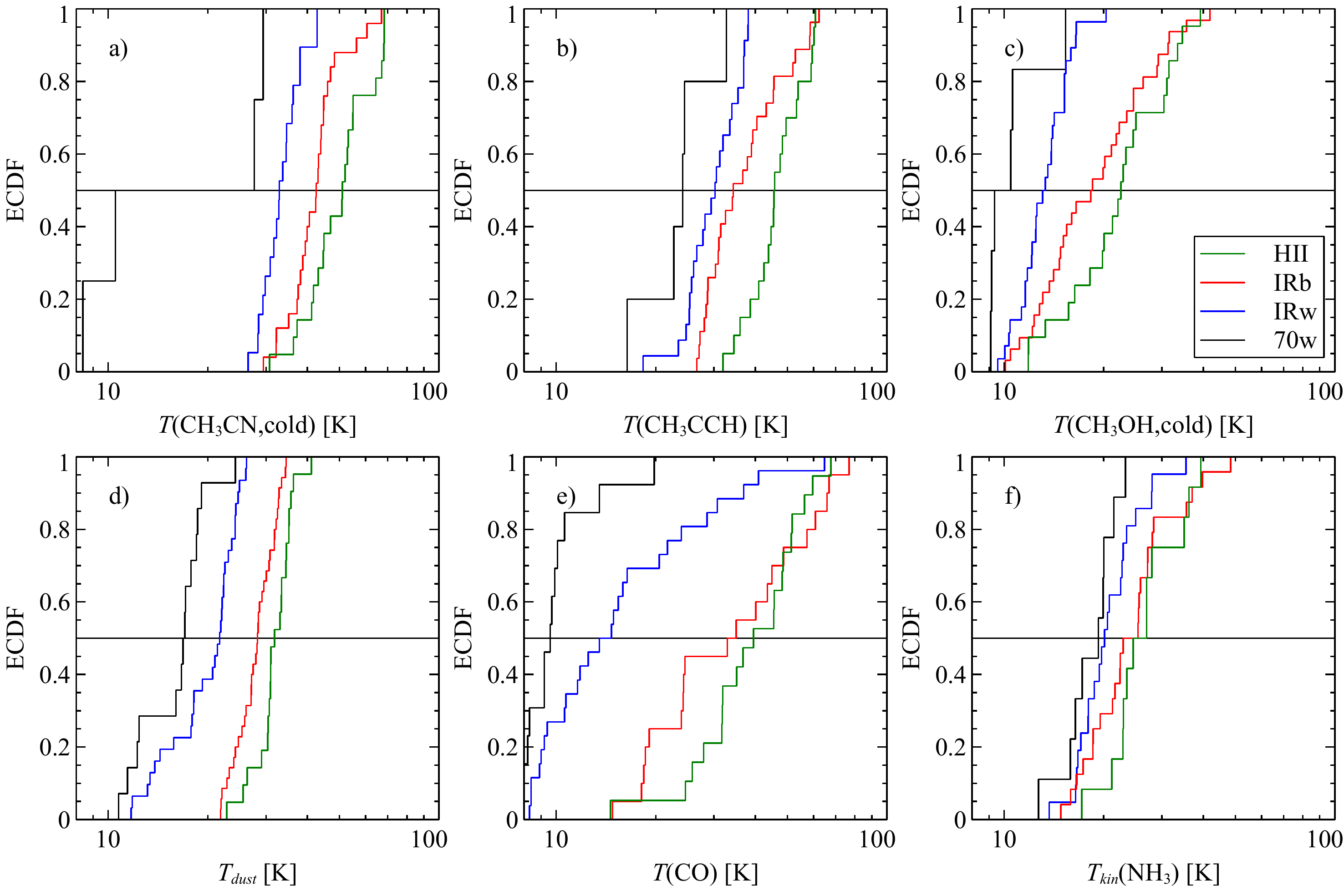}
				\caption{Comparison between empirical cumulative distribution functions of the temperature for the four evolutionary classes defined for the TOP100. Each panel show one of the species analised in this work. The four evolutionary classes are indicated in a different colour, according to the legend shown in the lower right corner of panel c). The solid line at $p=0.5$ shows the median of the distributions.}\label{fig:ecdf_temp_class}
			\end{figure*}
			\begin{figure*}
				\centering
				\includegraphics[width=0.66\columnwidth]{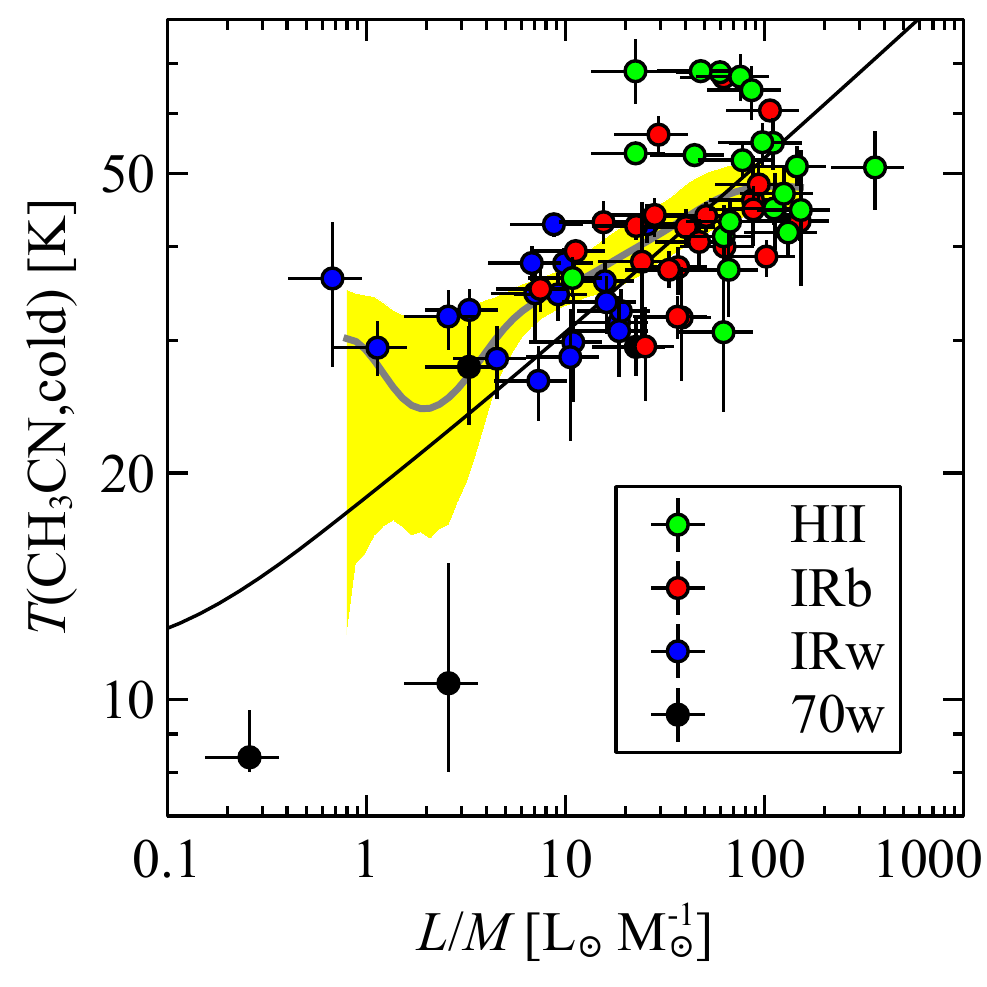}
				\includegraphics[width=0.66\columnwidth]{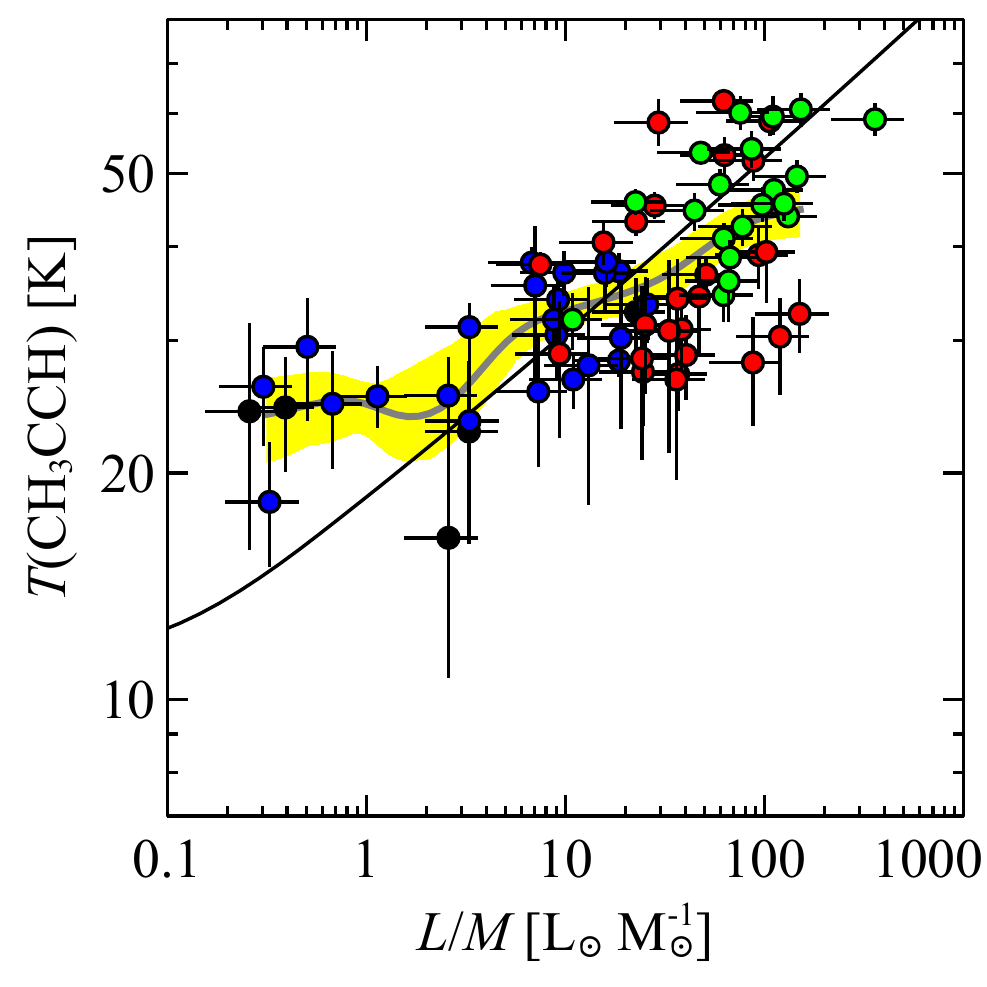}
				\includegraphics[width=0.66\columnwidth]{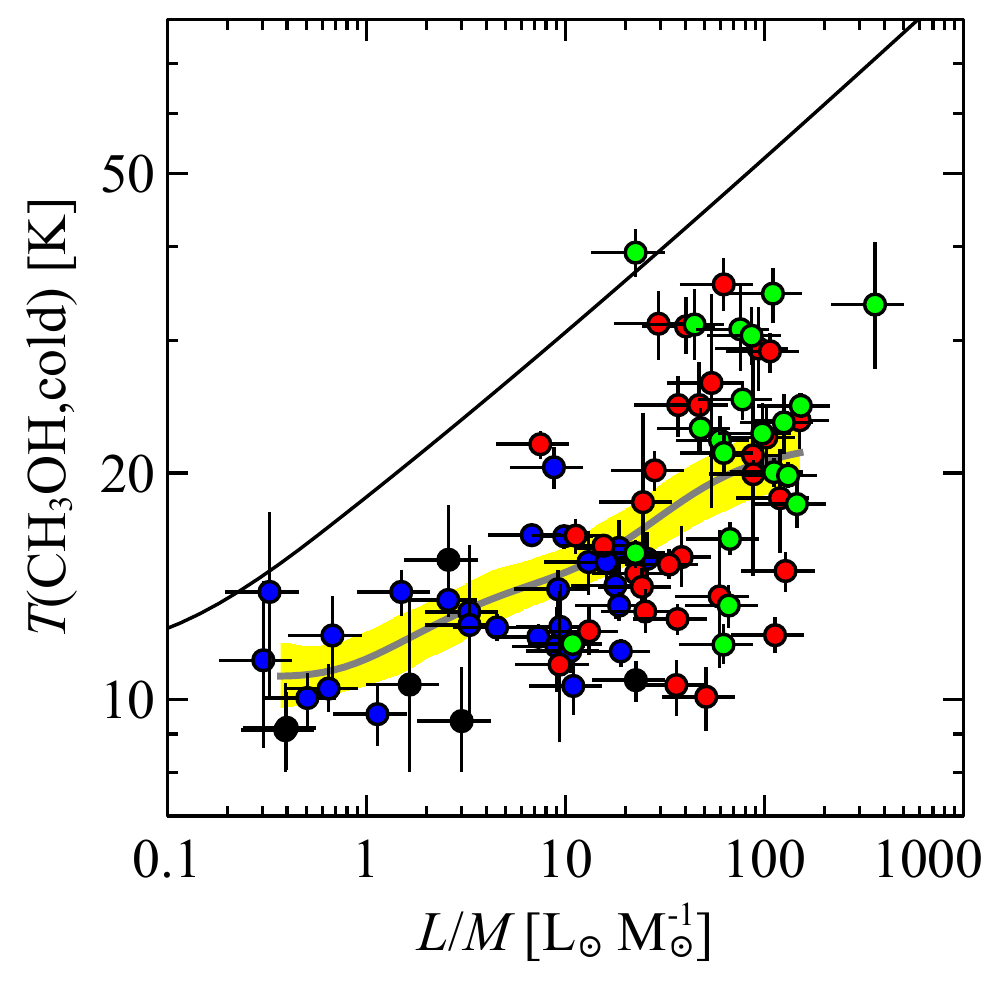}\\
				\includegraphics[width=0.66\columnwidth]{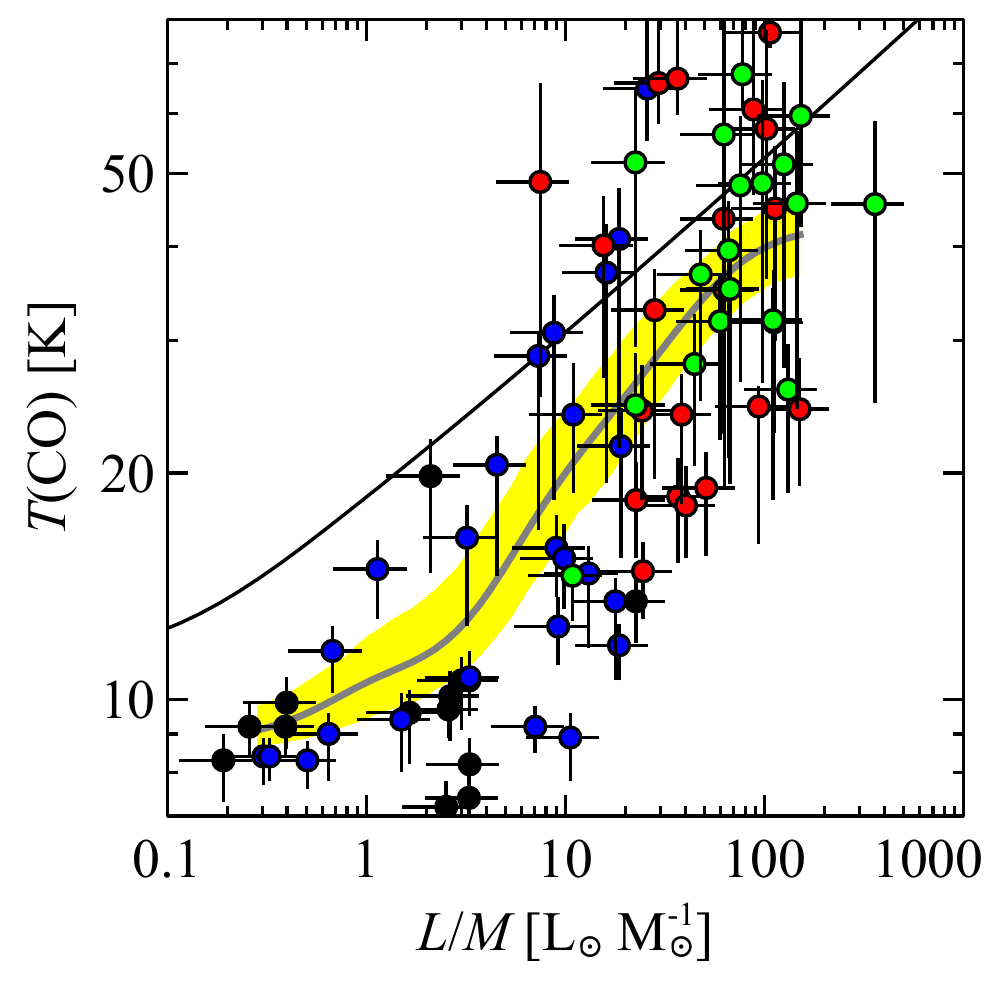}
				\includegraphics[width=0.66\columnwidth]{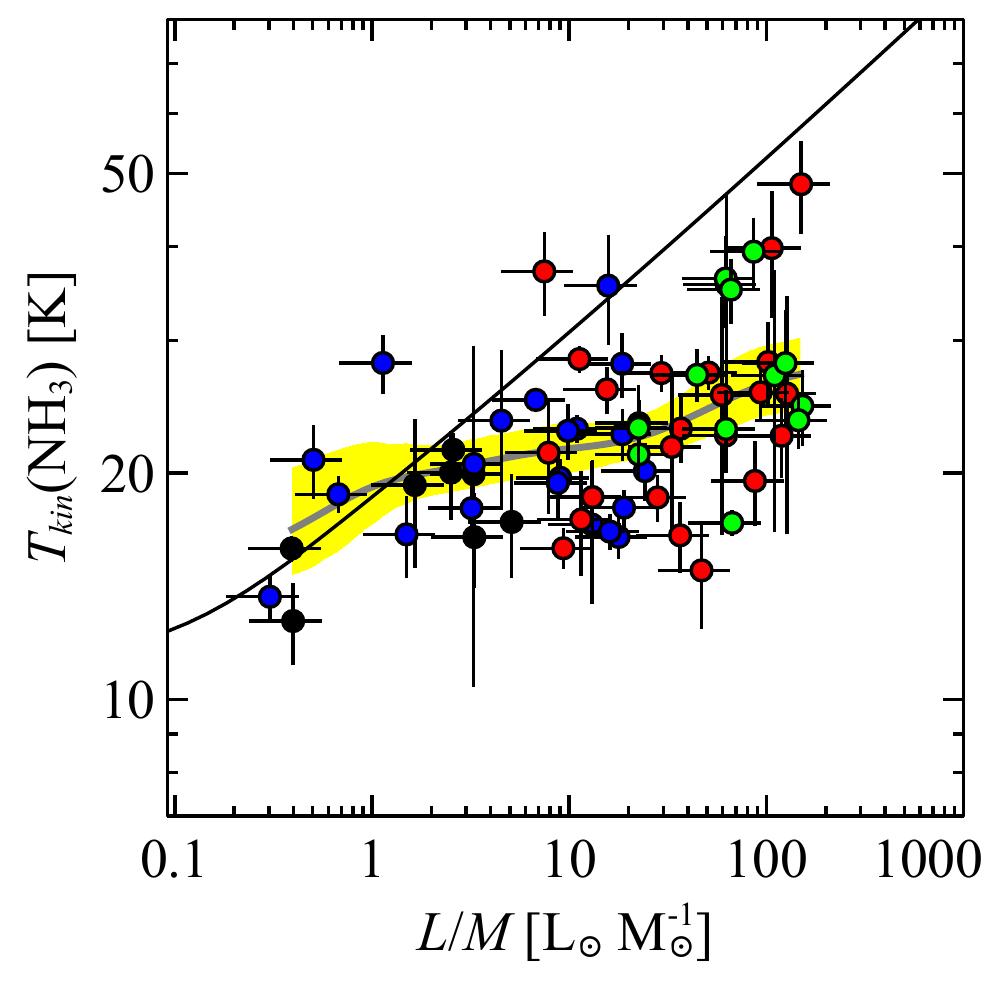}
				\includegraphics[width=0.66\columnwidth]{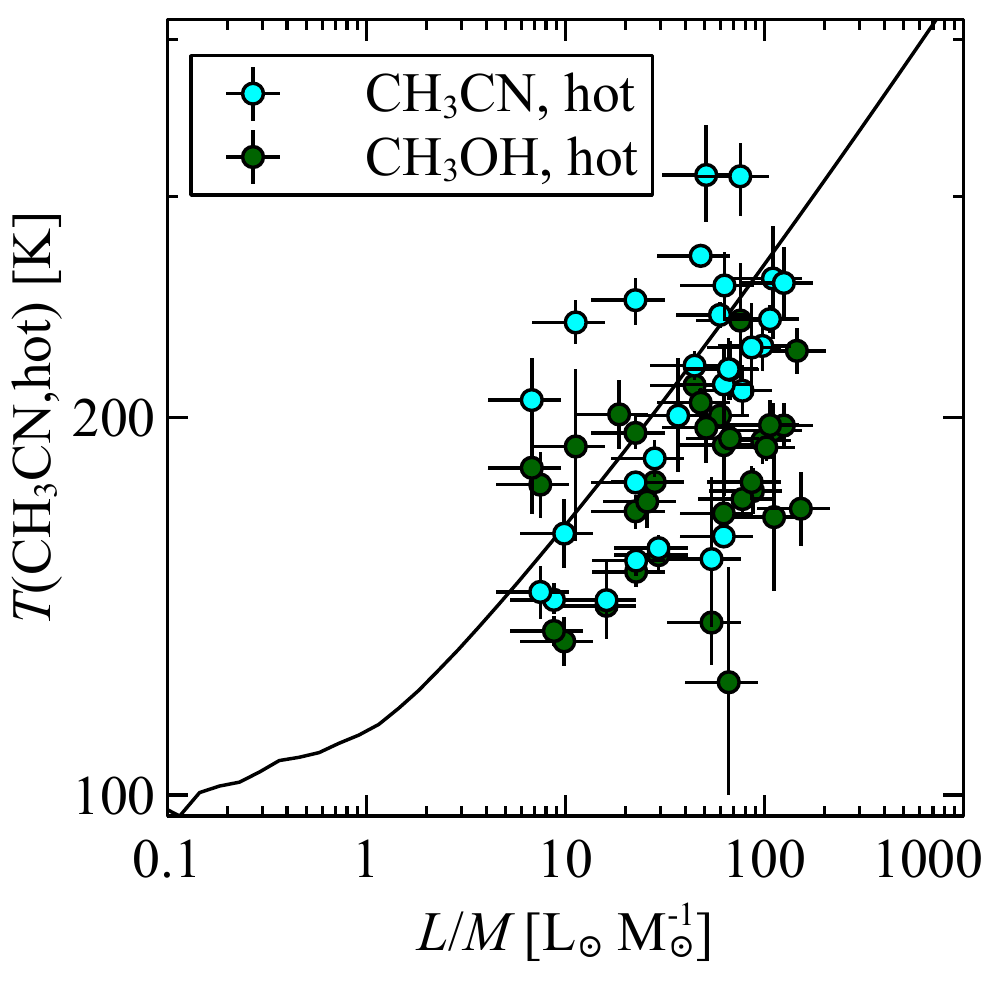}
				\caption{$L/M$ ratio vs. temperature. The black solid line indicates the line-of-sight- and beam-averaged temperature for a spherical clump model passively heated by a central object accounting for the total luminosity. The thick grey line shows the non-parametric regression of the data, with uncertainties indicated by the yellow-shaded area.}\label{fig:lm_temperatures}
			\end{figure*}

        \subsection{Feedback from high-mass YSOs: Warm-up of the gas}\label{sec:warm_up}
                    
            The feedback from forming stars causes the progressive heating of the surrounding gas. Thanks to the fact that the TOP100 is a representative sample of high-mass star-forming regions in different evolutionary stages we can evaluate this effect by comparing all the available tracers for the different groups of objects, to validate the scheme used to classify them.
            To test whether we see the gas being heated by the feedback from the nascent massive young stellar objects, and determine to what degree the various species are influenced by it, we divided the TOP100 in the four classes described in Sect.~\ref{sec:sample}, to compare the ECDFs of the temperature for each molecule (Fig.~\ref{fig:ecdf_temp_class}). We then made use of the Anderson-Darling test \citep{ScholzStephens87_JASA82_918}, implemented in the R package kSamples\footnote{\url{https://cran.r-project.org/web/packages/kSamples/index.html}}, to assess whether or not they are drawn from a common distribution.
            The tests show that the temperatures found in consecutive classes (70w, IRw, IRb, \hii) increase, with p-values always below $0.9\%$ (cf. Table~\ref{tab:ad_tests_temps}), allowing to reject the null hypothesis that the temperature is not increasing in more evolved sources at a level $>2.5\sigma$, for methyl acetylene and acetonitrile, the best thermometers available to us. The typical temperature of each class also appear different and increasing from methanol.
            This confirms the trends seen in dust temperature \koenigSed, ammonia \citep[e.g.][]{Wienen+12_aap544_146} and CO \citep{Giannetti+14_aa570_65}.
            
            In conclusion, we find that all probes indicate a progressive warm-up of the gas with evolution, albeit with significantly different behaviours which are determined by the properties of the species and of the transitions. 
            Methyl acetylene and acetonitrile are the most sensitive to the warm-up process, methanol and ammonia (when using the two lowest inversion transitions) show only a marginal increase in temperature along the evolutionary sequence.
            
            Whereas the classes defined for the TOP100 on the basis of the IR- and radio continuum properties of the clumps only allow for a comparison between their average properties, the $L/M$ ratio defines a continuous sequence in time. We can thus investigate directly whether the temperature and the $L/M$ ratio are correlated.
            In Fig.~\ref{fig:lm_temperatures} we show the temperature traced by all molecular probes available for the TOP100 as a function of the luminosity-to-mass ratio. A non-parametric regression\footnote{We used a Gaussian kernel, with a bandwidth selected with Kullback-Leibler cross-validation \citep{Hurvich+98_JRSS60_271}, and a local constant estimator.} with bootstrapped uncertainties (estimated using the $2.5\%$ and $97.5\%$ quantiles) was performed for each tracer with the R package \textit{np} \citep{HayfieldRacine08_jstatsoft27_5}: it indicates that a correlation between these parameters exists for all tracers, 
            confirming that $L/M$ is a good indicator of evolution as well \citep[e.g.][]{Molinari+08_aap481_345,Molinari+16_apjl826_8}, and that the forming cluster has a strong impact on the physical properties of the surrounding environment on a clump scale.            
                
			A simple toy model was constructed, considering a spherical clump with power-law density and temperature profiles; the line-of-sight and beam-averaged temperature was calculated at the peak of column density, to be compared with the available tracers for the TOP100.
			The density gradient was assumed to be proportional to $r^{-1.5}$. Consistently with the analysis performed through MCWeeds, the $T$-profile is described by Eq.~\ref{eq:t_str}. The average temperature was calculated by weighting the values in individual pixels by the mass in that pixel and beam response, assuming it is Gaussian.
			The results are superimposed to the data points in the panels of Fig.~\ref{fig:lm_temperatures}. 

			\begin{table*}
\centering
\caption{Median temperature per class, as measured with all molecular tracers considered. The 90\% CI for each class and tracer is also listed. The significance of the differences in the measured properties is summarised in Table~\ref{tab:ad_tests_temps}.\label{tab:med_temps}}
\begin{tabular}{l*{8}{r}}
\hline \hline
Species & 70w &$90\%\,$CI& IRw &$90\%\,$CI& IRb &$90\%\,$CI& \hii &$90\%\,$CI\\
 & [K] & [K] & [K] & [K] & [K] & [K] & [K] & [K] \\
\hline
\an(cool) & 20.5 & 9.1-29.4 & 33.3 & 26.7-42.8 & 42.5 & 32.4-59.9 & 50.8 & 36.1-67.8 \\
\ma & 24.1 & 17.7-31.1 & 30.2 & 23.6-37.9 & 35.5 & 27.1-58.6 & 45.7 & 34.3-60.2 \\
\mt(cool) & 9.9 & 9.1-14.1 & 13.2 & 10.1-16.5 & 18.3 & 10.8-31.4 & 22.6 & 11.8-34.6 \\
C$^{17}$O & 9.6 & 7.3-16.0 & 14.1 & 8.4-39.9 & 34.0 & 17.9-67.3 & 39.5 & 23.6-60.4 \\
NH$_{3}$ & 18.3 & 14.1-22.5 & 20.1 & 16.6-28.0 & 25.4 & 16.0-41.0 & 25.8 & 19.4-37.6 \\
\an(hot) & \dots & \dots & 146.7 & 133.9-194.6 & 196.6 & 153.2-285.6 & 232.4 & 194.3-299.9 \\
\mt(hot) & \dots & \dots & 156.5 & 133.3-196.4 & 176.8 & 144.0-196.7 & 192.4 & 157.9-228.6 \\
\hline
\end{tabular}
\end{table*}

            \begin{table*}
\centering
\caption{Resulting p-values of Anderson-Darling tests for the three species considered in this work.\label{tab:ad_tests_temps}}
\begin{tabular}{l*{6}{r}}
\hline \hline
Species & HII--IRb & HII--IRw & HII--70w & IRb--IRw & IRb--70w & IRw--70w \\
\hline
CH$_{3}$CCH & $7.0\times 10^{-04}$ & $<1.0\times 10^{-05}$ & $3.9\times 10^{-05}$ & $1.7\times 10^{-03}$ & $3.3\times 10^{-04}$ & $8.6\times 10^{-03}$ \\
CH$_{3}$CN & $4.3\times 10^{-03}$ & $<1.0\times 10^{-05}$ & $7.6\times 10^{-05}$ & $1.8\times 10^{-05}$ & $1.9\times 10^{-05}$ & $1.8\times 10^{-03}$ \\
CH$_{3}$OH & $6.0\times 10^{-02}$ & $<1.0\times 10^{-05}$ & $1.8\times 10^{-05}$ & $8.2\times 10^{-05}$ & $1.8\times 10^{-04}$ & $3.0\times 10^{-03}$ \\
\hline
\end{tabular}
\end{table*}

			\paragraph{\ma\ and \an:} Methyl acetylene and acetonitrile are in good agreement with this simple model for $L/M$ ratios above a few $\lsun \msun^{-1}$, confirming that these molecules are reliable tracers of physical properties and passive heating in the dense layers of the clump. Changing the slope of the density profile by $\pm0.5$ moves the curve up or down, enveloping the data points for these two molecules. 
			As discussed in Sect.~\ref{sec:selection_tracers}, CH$_{3}$OH and CH$_{3}$CN are formed mainly on dust grains and are released in the gas phase for temperatures above $80-100\kel$. With this in mind, to compare the results of the hot components to the model, we only considered in the average the regions where the temperature exceeds $80\kel$, in which the abundance is substantially increased. The bottom right panel of Fig.~\ref{fig:lm_temperatures} shows the results of this procedure: the trend and average $T$ are well reproduced for acetonitrile.
			
			The good match of the observed $T \, \mathrm{vs.} \, L/M$ relation for $L/M\gtrsim2-10\lsun \msun^{-1}$ for CH$_3$CCH, CH$_3$CN, and, to a lesser degree, C$^{17}$O from \citet{Giannetti+14_aa570_65} with the predictions of our spherical clump model, centrally heated by a star having a luminosity equal to the observed bolometric luminosity, indicates that the heating of the gas in the cluster environment is dominated by its most massive member. This supports the findings by \citet{Urquhart+14_mnras443_1555}, showing that the observed bolometric luminosity agrees better with the luminosity of the most massive star, rather than the luminosity of the entire cluster, for clumps with $M>\mathrm{few}\times\pot{3} \msun$ (adopting the ATLASGAL fluxes from \citealt{Contreras+13_aa549_45}; the masses obtained with these fluxes are typically a factor of $\sim3$ higher than obtained from the SEDs, K\"onig, priv. comm.). The authors argue that this lends support to the idea that high-mass stars form first, while the rest of the cluster members must yet reach the ZAMS, thus having a minor contribution to $L_{bol}$. A similar conclusion is obtained by \citet{Zhang+15_apj804_141}, and \citet{Sanna+14_aap565_34} show that at least in the case of the high-mass star-forming region G023.01--00.41, the luminosity and the feedback are indeed dominated by the most massive star.

			Acetonitrile is vastly undetected in cold and young sources with $L/M\lesssim2\lsun\msun^{-1}$ and this is not simply a column density effect. Assuming that CH$_{3}$CN has an abundance equal to the median in the detected sources for the cool component, the column density should be sufficient to detect this molecule: the predicted \an\ column densities are found to range between $1-6\times\pot{13}\cm^{-2}$, suggesting a lower abundance or central density, or a smaller emitting region.
			In contrast to acetonitrile, all other molecules are detected in several sources in the interval $0.1\lsun\msun^{-1} \lesssim L/M \lesssim 2\lsun\msun^{-1}$.
			The temperature derived from methyl acetylene in this interval of $L/M$ is approximately $20\kel$ and relatively constant. Our toy model predicts a temperature lower than observed, possibly suggesting that some mechanism providing additional heating is present. 
			It is interesting to note that, according to \citet{Urquhart+14_mnras443_1555}, the distribution of sources associated with a massive YSO or an UC\hii\ region in a $M-L$ plot is enveloped by two lines of constant luminosity-to-mass ratios, between $1\lsun\msun^{-1}$ and $100\lsun\msun^{-1}$. The lower threshold in $L/M$ well matches the flattening of the temperature as a function of $L/M$ for CH$_{3}$CCH, as well as the detection threshold for CH$_{3}$CN.

			\citet{Molinari+16_apjl826_8} find a similar behaviour for the temperature traced by \ma, using the $J = 12 \rightarrow 11$ rotational transition, but for sources in the range $1\lsun\msun^{-1} \lesssim L/M \lesssim 10\lsun\msun^{-1}$, levelling to $\sim35\kel$. The authors discuss how this can be caused by the excess heating of the forming cluster. Our data show no such plateau in this regime, and the data points are already broadly consistent with our toy model. The use of lower $J$ transitions, which are also excited in colder and less dense gas, may be the cause of this discrepancy. On the one hand, thanks to the lines at 3~mm, we may be more sensitive to this cold gas, reducing the measured average temperatures, on the other hand, because thermalisation of higher $J$ levels is more difficult, this could be due to sub-thermal excitation \citep[cf. the results of][]{Bergin+94_apj431_674}. Finally, a low signal-to-noise ratio for the (12--11) series can lead to overestimates of the temperature (cf. Sect.~\ref{sec:fit_results}) as well. 
			\citet{Molinari+16_apjl826_8} also find that high-mass star-forming regions with $L/M\lesssim1\lsun\msun^{-1}$ are not detected in CH$_{3}$CCH~(12--11), attributing this to an inefficient evaporation of this species from the grains in cold gas. We detect sources in this regime, seven of which also have reliable temperatures (five more clumps were detected; the fits yield consistent temperatures, but the uncertainties exceed $20\kel$), which is not consistent with an inefficient evaporation from the grains, rather pointing to an effect of excitation. Their hypothesis that star-formation activity is not strong enough to excite the observed transitions well matches our results for CH$_{3}$CN, too. 
			The clumps in the TOP100 have been observed as part of the Methanol Multibeam (MMB) survey \citep{Green+09_mnras392_783,Breen+15_mnras450_4109}. Sixty-six clumps are found to be associated with Class-II methanol maser emission \koenigSed. 
			Only three sources associated with masers are undetected in \an, and only one 70w source is found to show methanol maser emission. 
			This, combined with the increasing detection rate of \an\ with $L/M$ and evolutionary class, and the larger detection rate in 70w, suggest that, in fact, the detection of \an\ can be considered an early signpost of the beginning of the warm-up phase.
			The forming high-mass YSOs raise the temperature to the point where HCN is released from grains, leading to the formation of acetonitrile. 
			As time proceeds, the gas is continuously heated up until hot cores are formed and they become detectable in single-dish observations.
			Therefore, as also indicated by the larger linewidths with respect to CH$_{3}$CCH and ammonia, CH$_{3}$CN is a good tracer of active star-formation in general and of the warm-up process in particular, from its very beginning to the development of a hot core.            
			The observed threshold in temperature for this species could be also the result of excitation, rather than chemistry. The potential of this species as a tracer of star-formation activity remains unaltered: instead of marking the point where $T$ exceeds a given value, it would indicate sources with densities high enough to excite the selected $3\mm$ transitions. The detection rate as a function of evolutionary class would then indicate a density increasing with time (see also Sect.~\ref{sec:evol_DV_nH2}).
			\paragraph{\mt:} Methanol traces both cold and hot gas. For the cool component the linewidths are broad, even exceeding those of CH$_{3}$CN (see Fig.~\ref{fig:ecdf_T_DV_theta_N}), despite the temperatures derived are lower. The different behaviour of methanol with respect to \an\ and \ma, and the toy model, as well as the small range in temperature probed by this species, confirm that \mt\ is less sensitive to the warm-up process with respect to acetonitrile and methyl acetylene. 
			As discussed in Sect.~\ref{sec:results_MT}, methanol is sensitive to both temperature and density and could be subthermally excited, which may explain at least in part this behaviour, but the results from non-LTE modelling of this molecule show that it indeed traces and is abundant in cold gas \citep[e.g.][]{Leurini+04_aap422_573,Leurini+07_aap466_215}.
			Because CH$_{3}$OH has a high abundance also in cold gas (cf. Fig.~\ref{fig:ECDF_abundances}) and it is commonly observed in outflows, the large observed linewidths are most likely the result of this molecule tracing outflows and a large volume of gas. 
			In a recent work \citet{Leurini+16_aap595_4} show that around a low-mass class 0 protostar \mt\ is likely entrained in the outflow or in a disk wind.
			For the hot component we find that our toy model is in qualitative agreement with the observations, despite the lower degree of correlation with $L/M$ with respect to \an.
						
			\paragraph{CO isotopologues:} C$^{17}$O shows a large range of temperatures, from very cold to hot gas, with a weak dependence of the measured $\tex$ also for low $L/M$ ratios, contrary to CH$_{3}$CCH. This behaviour can be explained by depletion in dense and cold gas. 
			For temperatures below $\sim20-25\kel$ CO is efficiently removed from the gas phase and locked onto dust grains for densities exceeding $\mathrm{few}\times\pot{4}\cm^{-3}$ \citep[][for low- and high-mass objects, respectively]{Bacmann+02_aap389_6,Giannetti+14_aa570_65}. In this case, a significant fraction of the emission comes from the external layers of the clumps, where the density is low and CO isotopologues can be subthermally excited; in addition, the volume of gas traced by the (3--2) transition may be smaller than that of the (1--0), leading to lower measured temperatures. All 70w sources and $\sim70\%$ of clumps classified as IRw have a temperature as measured by CO below $20\kel$, whereas the vast majority of IRb and \hii\ objects have temperatures in excess of this threshold.
			At $L/M \sim 10\lsun \msun^{-1}$ 
			the temperature of the dense gas in the inner envelope is raised above $20-25\kel$
			and CO is thus released from the grain surfaces. In this regime the emission is dominated by the warm, dense gas in the clump, and the temperatures are broadly consistent with methyl acetylene and acetonitrile, but are characterised by larger uncertainties. This is in agreement with the discussion in Sect.~\ref{sec:CO_T_distr}.

			\paragraph{NH$_{3}$:} Ammonia is not ideal to trace the gas warm-up throughout the star formation process, at least when using only the (1,1) and (2,2) inversion transitions. It is clear from Fig.~\ref{fig:ecdf_T_DV_theta_N} that only cold gas is traced, with temperatures similar to the average ones obtained by the SED fit, but with a less clear separation between the evolutionary classes. This is due to a combination of optical depth effects and the limited temperature sensitivity of the (1,1) and (2,2) ratio.

			\medskip
			
\begin{figure}
				\includegraphics[width=\columnwidth]{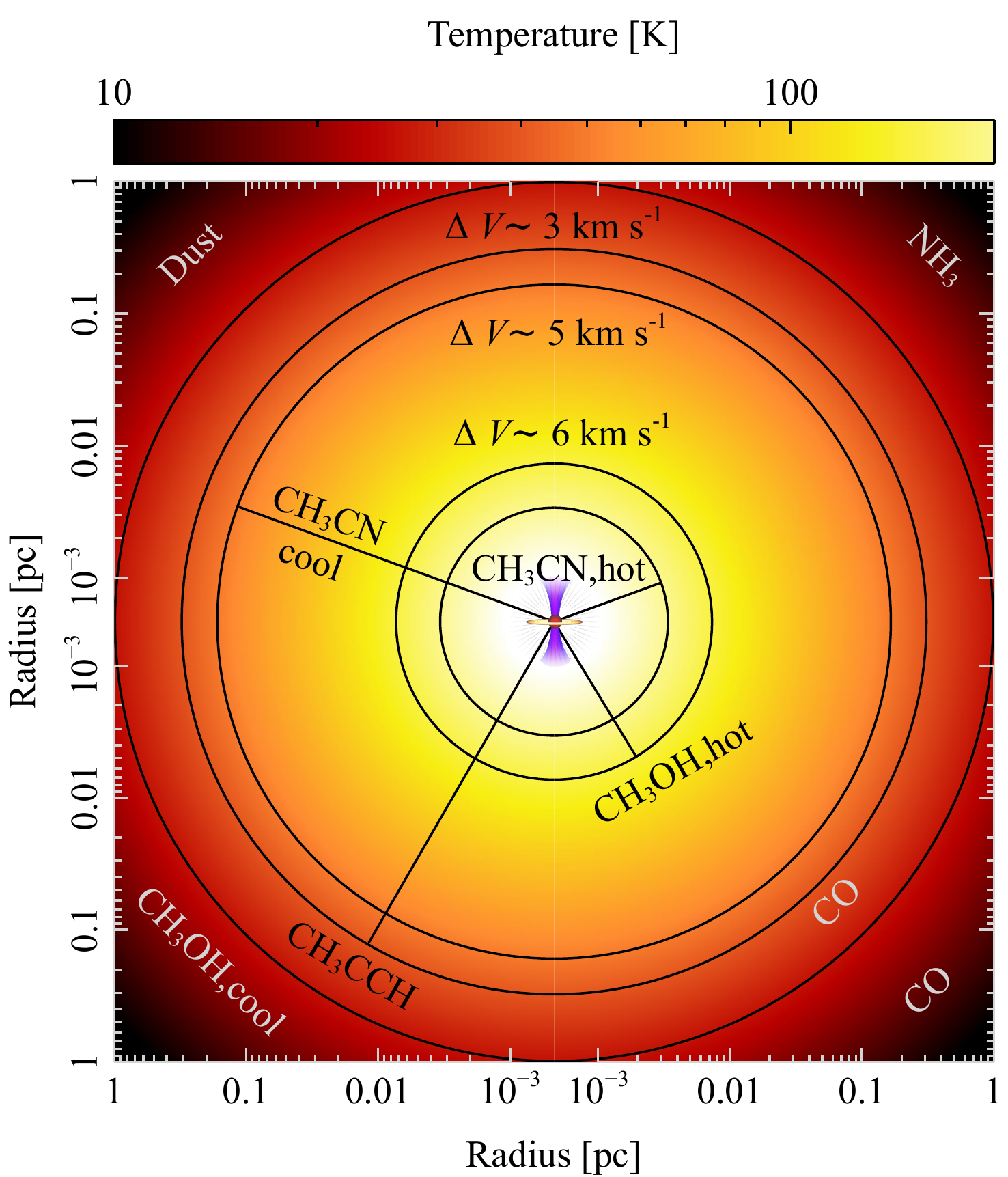}
				\caption{Schematic view of the region of the clump and the typical temperatures traced by different molecules. This represents a clump with the typical properties traced by the different molecules, as observed in this work (cf. Fig.~\ref{fig:ecdf_T_DV_theta_N}). The circles indicate the approximate radius at the median temperature traced by the species marked near to it (cf. Table~\ref{tab:typical_properties}), according to the colourscale. CO appears twice as it traces the outskirts of clumps when the temperature in the inner region is below $25\kel$, and the bulk of the dense gas when the temperature is above this threshold (see text).}\label{fig:summary_tracers}
			\end{figure}

            To summarise, CH$_{3}$CCH is the best tracer of the physical conditions in the dense envelope, closely followed by CH$_{3}$CN, if the source size and the contribution from the hot component are taken into account. To get a mass-averaged temperature for the entire clump, dust is a good option. Both methanol and acetonitrile are equally efficient in revealing the presence of hot gas. Because we find that high-$J$ and high-$K$ lines of \an\ and torsionally excited lines of \mt\ are not significantly contaminated by the presence of warm and cold gas, we propose that these lines are well suited to study the kinematics of the hot gas in the close proximity of young stellar objects. Table~\ref{tab:summary_tracers} and Fig.~\ref{fig:summary_tracers} provide quick references to this summary.

        \subsection{Dependence of temperature on the $L/M$ ratio: an evolutionary sequence}\label{sec:dependence_lm}
        
			Inspecting the relation between temperature and $L/M$ ratio (Fig.~\ref{fig:lm_temperatures}) and using the classification defined for the TOP100 we identify three intervals in the luminosity-to-mass ratio, corresponding to different stages.
			
			For $L/M\lesssim1-2\lsun\msun^{-1}$ the derived gas temperatures depend only weakly on the ratio, or remains constant. In this interval, only 70w and IRw sources are present.
			Depending on the tracer used, for $L/M\gtrsim2-10\lsun\msun^{-1}$ the relation between the temperature and the luminosity-to-mass ratio becomes steeper, revealing the direct correlation between these quantities in this regime, and showing that thermal feedback of the young stellar objects is now influencing the entire clump.

            A threshold is identified above which hot cores begin to appear, for $L/M\gtrsim5-10\lsun \msun^{-1}$.
            In this work we define a hot core as a source that was detected in the $J = 19 \rightarrow 18$ line series of \an\ and/or in the $\nu_t=1$ band of methanol. The fit results described in Sections~\ref{sec:results_AN} and \ref{sec:results_MT} indicate that hot gas with $T\gtrsim150\kel$ must be present in these sources. 

            In the TOP100 37 objects fulfil these criteria, mainly in the IRb and \hii\ classes. The presence of a hot core is reported in Tables~\ref{tab:fit_res_an_hot} and \ref{tab:fit_res_mt_hot}. There is a very good correspondence between the two tracers, with only one clump detected in \an\ not detected in \mt\ (AGAL305.209+00.206 was not observed in methanol) and three sources detected in \mt, but not in acetonitrile, due to the contamination of CCH. As for the other tracers, the detection rate increases with evolution: no 70w source is detected, $20\%$ of IRw clumps, $41\%$ of IRb and $82\%$ of \hii\ are found to harbour a hot core. Clearly, hot molecular gas is still present in the expanding envelopes of compact \hii\ regions, whose high opacity shields it from the ionising photons of the central source. The higher detection rate is likely a combination of a larger volume of gas significantly heated by the young stellar objects and multiple unresolved sources.
            The temperature in the envelope from \an\ and \ma\ of the $19$ sources that host a hot core, but do not show radio-continuum emission yet, is marginally larger than that found in the remaining IRb, possibly suggesting that these clumps are more evolved.

            The position of the sources with $L/M \gtrsim 10 \lsun \msun^{-1}$ in a $M-L$ plot, as well as their bolometric luminosities, indicate that the first intermediate- or high-mass ZAMS stars have appeared, as discussed also in \citet{Molinari+16_apjl826_8}. Further evidence that this change is caused by the substantial heating from luminous ZAMS stars comes from the work of \citet{Urquhart+13_mnras431_1752,Urquhart+13_mnras435_400,Urquhart+14_mnras443_1555}, investigating in detail the properties of ATLASGAL sources associated with methanol masers and UC\hii\ regions: from Fig.~21 in \citet{Urquhart+14_mnras443_1555} it appears evident that massive YSOs and UC\hii\ are consistent with this value in $L/M$. 
            
            Finally, for $L/M$ values in excess of $\sim40\lsun\msun^{-1}$ \hii\ regions become predominant, marking the point where the dissipation of the parental clump starts to dominate.

			\begin{table*}
				\centering
				\caption{Summary of the properties of the tracers.}\label{tab:summary_tracers}
				\begin{tabular}{lccc}
					\hline
					\hline
					Species                 & Region traced            & Gas traced          & Warm-up sensitivity \\
					\hline    
					\an                     & Dense envelope, hot core & Warm, hot           & Good                \\
					\ma                     & Dense envelope           & Warm                & Good                \\
					\mt                     & Envelope, hot core       & Cold, hot           & Fair                \\
					C$^{17}$O               & Envelope, dense envelope & Cold, warm          & Fair                \\
					NH$_{3}$ (1,1) \& (2,2) & Envelope                 & Cold                & Low                 \\
					\hline
				\end{tabular}
			\end{table*}

		\subsection{Emitting regions and qualitative comparison with chemical models}\label{sec:em_reg_chem_mods}
            
            In Sect.~\ref{sec:results_AN} we find that acetonitrile is characterised by two distinct components, one warm and extended and the other hot and compact, showing that this species is not tracing exclusively the gas in the immediate surroundings of high-mass young stellar objects, in agreement with the findings for NGC 7538 IRS9 by \citet{Oberg+13_apj771_95}. CH$_{3}$CN is therefore not a pure hot molecule, as usually assumed \citep[e.g.][]{Bisschop+07_aap465_913}. As for the origin of the extended emission, astrochemical models predict a significant abundance of acetonitrile in warm gas, following the release of HCN from grains, leading to the gas-phase formation of CH$_{3}$CN, for $T\gtrsim30\kel$ \citep{Garrod+08_apj682_283}.
            If CH$_{3}$CN is formed in the gas phase after evaporation of HCN from dust grains, the temperature traced by acetonitrile should show a minimum above the evaporation temperature of HCN. As discussed in Sect.~\ref{sec:warm_up}, such a threshold effect is seen in Figs.~\ref{fig:ecdf_T_DV_theta_N} and \ref{fig:lm_temperatures}; excluding two detected 70w sources, all clumps have temperature in excess of $20-30\kel$, which also corresponds to a threshold in $L/M\sim2\lsun/\msun$. This minimum average $T$ for the warm component falls below the models predictions, possibly indicating a lower evaporation temperature for HCN or a different desorption mechanism.
            The observations reported in this work are consistent with the predictions of the chemical models concerning the presence of an early and a late abundance peaks, characterising an extended warm component and the hot core, respectively. The measured median abundances of acetonitrile are a factor of $\sim 5-10$ higher than predicted by \citet{Garrod+08_apj682_283} (we consider models with a medium warm-up timescale for this comparison), and their ratio is consistent with the model. Considering the uncertainties in the model and in the observations, predicted and observed abundances for \an\ are in broad agreement.

            \citet{Gratier+13_aap557_101} argue that bright acetonitrile lines can be associated with photon-dominated regions (PDRs), using deep spectra of the Horsehead Nebula taken at the DCO$^{+}$ and HCO emission peaks, associated with dense, relatively cold gas and the PDR, respectively. The authors find abundances 30 times larger in the diffuse and UV-illuminated gas than in the dense core, concluding that the detection of CH$_{3}$CN does not automatically imply high-densities.
            PDRs are associated with strong emission of polycyclic aromatic hydrocarbons (PAHs), which have strong bands at $8\mum$. Using the fluxes at this wavelength from the SEDs, taken from \koenigSedt, we can verify whether the extended component is coming from PDRs, rather than warm, dense gas in the envelope, passively heated by the forming high-mass stars. If this were the case, a correlation should exist between eight micron fluxes and integrated line fluxes. Spearman's and Kendall's correlation tests reveal no correlation among these two quantities; a stronger correlation is found between the line flux and the bolometric luminosity of the source ($\rho=0.4$, $\tau=0.3$, with p-values below $5\times\pot{-4}$), dominated by the high-mass YSOs, indicating that passive heating of the envelope is connected to the presence of acetonitrile in the gas phase.

\begin{figure}
				\includegraphics[width=\columnwidth]{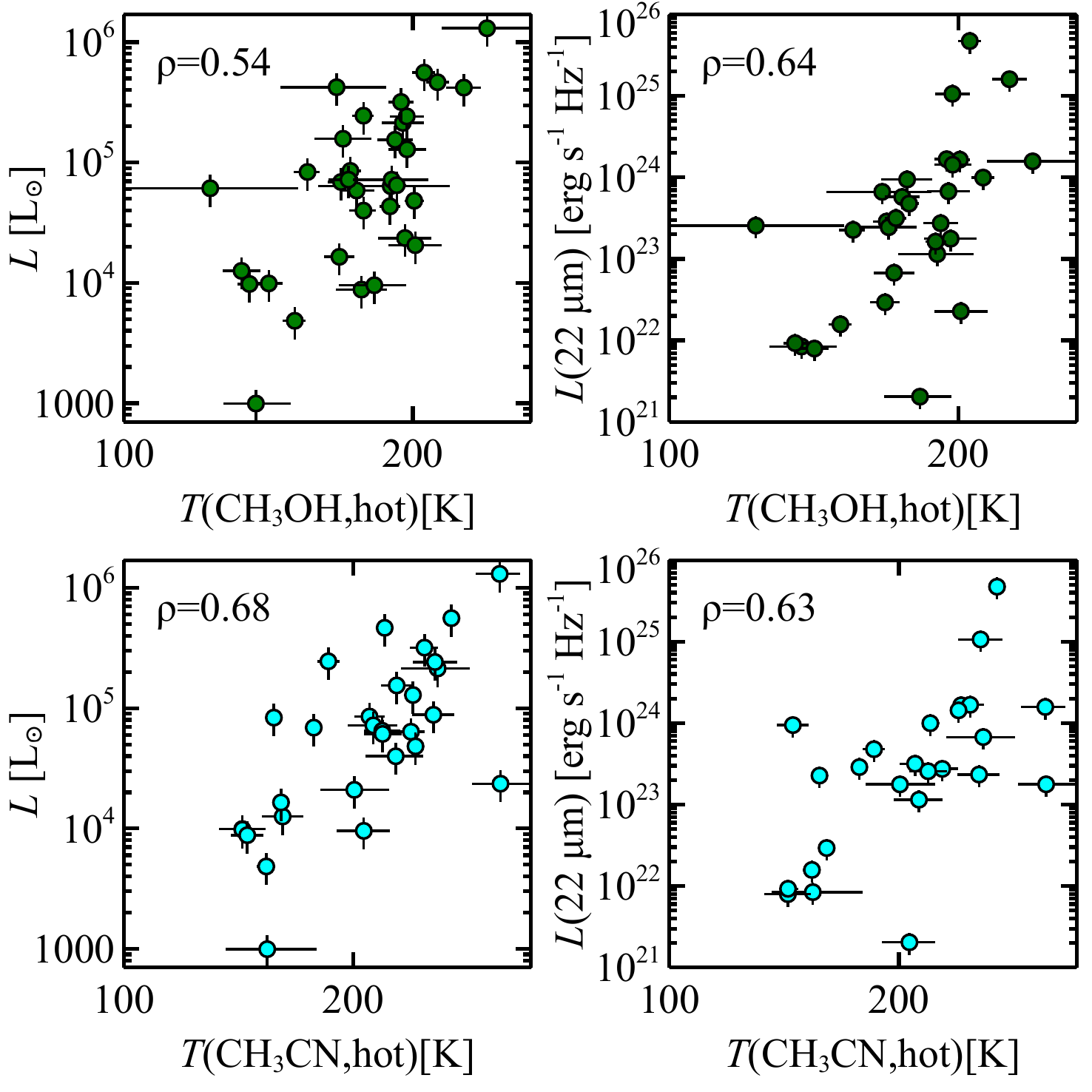}
				\caption{Relations between the temperature of the hot components of \an\ and \mt, and the bolometric- or $22\mum$ luminosity. In the top right corner of each panel the Spearman $\rho$ is indicated.}\label{fig:Lnu_Thot}
			\end{figure}

			\citet{Leurini+07_aap466_215} show that to effectively populate the \mt\ torsionally-excited states radiative pumping is required. The necessity of an intense radiation field to excite these transitions makes them reliable tracers of hot cores that need the radiative and mechanical feedback from nascent high-mass YSOs. 
			The energy of the first torsionally-excited transitions of methanol corresponds to a wavelength range of $\sim20-35\mum$. If the photons at this wavelength are responsible for the excitation of the $\nu_t=1$ band at $\sim337~\giga\hertz$ band, and CH$_3$OH, $\nu_t=1$ is thermalised with the radiation, an increase in the luminosity in the previously mentioned wavelength range leads to a higher temperature, as traced by these transitions.
			The closest measured continuum fluxes are those of WISE, MIPSGAL and MSX ($21-24\mum$), and a Spearman test demonstrates that the temperature of the hot methanol component is more strongly correlated with the luminosity at these wavelengths, than with the bolometric luminosity of the source ($\rho\sim0.55$ and $\rho\sim0.65$, respectively, see Fig.~\ref{fig:Lnu_Thot}). In contrast, high-excitation transitions from CH$_3$CN are more tightly correlated with $L_{bol}$, suggesting that passive heating plays a more important role in this case. 
			To conclude, the data indicate that methanol torsionally-excited lines and high-excitation transitions of CH$_3$CN trace the hot core, but they are sensitive to the radiation field and to the passive heating from the massive YSO, respectively.
			For \mt, the observations are not in agreement with the models of \citet{Garrod+08_apj682_283}. In cold gas, the abundance of CH$_{3}$OH is much larger than predicted by the model (even considering subthermal excitation, see Sections~\ref{sec:results_MT} and \ref{sec:warm_up}) and lower for the gas in the hot core, reflecting known problems related to desorption of this molecule and its overproduction on the grains \citep[][]{Garrod+08_apj682_283}. However, in the hot gas, the measured abundance is consistent with the reduced ice abundance models described in \citet[][]{Garrod+08_apj682_283}.
			
            Our results indicate that the emitting region of \ma\ is extended, and that it traces predominantly cold and warm gas in the envelope; as discussed in Sect.~\ref{sec:results_MA}, we see no indications for a hot component suggesting a strong increase in abundance in hot cores, in contrast to CH$_{3}$CN and CH$_{3}$OH. A jump in abundance at temperatures higher than $\approx 20\kel$, analogous to CO, is also excluded by the gradual, regular increase in temperature observed from the ECDFs (cf. Fig.~\ref{fig:ecdf_temp_class} and Sect.~\ref{sec:dependence_lm}). For temperatures lower than this value, the possibility remains that methyl acetylene is mostly locked in grains.
            The extended nature of the emitting region and the absence of a strong jump in abundance for CH$_{3}$CCH in hot gas is confirmed by interferometric observations of a small sample of high-mass star-forming regions in \citet{Oberg+13_apj771_95} and \citet{Fayolle+15_aap576_45}. These works find that only a minor fraction ($10-15\%$) of the single-dish line flux is recovered by interferometric data on scales of $4\arcsec$, showing that most of the emission is originated in the envelope, also for high-excitation lines.
            A strong debate is ongoing in literature about how CH$_{3}$CCH is formed, focusing on whether or not surface chemistry is necessary to explain the observed abundances of this species \citep[e.g.][]{Occhiogrosso+13_mnras432_3423, Hickson+16_MAs3_1}. 
            We find that the detection rate of CH$_{3}$CCH increases with evolutionary class, but we find no correlation between the temperature traced by methyl acetylene and its abundance, even when considering only sources with distances below $6\kpc$, which would indicate a progressively more efficient evaporation from the grains due to the warm-up.
            This finding is in contrast with the results form \citet{Miettinen+06_aap460_721}, but is based on a much larger number of sources.  
            Therefore, if methyl acetylene is formed onto dust grains, it needs efficient routes of formation at low temperatures, such as the progressive hydrogenation reactions of $\mathrm{C}_{3}$ proposed by \citet{Hickson+16_MAs3_1}, as well as desorption mechanisms able to release a significant quantity of CH$_{3}$CCH already for $T\sim20\kel$, to explain the scales and temperatures traced.
            Because no clear jump in abundance is seen in hot cores for this species also when the mantle is sublimated, only a small quantity of methyl acetylene must remain in the ice when a temperature of $\sim110\kel$ is reached, either due to reprocessing or nearly complete sublimation at lower temperatures.

			\citet{Giannetti+13_aa556_16} previously compared NH$_3$ (1,1) and (2,2) emission with dust continuum for a sample of massive clumps, using data with comparable spatial resolution. In their maps the emitting regions of these tracers are very similar, and the temperatures are closely related. An extended emitting region for ammonia is in agreement with the properties derived from this species in the TOP100. 

		\subsection{Reliability of \an\ as a thermometer}

			In the previously-mentioned study of the Horsehead nebula, \citet{Gratier+13_aap557_101} find that CH$_{3}$CN is subthermally excited with relevant implications for the derivation of column densities. 
			The authors compare results from large velocity gradient models with traditional rotational diagrams distinguishing three regimes: LTE, moderate- and strong subthermal excitation. The volume densities to have $3\mm$ transitions in LTE should be of the order of a few $\pot{5}\cm^{-3}$, whereas $n_\mathrm{H_{2}}\gtrsim4\times\pot{4}\cm^{-3}$ is sufficient to enter the regime of moderate subthermal excitation, for which the ratio of different $K$ lines is still a good indicator of $\tk$. This is of the order of the median volume density averaged along the line of sight for the TOP100. Independent tests performed with RADEX \citep{VanderTak+07_aap468_627} confirm that $\sim\pot{4}\cm^{-3}$ is sufficient to obtain reliable estimates of $\tk$ in this sample.  
			The tight correlation between temperatures derived with \an\ and \ma\ give additional support to the idea that acetonitrile is a good thermometer in massive and dense clumps.
			
			The $J = 19 \rightarrow 18$ lines have a much higher critical density, however; the high temperatures required to excite these lines ensure that the emission comes predominantly from the hot core, where densities are much higher \citep[$\gtrsim\pot{7-8}\cm^{-3}$,][]{GarrodWeaver13_ChRv113_8939}, again providing a reliable temperature probe for these regions.
			
			Although the derived temperatures are reliable, column densities may be underestimated by a factor of a few, according to the discussion in \citet{Gratier+13_aap557_101}. 

        \subsection{Evolution of linewidths and volume densities}\label{sec:evol_DV_nH2}
    
            \begin{figure}
                \centering
                \includegraphics[width=0.66\columnwidth]{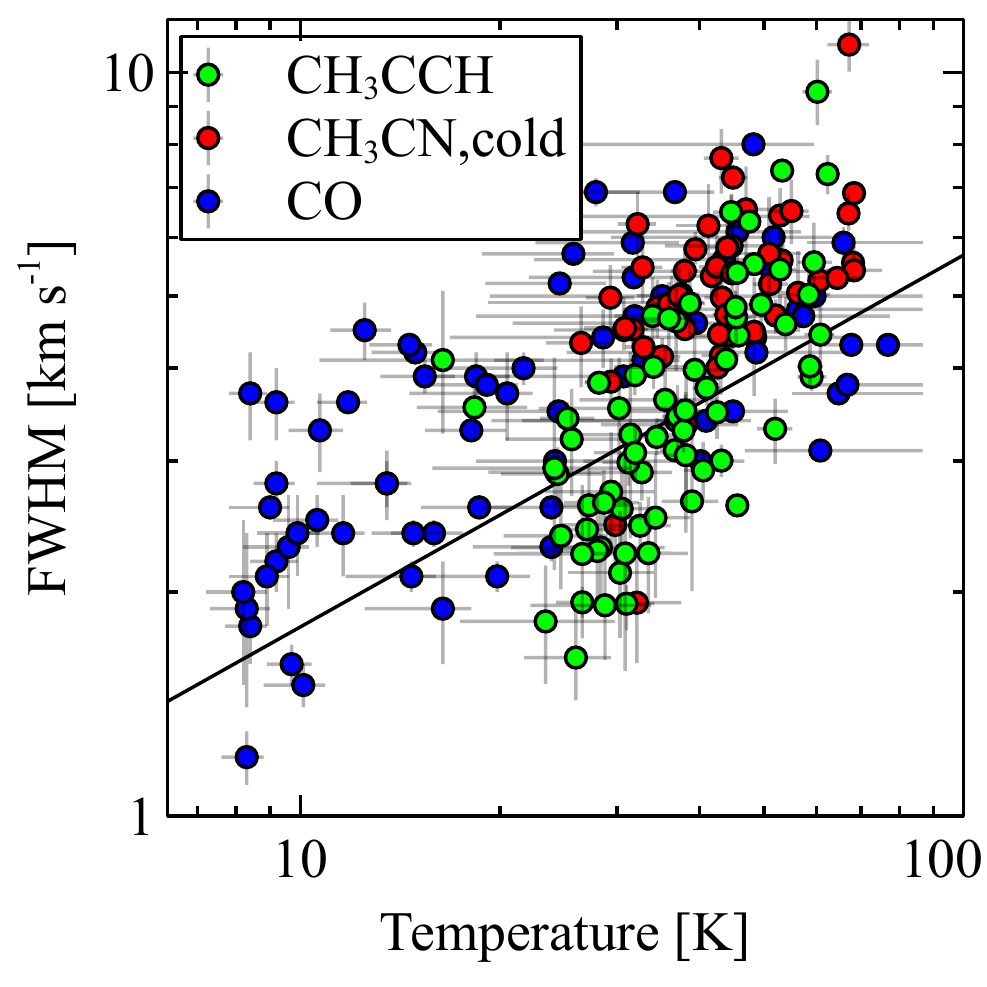}   
                \caption{Relation between temperature and line FWHM for CH$_{3}$CN, CH$_{3}$CCH and CO. The solid line indicate a mach number of 4.}\label{fig:temperatures_DV}
            \end{figure}
         
            The forming stars continuously inject turbulence in the surrounding medium through outflows and winds. This leads to a local increase of the turbulent linewidth. Thus, as time passes and temperature increases, lines also become broader. Figure~\ref{fig:temperatures_DV} shows that for CH$_{3}$CN, CH$_{3}$CCH, and C$^{17}$O the measured gas temperatures also correlate with linewidths, that are always supersonic. \an\ and \ma\ linewidths are less contaminated by external layers in the envelope, and have a steeper dependence on the temperatures, becoming progressively more supersonic. The linewidths can also be broadened by infall motions: \citet{Wyrowski+16_aap585_149} use ammonia absorption to investigate the presence of infall in a sample of high-mass star-forming regions, finding that $72\%$ of the sources show infall signatures and that the associated velocity is $3-30\%$ of the free-fall value. Assuming a typical $10\%$ of the free-fall velocity, we obtain an infall velocity of $\approx0.3\kms$ for the clumps in the TOP100, which has a negligible effect in terms of broadening. \citet{Hajigholi+16_aap585_158}, however, show that in the inner regions of the envelope the infall proceeds with a velocity $0.4-0.7$ of the free-fall value, which is also in line with a progressive broadening of the linewidth on smaller scales.

            \begin{figure}
                \centering
                \includegraphics[width=0.66\columnwidth]{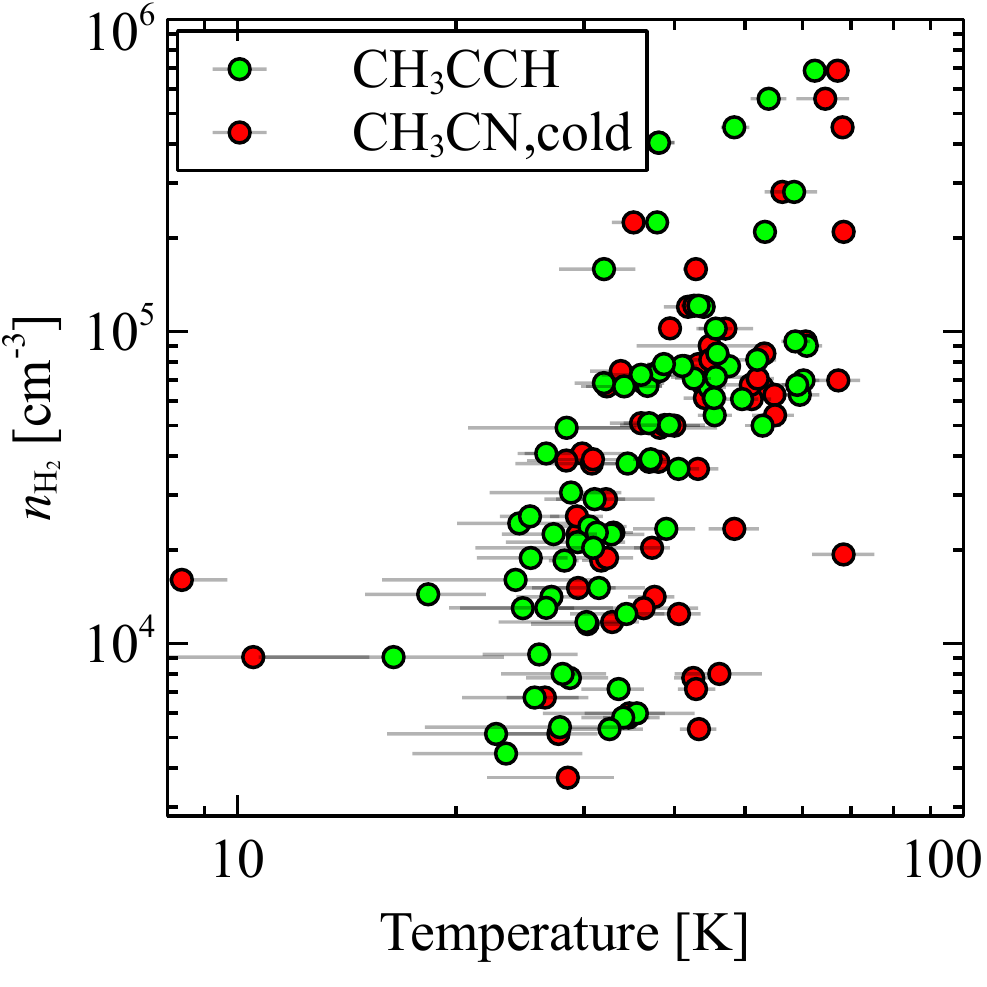}
                \caption{Relation between volume density of molecular hydrogen at the peak of submm emission and temperature. This relation also shows that $n_\mathrm{H_2}$ increases with evolution.}\label{fig:temperatures_nh2}
            \end{figure}

            Figure~\ref{fig:temperatures_nh2} illustrates that a relation exists between the temperature in the inner part of the envelope and the volume density of molecular hydrogen, calculated at the peak of continuum emission as $n_\mathrm{H_2} = N_\mathrm{H_2}/2R$, where $R$ is the source radius, as measured from dust. Following the discussion in Sect.~\ref{sec:dependence_lm}, we deduce that the density in the central part of the source increases with evolution, due to the progressive accumulation of material at the bottom of the clump potential well.
            
            These trends lend further support to the statistical validity of the evolutionary schemes analysed in this work.

        \subsection{Gravitational stability: best tracers}

            The virial stability of the clumps in the TOP100 was discussed in detail in \citet{Giannetti+14_aa570_65}, and successively re-analised in light of the more accurate masses from the SED fit \koenigSed, using C$^{17}$O~(3--2) to derive linewidths. In this work we compared several molecular tracers; depending on which species is used to measure the linewidth, variations of a factor of $\sim 2$ are found in the median values for the FWHM in cold and warm gas. As a consequence, the virial mass varies of a factor of 4, as it depends on the square power of the linewidth, an uncertainty that has to be added to those for size and density profile.
			
            Our results show that when investigating the stability of whole clumps, methyl acetylene is the most reliable tracer, as it comes from the dense part of the clump envelope, similar to that traced by dust continuum; CH$_3$CCH also show no strong jumps in abundance across the evolutionary sequence.
            C$^{17}$O~(3--2) linewidths are in general well correlated with those of CH$_3$CCH, however, it tends to slightly overestimate the linewidths in cold sources. Therefore, this transition also allows to reliably derive virial masses, except in the earliest evolutionary stages.
            CH$_3$CN is more appropriate for the region surrounding the high-mass YSOs, and shows linewidths that are, on average, a factor $1.5$ larger than those measured with methyl acetylene, due to its sharp variations in abundance connected to star-formation activity.
            The FWHM measured by the low-excitation methanol lines have very little correlation with clump properties, and are likely strongly influenced by the presence of outflows \citep[cf.][]{Leurini+16_aap595_4}. Therefore, they are not reliable for this kind of analysis.
            The linewidths of methanol torsionally-excited lines and from CH$_3$CN~(19--18) are well correlated, and both these species can be used to test the stability of the hot core, provided that a measure of the mass and of the emitting region is available.             
                        
    \section{Conclusions and summary}
            
        The physical conditions traced by several commonly-used tracers were compared for the TOP100, a representative sample of high-mass star-forming regions in the disk of our Galaxy, identifying the most reliable probes for the envelope and for hot cores.
        
        \begin{figure*}
			\centering
			\includegraphics[width=\textwidth]{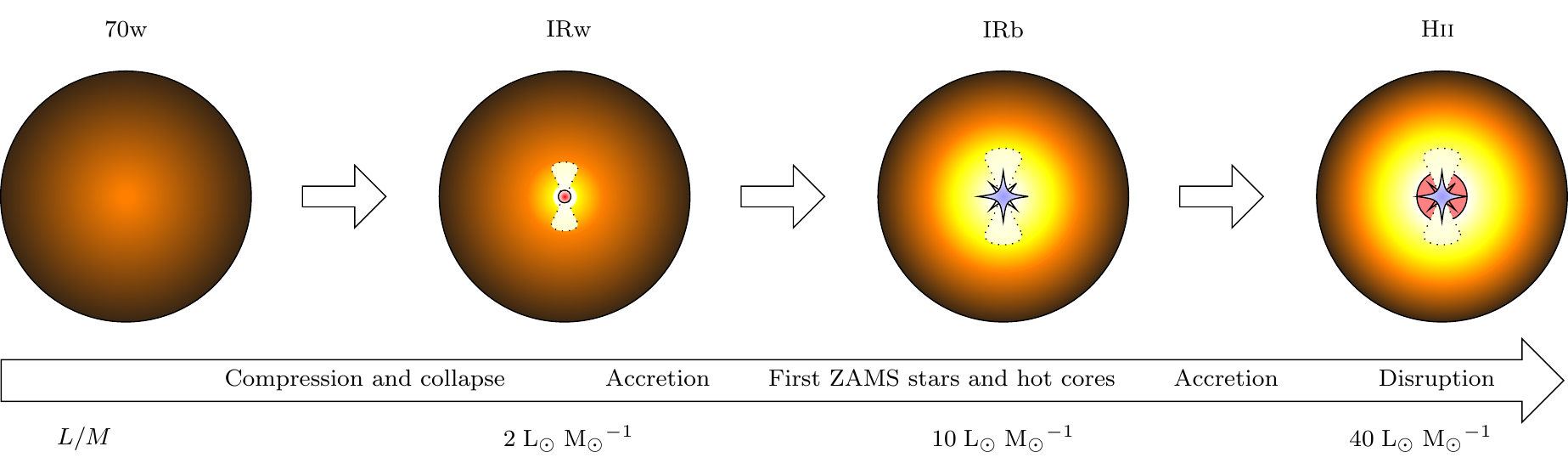}   
			\caption{Simplified representation of the different phases identified for the TOP100, summarising our results. The colourscale indicates the progressive warm-up of the clump around a young stellar object. The physical processes dominating in the corresponding interval of $L/M$ are also shown.}\label{fig:summary_evol}
		\end{figure*}

        The use of this sample to study the physical conditions of the gas and investigate whether the common paradigms used to classify high-mass star-forming regions in terms of evolution are consistent with the expected warm-up in time.
        From the comparison of the different tracers we find that:
            different tracers can probe widely different volumes of gas and physical properties (cf. Fig.~\ref{fig:summary_tracers}), spanning temperatures between $\sim10\kel$ and $\sim300\kel$, and linewidths in the range $1-15\kms$. 
            A major role in this sense is played by the specific chemistry of each species (formation routes, depletion, role of surface chemistry and evaporation from grains, etc.), which also influences their sensitivity to the warm-up process.
            In detail, CH$_{3}$CCH show no significant jump in abundance and has narrow linewidths, indicating that it is tracing the dense and relatively quiescent material in the envelope. CH$_{3}$CN is not exclusively a hot core tracer, but it is abundant in warm gas, associated with star-forming clumps as well. Because of its chemical properties, this molecule is an efficient tracer of the warm-up process, from its beginning to the development of a hot core. CH$_{3}$OH traces both hot cores and cold gas, but it is a reliable probe of temperature only in the first case. 
            C$^{17}$O is a fair probe of physical conditions, especially for evolved sources, but the uncertainties are much larger when compared to those obtained by methyl acetylene and acetonitrile. 
            NH$_{3}$ is not a good tracer of warm-up when using only the (1,1) and (2,2) inversion transitions.
            
            The detection of torsionally-excited methanol and high-excitation CH$_{3}$CN lines ($J=19\rightarrow18$ was used in this work) implies temperatures in excess of $150\kel$, presenting an easy way to identify hot cores. CH$_{3}$OH~$\nu_{t}=1$ levels are pumped by IR radiation from high-mass YSOs, whereas the CH$_{3}$CN~(19--18) line series is excited by collision. The negligible contamination of cold and warm gas to high-$J$ and high-$K$ lines of acetonitrile and to torsionally-excited lines of methanol make these transitions suitable to study the kinematics of the hot gas in the immediate surroundings of young stellar objects.
            We compared the abundance of \an\ and \mt\ with those predicted by the models described in \citet{Garrod+08_apj682_283}. The hot core and envelope abundances for CH$_{3}$CN are $\sim4\times\pot{-10}$ and $\pot{-7}$, respectively; the abundance ratio in hot and warm gas is in broad agreement with the models. 
            On the contrary, the abundance of CH$_{3}$OH is much larger than predicted by the model for cold gas; for hot gas, \mt\ abundance is consistent with the predictions of \citet[][]{Garrod+08_apj682_283} only when using the reduced ice abundance models.

        Linking these findings with potential ways of defining an evolutionary sequence, we conclude that both the $L/M$ ratio and the classification based on the IR and radio continuum properties of the source define a statistically valid evolutionary sequence, demonstrated by the progressive warm-up of the gas in the clump, as well as increase in linewidths and volume density of molecular hydrogen. 
        The different phases of the process can thus be separated as follows (Fig.~\ref{fig:summary_evol}):
        \begin{itemize}
            \item $L/M\lesssim2\lsun\msun^{-1}$: material is still being accumulated and compressed (see Sect.~\ref{sec:warm_up}, \ref{sec:dependence_lm} and the discussion in \citealt{Molinari+16_apjl826_8}), as suggested by the low detection rate of CH$_{3}$CN and, to a lower extent, CH$_{3}$CCH. This is in agreement with the higher volume density needed to excite acetonitrile lines, and with the subthermal excitation of CO. In this interval of $L/M$ ratios (proto-)stellar activity is just starting: only two sources ($17\%$) are associated with methanol masers, and a limited fraction show weak IR emission. We are therefore observing the compression and collapse phases.
            \item $2\lsun\msun^{-1}\lesssim L/M \lesssim 40\lsun\msun^{-1}$: the young stellar objects accrete material and grow in mass, reaching the ZAMS at $L/M\sim10\lsun\msun^{-1}$. When they begin to burn hydrogen, they significantly increase their feedback on the surrounding environment (see Sect.~\ref{sec:dependence_lm}); hot cores and compact \hii\ regions start to appear around this value in $L/M$ as well. After reaching the ZAMS the young stellar object is still accreting mass. The fraction of sources associated with a maser increases substantially in this interval of $L/M$: it is $33\%$ below $L/M\sim10\lsun\msun^{-1}$ and $80\%$ above this threshold.
            \item $L/M\gtrsim40\lsun\msun^{-1}$: radio continuum becomes common, marking the point where dissipation starts to dominate. The detection rate of methanol masers remains stable at $\sim 80\%$.
        \end{itemize}

        These findings constitute the foundations for a complete study of the process of high-mass star formation throughout our Galaxy: 
        it is now possible to identify virtually all massive star-forming regions in the Galaxy (as seen by ATLASGAL, Urquhart et al., in prep.) and classify them in terms of evolution, as an example, to investigate the timescales of the process and the statistical lifetime of each phase.

\begin{acknowledgements}
This work was carried out within the Collaborative Research Council 956, sub-project A6, funded by the Deut\-sche For\-schungs\-ge\-mein\-schaft (DFG). This paper is based on data acquired with the Atacama Pathfinder EXperiment (APEX). APEX is a collaboration between the Max Planck Institute for Radioastronomy, the European Southern Observatory, and the Onsala Space Observatory. 
This work also make use of observations carried out with the Mopra telescope and with the IRAM 30m Telescope, under project number 181-10. The Mopra radio telescope is part of the Australia Telescope National Facility which is funded by the Australian Government for operation as a National Facility managed by CSIRO. IRAM is supported by INSU/CNRS (France), MPG (Germany) and IGN (Spain).
T.Cs. acknowledges support from the DFG via the SPP (priority programme) 1573 'Physics of the ISM'. 
This research made use of Astropy, a community-developed core Python package for Astronomy \citep[][\url{http://www.astropy.org}]{astropy_2013}, of NASA's Astrophysics Data System, of Matplotlib \citep{Hunter_2007_matplotlib}, and of the Veusz plotting package (\url{http://home.gna.org/veusz/}).
We thank the anonymous referee for their constructive and useful report that allowed to improve the clarity of this work.
\end{acknowledgements}

\bibliographystyle{bibtex/aa}
\bibliography{bibtex/biblio.bib}

\Online
    
\begin{appendix}
    \section{Tables}\label{apdx:tables}

\begin{longtab}
\tiny
\onecolumn
\begin{landscape}
\begin{longtable}{l*{10}{r}}
\caption{Properties of the sources in the TOP100.\label{tab:source_properties}}\\
\hline \hline
Source & $l$ & $b$ & Distance & $L_{bol}$ & $M$ & $N_\mathrm{H_2}$ & Class & \an & \ma & \mt \\
 & [$\deg$] & [$\deg$] & [kpc] & [$\lsuntab$] & [$\pot{2}\times\msuntab$] & [$\pot{22}\times\cm^{-2}$] &  &  &  &  \\
\hline
\endfirsthead
\caption{Continued.}\\
\hline \hline
Source & $l$ & $b$ & Distance & $L_{bol}$ & $M$ & $N_\mathrm{H_2}$ & Class & \an & \ma & \mt \\
 & [$\deg$] & [$\deg$] & [kpc] & [$\lsuntab$] & [$\pot{2}\times\msuntab$] & [$\pot{22}\times\cm^{-2}$] &  &  &  &  \\
\hline
\endhead
\hline
\endfoot
\hline
\endlastfoot
AGAL008.684--00.367 & 8.6834 & --0.3675 & 4.78 & 27500 & 14.8 & 8.7 & IRw & Observed & Observed & Observed \\
AGAL008.706--00.414 & 8.7064 & --0.4136 & 4.78 & 504 & 16.6 & 6.1 & IRw & Observed & Observed & Observed \\
AGAL010.444--00.017 & 10.4446 & --0.0178 & 8.55 & 11314 & 16.1 & 5.0 & IRw & Observed & Observed & Observed \\
AGAL010.472+00.027 & 10.4722 & 0.0277 & 8.55 & 465417 & 104.6 & 46.8 & HII & Observed & Observed & Observed \\
AGAL010.624--00.384 & 10.6237 & --0.3833 & 4.95 & 422757 & 37.9 & 37.3 & HII & Observed & Observed & Observed \\
AGAL012.804--00.199 & 12.8057 & --0.1994 & 2.4 & 246869 & 18.8 & 36.4 & HII & Observed & Observed & Observed \\
AGAL013.178+00.059 & 13.1768 & 0.0599 & 2.4 & 8318 & 3.7 & 7.0 & 70w & Observed & Observed & Observed \\
AGAL013.658--00.599 & 13.657 & --0.5992 & 4.47 & 20956 & 5.7 & 6.3 & IRb & Observed & Observed & Observed \\
\end{longtable}
\end{landscape}
\end{longtab}

\clearpage

\setcounter{table}{1}

\begin{longtab}
\tiny
\onecolumn
\begin{longtable}{l*{8}{r}}
\caption{Physical parameters for \ma.\label{tab:fit_res_ma}}\\
\hline \hline
Source & $T$ & $T 95\%$~HPD & $log(N)$ & $log(N) 95\%$~HPD & $\vartheta$ & $\vartheta 95\%$~HPD & $\Delta V$ & $\Delta V 95\%$~HPD \\
 & [K] & [K] & $[log(\cm^{-2})]$ & $[log(\cm^{-2})]$ & $[\arcsec]$ & $[\arcsec]$ & $[\kmstab]$ & $[\kmstab]$ \\
\hline
\endfirsthead
\caption{Continued.}\\
\hline \hline
Source & $T$ & $T 95\%$~HPD & $log(N)$ & $log(N) 95\%$~HPD & $\vartheta$ & $\vartheta 95\%$~HPD & $\Delta V$ & $\Delta V 95\%$~HPD \\
 & [K] & [K] & $[log(\cm^{-2})]$ & $[log(\cm^{-2})]$ & $[\arcsec]$ & $[\arcsec]$ & $[\kmstab]$ & $[\kmstab]$ \\
\hline
\endhead
\hline
\endfoot
\hline
\endlastfoot
AGAL008.684--00.367 & 28.2 & 26.8 - 29.5 & 15.11 & 15.06 - 15.17 & 52.4 & 46.6 - 59.1 & 3.8 & 3.7 - 3.9 \\
AGAL008.706--00.414 & 26.0 & 21.7 - 29.4 & 15.57 & 15.12 - 15.92 & 5.3 & 3.8 - 8.0 & 1.6 & 1.4 - 1.9 \\
AGAL010.444--00.017 & 35.5 & 26.3 - 42.6 & 15.03 & 14.42 - 15.45 & 9.5 & 5.6 - 19.0 & 3.6 & 3.1 - 4.2 \\
AGAL010.472+00.027 & 44.6 & 41.9 - 47.0 & 15.22 & 15.16 - 15.27 & 54.5 & 46.9 - 62.6 & 6.5 & 6.1 - 6.8 \\
AGAL010.624--00.384 & 47.5 & 46.1 - 49.0 & 15.43 & 15.40 - 15.47 & 75.8 & 70.3 - 81.9 & 6.3 & 6.1 - 6.5 \\
AGAL012.804--00.199 & 43.8 & 42.4 - 45.2 & 15.36 & 15.33 - 15.39 & 136.6 & 126.4 - 148.0 & 4.1 & 4.0 - 4.2 \\
AGAL013.178+00.059 & 32.7 & 28.4 - 36.3 & 14.62 & 14.47 - 14.76 & 34.1 & 24.9 - 46.7 & 2.9 & 2.7 - 3.1 \\
AGAL013.658--00.599 & 27.1 & 24.2 - 30.0 & 14.55 & 14.45 - 14.64 & 52.5 & 39.6 - 67.9 & 2.4 & 2.2 - 2.6 \\
\hline
\end{longtable}
\end{longtab}

\clearpage
\setcounter{table}{2}

\begin{longtab}
\tiny
\onecolumn
\begin{longtable}{l*{8}{r}}
\caption{Velocity and calibration factors for \ma.\label{tab:fit_res_ma_cf}}\\
\hline \hline
Source & $\vlsr$ & $95\%$~HPD & $CF(5-4)$ & $95\%$~HPD & $CF(6-5)$ & $95\%$~HPD & $CF(20-19)$ & $95\%$~HPD \\
 & $[\kmstab]$ & $[\kmstab]$ &  &  &  &  &  &  \\
\hline
\endfirsthead
\caption{Continued.}\\
\hline \hline
Source & $\vlsr$ & $95\%$~HPD & $CF(5-4)$ & $95\%$~HPD & $CF(6-5)$ & $95\%$~HPD & $CF(20-19)$ & $95\%$~HPD \\
 & $[\kmstab]$ & $[\kmstab]$ &  &  &  &  &  &  \\
\hline
\endhead
\hline
\endfoot
\hline
\endlastfoot
AGAL008.684--00.367	&	37.70	&	37.64	-	37.76	&	1.08	&	0.98	-	1.19	&	0.91	&	0.82	-	1.00	&	1.00	&	0.87	-	1.14	\\
AGAL008.706--00.414	&	38.69	&	38.60	-	38.77	&	1.17	&	1.07	-	1.28	&	0.76	&	0.70	-	0.84	&	1.00	&	0.87	-	1.14	\\
AGAL010.444--00.017	&	75.81	&	75.48	-	76.10	&	1.00	&	0.88	-	1.13	&	0.99	&	0.86	-	1.12	&	1.01	&	0.88	-	1.15	\\
AGAL010.472+00.027	&	66.59	&	66.46	-	66.74	&	1.00	&	0.89	-	1.10	&	0.96	&	0.86	-	1.06	&	1.05	&	0.92	-	1.17	\\
AGAL010.624--00.384	&	--2.91	&	--2.98	-	--2.85	&	0.93	&	0.84	-	1.02	&	0.93	&	0.84	-	1.02	&	1.12	&	1.02	-	1.23	\\
AGAL012.804--00.199	&	34.79	&	34.75	-	34.83	&	1.19	&	1.09	-	1.28	&	0.80	&	0.74	-	0.87	&	0.94	&	0.82	-	1.04	\\
AGAL013.178+00.059	&	48.97	&	48.86	-	49.08	&	1.05	&	0.94	-	1.17	&	0.93	&	0.82	-	1.03	&	1.02	&	0.88	-	1.15	\\
AGAL013.658--00.599	&	47.46	&	47.40	-	47.53	&	1.07	&	0.96	-	1.18	&	0.93	&	0.82	-	1.02	&	0.99	&	0.85	-	1.12	\\
\hline
\end{longtable}
\end{longtab}

\clearpage
\setcounter{table}{3}
	
\begin{longtab}
\tiny
\onecolumn
\begin{longtable}{l*{8}{r}}
\caption{Physical parameters of the cool component for \an.\label{tab:fit_res_an_cold}}\\
\hline \hline
Source & $T(\mathrm{cool})$ & $95\%$~HPD & $log[N(\mathrm{cool})]$ & $95\%$~HPD & $\vartheta(\mathrm{cool})$ & $95\%$~HPD & $\Delta V(\mathrm{cool})$ & $95\%$~HPD \\
 & [K] & [K] & $[log(\cm^{-2})]$ & $[log(\cm^{-2})]$ & $[\arcsec]$ & $[\arcsec]$ & $[\kmstab]$ & $[\kmstab]$ \\
\hline
\endfirsthead
\caption{Continued.}\\
\hline \hline
Source & $T(\mathrm{cool})$ & $95\%$~HPD & $log[N(\mathrm{cool})]$ & $95\%$~HPD & $\vartheta(\mathrm{cool})$ & $95\%$~HPD & $\Delta V(\mathrm{cool})$ & $95\%$~HPD \\
 & [K] & [K] & $[log(\cm^{-2})]$ & $[log(\cm^{-2})]$ & $[\arcsec]$ & $[\arcsec]$ & $[\kmstab]$ & $[\kmstab]$ \\
\hline
\endhead
\hline
\endfoot
\hline
\endlastfoot
AGAL008.684--00.367 & 31.7 & 30.0 - 33.5 & 13.49 & 13.42 - 13.56 & 39.0 & 34.0 - 44.8 & 4.5 & 4.3 - 4.8 \\
AGAL008.706--00.414 & 25.7 & 11.6 - 36.4 & 13.69 & 12.36 - 14.53 & 5.4 & 1.5 - 29.7 & 4.8 & 2.9 - 7.7 \\
AGAL010.444--00.017 & 35.1 & 30.5 - 39.2 & 13.78 & 13.49 - 14.05 & 9.8 & 7.2 - 13.6 & 6.4 & 4.9 - 8.1 \\
AGAL010.472+00.027 & 52.6 & 50.9 - 54.3 & 14.04 & 13.99 - 14.09 & 36.0 & 33.2 - 39.0 & 6.3 & 5.9 - 6.8 \\
AGAL010.624--00.384 & 44.8 & 42.6 - 47.1 & 13.87 & 13.83 - 13.91 & 80.5 & 70.9 - 91.0 & 7.2 & 6.9 - 7.5 \\
AGAL012.804--00.199 & 42.0 & 39.0 - 44.6 & 13.43 & 13.39 - 13.48 & 148.6 & 126.0 - 176.3 & 5.3 & 4.9 - 5.8 \\
AGAL013.178+00.059 & 29.6 & 27.0 - 32.0 & 13.26 & 13.17 - 13.35 & 42.2 & 33.7 - 52.1 & 3.8 & 3.5 - 4.1 \\
AGAL013.658--00.599 & 37.7 & 35.2 - 40.6 & 13.65 & 13.52 - 13.76 & 22.4 & 18.5 - 26.5 & 5.0 & 4.6 - 5.5 \\
\end{longtable}
\end{longtab}

\clearpage
\setcounter{table}{4}

\begin{table}
\tiny
\centering
\caption{Physical parameters of the hot component for \an.\label{tab:fit_res_an_hot}}
\begin{tabular}{l*{8}{r}}
\hline \hline
Source & $T(\mathrm{hot})$ & $95\%$~HPD & $log[N(\mathrm{hot})]$ & $95\%$~HPD & $\vartheta(\mathrm{hot})$ & $95\%$~HPD & $\Delta V(\mathrm{hot})$ & $95\%$~HPD \\
 & [K] & [K] & $[log(\cm^{-2})]$ & $[log(\cm^{-2})]$ & $[\arcsec]$ & $[\arcsec]$ & $[\kmstab]$ & $[\kmstab]$ \\
\hline
AGAL010.472+00.027 & 219.8 & 215.2 - 223.8 & 18.68 & 18.57 - 18.77 & 1.01 & 0.96 - 1.06 & 6.2 & 5.9 - 6.4 \\
AGAL010.624--00.384 & 285.2 & 191.0 - 388.4 & 16.08 & 15.03 - 16.95 & 0.79 & 0.21 - 1.76 & 8.3 & 6.6 - 10.0 \\
AGAL013.658--00.599 & 198.7 & 180.8 - 220.7 & 17.17 & 16.73 - 17.53 & 0.35 & 0.27 - 0.44 & 6.2 & 4.6 - 8.3 \\
AGAL023.206--00.377 & 132.0 & 126.7 - 138.5 & 17.01 & 16.80 - 17.22 & 0.71 & 0.62 - 0.78 & 2.4 & 2.0 - 2.9 \\
AGAL030.818--00.056 & 235.6 & 225.7 - 245.0 & 17.94 & 17.77 - 18.13 & 0.43 & 0.38 - 0.47 & 8.0 & 6.9 - 9.3 \\
AGAL031.412+00.307 & 167.6 & 163.3 - 171.7 & 17.78 & 17.63 - 17.93 & 1.04 & 0.98 - 1.11 & 4.1 & 3.8 - 4.4 \\
\hline
\end{tabular}
\end{table}

\clearpage

\begin{longtab}
\tiny
\onecolumn
\begin{longtable}{l*{8}{r}}
\caption{Velocity and calibration factors for \an.\label{tab:velo_cf_an}}\\
\hline \hline
Source & $\vlsr$ & $95\%$~HPD & $CF(5-4)$ & $95\%$~HPD & $CF(6-5)$ & $95\%$~HPD & $CF(19-18)$ & $95\%$~HPD \\
 & $[\kmstab]$ & $[\kmstab]$ &  &  &  &  &  &  \\
\hline
\endfirsthead
\caption{Continued.}\\
\hline \hline
Source & $\vlsr$ & $95\%$~HPD & $CF(5-4)$ & $95\%$~HPD & $CF(6-5)$ & $95\%$~HPD & $CF(19-18)$ & $95\%$~HPD \\
 & $[\kmstab]$ & $[\kmstab]$ &  &  &  &  &  &  \\
\hline
\endhead
\hline
\endfoot
\hline
\endlastfoot
AGAL008.684--00.367 & 38.08 & 38.01 - 38.17 & 1.13 & 1.02 - 1.23 & 0.91 & 0.82 - 1.00 & 0.93 & 0.80 - 1.08 \\
AGAL008.706--00.414 & 39.13 & 38.12 - 39.98 & 1.04 & 0.91 - 1.17 & 0.96 & 0.83 - 1.10 & 0.99 & 0.86 - 1.13 \\
AGAL010.444--00.017 & 75.67 & 75.07 - 76.21 & 1.09 & 0.97 - 1.22 & 0.91 & 0.79 - 1.04 & 0.97 & 0.84 - 1.12 \\
AGAL010.472+00.027 & 66.67 & 66.57 - 66.77 & 1.30 & 1.29 - 1.30 & 1.28 & 1.25 - 1.30 & 0.70 & 0.70 - 0.71 \\
AGAL010.624--00.384 & --3.15 & --3.20 - --3.11 & 1.14 & 1.05 - 1.24 & 1.13 & 1.04 - 1.22 & 0.73 & 0.70 - 0.80 \\
AGAL012.804--00.199 & 34.68 & 34.64 - 34.72 & 1.10 & 0.99 - 1.20 & 1.06 & 0.96 - 1.15 & 0.80 & 0.70 - 0.91 \\
AGAL013.178+00.059 & 49.13 & 48.99 - 49.25 & 1.12 & 1.00 - 1.22 & 0.88 & 0.79 - 0.98 & 0.96 & 0.83 - 1.11 \\
AGAL013.658--00.599 & 47.54 & 47.40 - 47.72 & 1.16 & 1.05 - 1.27 & 0.98 & 0.88 - 1.06 & 0.77 & 0.70 - 0.90 \\
\end{longtable}
\end{longtab}
\setcounter{table}{6}

\begin{longtab}
\tiny
\onecolumn
\begin{landscape}
\begin{longtable}{l*{10}{r}}
\caption{Physical parameters of the cool component for \mt. The velocity and calibration factor are also included.\label{tab:fit_res_mt_cold}}\\
\hline \hline
Source & $T(\mathrm{cool})$ & $95\%$~HPD & $log[N(\mathrm{cool})]$ & $95\%$~HPD & $\Delta V(\mathrm{cool})$ & $95\%$~HPD & $\vlsr$ & $95\%$~HPD & $CF$ & $95\%$~HPD \\
 & [K] & [K] & $[log(\cm^{-2})]$ & $[log(\cm^{-2})]$ & $[\kmstab]$ & $[\kmstab]$ & $[\kmstab]$ & $[\kmstab]$ &  &  \\
\hline
\endfirsthead
\caption{Continued.}\\
\hline \hline
Source & $T(\mathrm{cool})$ & $95\%$~HPD & $log[N(\mathrm{cool})]$ & $95\%$~HPD & $\Delta V(\mathrm{cool})$ & $95\%$~HPD & $\vlsr$ & $95\%$~HPD & $CF$ & $95\%$~HPD \\
 & [K] & [K] & $[log(\cm^{-2})]$ & $[log(\cm^{-2})]$ & $[\kmstab]$ & $[\kmstab]$ & $[\kmstab]$ & $[\kmstab]$ &  &  \\
\hline
\endhead
\hline
\endfoot
\hline
\endlastfoot
AGAL008.684--00.367 & 15.9 & 14.6 - 17.3 & 15.15 & 15.03 - 15.28 & 4.8 & 4.5 - 5.1 & 37.99 & 37.88 - 38.12 & 1.01 & 0.87 - 1.14 \\
AGAL008.706--00.414 & 11.2 & 8.4 - 14.3 & 15.14 & 14.82 - 15.62 & 5.2 & 3.7 - 6.8 & 39.07 & 38.64 - 39.53 & 1.00 & 0.86 - 1.13 \\
AGAL010.444--00.017 & 13.4 & 8.0 - 21.4 & 14.77 & 14.25 - 15.56 & 5.9 & 3.8 - 8.1 & 76.32 & 75.48 - 77.08 & 1.00 & 0.87 - 1.13 \\
AGAL010.472+00.027 & 31.5 & 28.1 - 35.0 & 15.12 & 15.07 - 15.19 & 5.9 & 5.6 - 6.3 & 66.32 & 66.21 - 66.43 & 1.15 & 1.03 - 1.26 \\
AGAL010.624--00.384 & 20.1 & 19.1 - 20.9 & 15.87 & 15.78 - 15.96 & 8.4 & 8.2 - 8.6 & --2.71 & --2.75 - --2.66 & 0.92 & 0.78 - 1.06 \\
AGAL012.804--00.199 & 19.8 & 19.1 - 20.6 & 15.24 & 15.17 - 15.31 & 5.4 & 5.4 - 5.5 & 35.66 & 35.60 - 35.73 & 1.01 & 0.87 - 1.14 \\
AGAL013.178+00.059 & 10.6 & 9.9 - 11.3 & 16.02 & 15.86 - 16.18 & 5.4 & 5.1 - 5.6 & 48.85 & 48.74 - 48.95 & 1.01 & 0.88 - 1.14 \\
AGAL013.658--00.599 & 24.6 & 22.5 - 27.1 & 14.91 & 14.84 - 14.99 & 7.3 & 6.7 - 7.8 & 47.53 & 47.35 - 47.69 & 1.01 & 0.86 - 1.15 \\
\hline
\end{longtable}
\end{landscape}
\end{longtab}
\setcounter{table}{7}

\clearpage

\begin{table}
\tiny
\centering
\caption{Physical parameters of the hot component for \mt.\label{tab:fit_res_mt_hot}}
\begin{tabular}{l*{8}{r}}
\hline \hline
Source & $T(\mathrm{hot})$ & $95\%$~HPD & $log[N(\mathrm{hot})]$ & $95\%$~HPD & $\vartheta(\mathrm{hot})$ & $95\%$~HPD & $\Delta V(\mathrm{hot})$ & $95\%$~HPD \\
 & [K] & [K] & $[log(\cm^{-2})]$ & $[log(\cm^{-2})]$ & $[\arcsec]$ & $[\arcsec]$ & $[\kmstab]$ & $[\kmstab]$ \\
\hline
AGAL010.472+00.027 & 211.6 & 205.7 - 217.0 & 19.13 & 19.07 - 19.19 & 1.11 & 1.04 - 1.18 & 7.4 & 7.0 - 7.7 \\
AGAL010.624--00.384 & 166.8 & 147.4 - 188.2 & 16.8 & 16.51 - 17.09 & 3.01 & 2.19 - 4.03 & 5.8 & 5.4 - 6.3 \\
AGAL023.206--00.377 & 132.5 & 126.2 - 138.2 & 18.21 & 18.09 - 18.34 & 0.7 & 0.62 - 0.78 & 16.2 & 14.9 - 17.0 \\

AGAL030.818--00.056 & 189.6 & 159.3 - 218.4 & 18.05 & 17.57 - 18.47 & 0.74 & 0.49 - 1.08 & 14.3 & 12.0 - 16.9 \\

AGAL031.412+00.307 & 139.9 & 137.8 - 141.9 & 18.63 & 18.56 - 18.70 & 1.63 & 1.58 - 1.69 & 6.0 & 5.8 - 6.3 \\
AGAL034.258+00.154 & 199.4 & 194.9 - 204.5 & 18.48 & 18.43 - 18.53 & 1.68 & 1.58 - 1.78 & 6.1 & 5.9 - 6.2 \\
\hline
\end{tabular}
\end{table}

\end{appendix}

\end{document}